\documentstyle[graphicx,times]{mn}

\newif\ifAMStwofonts

\ifoldfss
  \newcommand{\rmn}[1] {{\rm #1}}

  \ifCUPmtlplainloaded \else
    \NewTextAlphabet{textbfit} {cmbxti10} {}
    \NewTextAlphabet{textbfss} {cmssbx10} {}
    \NewMathAlphabet{mathbfit} {cmbxti10} {} 
    \NewMathAlphabet{mathbfss} {cmssbx10} {} 
  \fi
  \ifAMStwofonts
    \ifCUPmtlplainloaded \else
      \NewSymbolFont{upmath} {eurm10}
      \NewSymbolFont{AMSa} {msam10}
      \NewMathSymbol{\upi}     {0}{upmath}{19}
      \NewMathSymbol{\umu}     {0}{upmath}{16}
      \NewMathSymbol{\upartial}{0}{upmath}{40}
      \NewMathSymbol{\leqslant}{3}{AMSa}{36}
      \NewMathSymbol{\geqslant}{3}{AMSa}{3E}

      \let\leq=\leqslant \let\le=\leqslant
       
    \fi
  \fi
\fi 

\ifnfssone
  \newmathalphabet{\mathit}
  \addtoversion{normal}{\mathit}{cmr}{m}{it}
  \addtoversion{bold}{\mathit}{cmr}{bx}{it}
  \newcommand{\rmn}[1] {\mathrm{#1}}

  \newmathalphabet{\mathbfit} 
  \addtoversion{normal}{\mathbfit}{cmr}{bx}{it}
  \addtoversion{bold}{\mathbfit}{cmr}{bx}{it}
  \newmathalphabet{\mathbfss} 
  \addtoversion{normal}{\mathbfss}{cmss}{bx}{n}
  \addtoversion{bold}{\mathbfss}{cmss}{bx}{n}
  \ifAMStwofonts
    \ifCUPmtlplainloaded \else
      \UseAMStwoboldmath
      \makeatletter
      \new@mathgroup\upmath@group
      \define@mathgroup\mv@normal\upmath@group{eur}{m}{n}
      \define@mathgroup\mv@bold\upmath@group{eur}{b}{n}
      \edef\UPM{\hexnumber\upmath@group}
      \new@mathgroup\amsa@group
      \define@mathgroup\mv@normal\amsa@group{msa}{m}{n}
      \define@mathgroup\mv@bold\amsa@group{msa}{m}{n}
      \edef\AMSa{\hexnumber\amsa@group}
      \makeatother
      \mathchardef\upi="0\UPM19
      \mathchardef\umu="0\UPM16
      \mathchardef\upartial="0\UPM40
      \mathchardef\leqslant="3\AMSa36
      \mathchardef\geqslant="3\AMSa3E

      \let\leq=\leqslant \let\le=\leqslant

    \fi
  \fi
\fi 

\ifnfsstwo
  \newcommand{\rmn}[1] {\mathrm{#1}}

  \DeclareMathAlphabet{\mathbfit}{OT1}{cmr}{bx}{it}
  \SetMathAlphabet\mathbfit{bold}{OT1}{cmr}{bx}{it}
  \DeclareMathAlphabet{\mathbfss}{OT1}{cmss}{bx}{n}
  \SetMathAlphabet\mathbfss{bold}{OT1}{cmss}{bx}{n}
  \ifAMStwofonts
    \ifCUPmtlplainloaded \else
      \DeclareSymbolFont{UPM}{U}{eur}{m}{n}
      \SetSymbolFont{UPM}{bold}{U}{eur}{b}{n}
      \DeclareSymbolFont{AMSa}{U}{msa}{m}{n}
      \DeclareMathSymbol{\upi}{0}{UPM}{"19}
      \DeclareMathSymbol{\umu}{0}{UPM}{"16}
      \DeclareMathSymbol{\upartial}{0}{UPM}{"40}
      \DeclareMathSymbol{\leqslant}{3}{AMSa}{"36}
      \DeclareMathSymbol{\geqslant}{3}{AMSa}{"3E}

      \let\leq=\leqslant \let\le=\leqslant

    \fi
  \fi
\fi 

\ifCUPmtlplainloaded \else
  \ifAMStwofonts \else 
    \def\upi{\pi}
    \def\umu{\mu}
    \def\upartial{\partial}
  \fi
\fi

\title[Structure of rich LMC clusters]
  {Surface brightness profiles and structural parameters for 53 rich stellar clusters in the Large Magellanic Cloud}
\author[A.~D.~Mackey \& G.~F.~Gilmore]
  {A.~D.~Mackey$^1$\thanks{E-mail: dmackey@ast.cam.ac.uk}
  and G.~F.~Gilmore$^1$\\
  $^1$Institute of Astronomy, University of Cambridge, Madingley Road,
  Cambridge CB3 0HA}
\date{Accepted --. Received --}
\pagerange{000--000}
\pubyear{2001}

\def\LaTeX{L\kern-.36em\raise.3ex\hbox{a}\kern-.15em
    T\kern-.1667em\lower.7ex\hbox{E}\kern-.125emX}

\begin{document}

\label{firstpage}

\maketitle

\begin{abstract}
We have compiled a pseudo-snapshot data set of two-colour observations
from the {\em Hubble Space Telescope} archive for a sample of 53 rich 
LMC clusters with ages $10^{6}-10^{10}$ yr. We present surface brightness
profiles for the entire sample, and derive structural parameters for each
cluster, including core radii, and luminosity and mass estimates. Because
we expect the results presented here to form the basis for several 
further projects, we describe in detail the data reduction and surface 
brightness profile construction processes, and compare our results with 
those of previous ground-based studies.

The surface brightness profiles show a large amount of detail, including
irregularities in the profiles of young clusters (such as bumps, dips,
and sharp shoulders), and evidence for both double clusters and post
core-collapse (PCC) clusters. In particular we find power-law profiles
in the inner regions of several candidate PCC clusters, with slopes of 
approximately $-0.7$, but showing considerable variation. We estimate 
that $20 \pm 7$ per cent of the old cluster population of the 
LMC has entered PCC evolution, a similar fraction to that for the 
Galactic globular cluster system. In addition, we examine the profile of R136
in detail and show that it is probably not a PCC cluster.

We also observe a trend in core radius with age that has been discovered
and discussed in several previous publications by different authors.
Our diagram has better resolution however, and appears to show a 
bifurcation at several hundred Myr. We argue that this observed 
relationship reflects true physical evolution in LMC clusters, with some
experiencing small scale core expansion due to mass loss, and others
large scale expansion due to some unidentified characteristic or 
physical process. 
\end{abstract}


\section{Introduction}
The star cluster system of the Large Magellanic Cloud (LMC) is unique in
containing rich star clusters of masses comparable to Galactic globular
clusters, but covering a wide age range ($10^{6}-10^{10}$ yr) and
being close enough for detailed observation. It therefore offers a
seemingly perfect opportunity for studies of all aspects of star cluster
astronomy, from cluster formation, evolution and dynamics, to luminosity
and mass function studies, as well as investigations which shed light on
the evolution history of the entire LMC. It is surprising then, that
while catalogues of high resolution surface brightness profiles and 
structural parameters exist for Galactic globular clusters 
(e.g., Trager, King \& Djorgovski \shortcite{trager}), no such uniform 
catalogue exists for a large sample of the rich LMC clusters.

There has however been some considerable activity in this field. In 
particular, significant numbers of surface brightness and/or density
profiles have been published for young and intermediate age 
clusters (Elson, Fall \& Freeman \shortcite{eff}, hereafter EFF87; 
Elson, Freeman \& Lauer \shortcite{efl}; Elson \shortcite{elsonbig}, 
hereafter E91; Elson \shortcite{elsonfive}); old clusters \cite{mateo}; 
and clusters with different spatial distributions -- in the LMC disk 
(Kontizas, Chrysovergis \& Kontizas \shortcite{kck}); in the LMC disk
and within 5 kpc of the rotation centre (Chrysovergis, Kontizas \&
Kontizas \shortcite{ckk}); and in the LMC halo (Kontizas, 
Hadjidimitriou \& Kontizas \shortcite{khk}; Metaxa, Kontizas \& Kontizas
\shortcite{mkk}). These studies are all ground-based and therefore 
suffer from problems -- primarily crowding and seeing related -- which 
limit their resolution, particularly in the inner regions of
clusters. This in turn renders the derivation of key parameters, such 
as the core radius, rather uncertain. In addition, each set of authors
uses a different data set and makes different measurements (e.g., 
Elson and collaborators construct $V$-band surface brightness profiles;
Mateo uses $B$-band profiles; and Kontizas and collaborators use 
number density profiles), so most of the important derived values 
are not strictly comparable between studies.

We have taken advantage of the presence of a large number of observations
of rich LMC clusters in the {\em Hubble Space Telescope} ({\em HST}) 
archive to compile a high resolution data set, unaffected by the problems
which beset ground-based studies. Rather than simply providing an atlas of
surface brightness profiles from this data, the aim of the present study is to 
obtain a statistically homogeneous set of profiles and key structural parameters
for as many clusters as possible, for the purposes of differential comparison.
To this end, we require a sample as free from bias as possible, and we have therefore 
applied a uniform selection and reduction process to the available observations. 
Because these procedures are rather detailed, and because we expect our measurements 
to form the basis of several additional projects, in this paper we discuss at length 
the data set and its selection (Section \ref{sample}) and the reduction and profile 
construction processes (Sections \ref{reduction} and \ref{sbprofile} respectively). 
In Section \ref{results} we present the surface brightness profiles and structural
parameter measurements, compare these results with those
of the authors previously mentioned, and examine some of the interesting
sub-groups present in the sample -- such as binary clusters and post
core-collapse clusters. Finally, in Section \ref{discussion} we 
observe and discuss the relationship between core radius and age for 
the LMC cluster system. This relationship has previously been studied
(Elson et al. \shortcite{efl}; E91; Elson \shortcite{elsonfive}); 
however our measurements are able to provide new insight into this 
problem.

The data presented in Tables \ref{data}, \ref{ages}, \ref{params}, and
\ref{luminmass}, and the surface brightness profiles (Fig. \ref{plots})
are available on-line at
{\em http://www.ast.cam.ac.uk/STELLARPOPS/LMC\_clusters/}.

\section{The cluster sample}
\label{sample}
\subsection{Observations}
The observational basis of this project is the presence
in the {\em HST} archive of a snapshot survey of LMC (and SMC) clusters
({\em HST} project 5475). This data set consists of two Wide Field 
Planetary Camera 2 (WFPC2) exposures per cluster, through the F450W and 
F555W filters respectively, and with exposure times covering the range 
20-300 s in F555W and 40-600 s in F450W, dependent on the age of the 
cluster under observation. The data were acquired between 1994 
January 27 and 1994 December 25, but mostly before the WFPC2 cool-down on 
1994 April 23.

Upon retrieval, we discovered that the sample was neither as
large nor as varied as desired, and consequently we returned to the 
archive to obtain observations of as many additional clusters 
as possible. We located a further six suitable studies ({\em HST} 
projects 5114, 5897, 5904, 5916, 7307, and 8134), consisting of 
detailed observations of several clusters each. To maintain the uniformity 
of our selection as far as possible, we attempted to mimic the snapshot data by 
using only two frames per cluster, with exposure times as close as possible to
the ranges described above, even though for many of the clusters from these
six studies, an abundance of archival frames is available. To avoid biases such
as that due to observer selection of ``interesting targets,'' it was necessary to limit
the use of these extra frames to match the least observed clusters. This fit with our 
primary aim of investigating the LMC cluster system as a whole. For this purpose 
we required data for as large a sample of clusters as possible, but to be 
reduced within a reasonable time frame. Therefore, we had to balance
ultimate accuracy (maximum data) for any given cluster in the sample
against overall reduction time. Had our aim been simply to provide an atlas of surface 
brightness profiles, or to study one or two clusters in great detail, the extra 
data would certainly have been included. Nonetheless, we were careful not to 
unnecessarily degrade the quality of our data. In cases where the neglect of 
additional data would have compromised our results (e.g., in extremely crowded 
clusters) we included minimal numbers of extra frames in order to proceed.

In the end, complete reproduction of the original snapshot data was not
possible, with the only uniform choice of filters between the
extra six studies being F555W and F814W. For nine clusters we
retrieved both long ($\sim 500$ s) and short ($\sim 10$ s) exposures in 
order to overcome data degradation due to severe crowding and/or 
saturation (as mentioned above). The final cluster sample is listed in Table 
\ref{data} together with the observational details\footnote{Notes to Table 1:\\ 
$^{a}$ The archived images of this cluster are incorrectly labelled as NGC 2156. \\
$^{b}$ The archived images of this cluster are incorrectly labelled as SL 633.}. 
The sample consists of 53 LMC clusters spanning the full age range. The entire sample
is observed through the F555W filter, providing a unique high resolution
data set. Thirty-one of the clusters are also observed through the
F450W filter, with the remaining 22 clusters having F814W as their second
filter -- we note however that for our surface brightness profile construction
process, the choice of the second colour is of little consequence. 

\begin{table*}
\begin{minipage}{162mm}
\caption{Cluster list and observation details.}
\begin{tabular}{@{}lccccccccccc}
\hline \hline
Cluster & Program & & \multicolumn{4}{c}{Principal Frame} & & \multicolumn{4}{c}{Secondary Frame} \\
\cline{4-7}  \cline{9-12} \vspace{-3mm} \\
Name & ID & & Filter & Data set & Date & Time (s) & & Filter & Data set & Date & Time (s) \\
\hline
NGC1466 & 5897 & & F555W & u2xj0103t & 22/10/1995 & 260 & & F814W & u2xj0101t & 22/10/1995 & 260 \\
NGC1651 & 5475 & & F555W & u26m0r02t & 01/02/1994 & 120 & & F450W & u26m0r01t & 01/02/1994 & 230 \\
NGC1711 & 5904 & & F555W & u2y80502r & 18/10/1997 & 300 & & F814W & u2y80504r & 18/10/1997 & 300 \\
NGC1718 & 5475 & & F555W & u26m0j02t & 06/02/1994 & 120 & & F450W & u26m0j01t & 06/02/1994 & 230 \\
NGC1754 & 5916 & & F555W & u2xq0103t & 21/10/1995 & 500 & & F814W & u2xq0109t & 21/10/1995 & 600 \\
        &      & & F555W & u2xq0101t & 21/10/1995 & 20 & & F814W & u2xq0106t & 21/10/1995 & 20 \\
NGC1777 & 5475 & & F555W & u26m0u02t & 23/11/1994 & 120 & & F450W & u26m0u01t & 23/11/1994 & 230 \\
NGC1786 & 5897 & & F555W & u2xj0203t & 19/10/1995 & 260 & & F814W & u2xj0201t & 19/10/1995 & 260 \\
NGC1805 & 7307 & & F555W & u4ax0206r & 25/07/1998 & 140 & & F814W & u4ax020br & 25/07/1998 & 300 \\
NGC1818 & 7307 & & F555W & u4ax3005r & 25/09/1998 & 140 & & F814W & u4ax300cr & 25/09/1998 & 300 \\
NGC1831 & 5475 & & F555W & u26m1002t & 27/01/1994 & 40 & & F450W & u26m1001t & 27/01/1994 & 80 \\
NGC1835 & 5916 & & F555W & u2xq0203t & 18/10/1995 & 500 & & F814W & u2xq0209t & 18/10/1995 & 600 \\
        &      & & F555W & u2xq0201t & 18/10/1995 & 20 & & F814W & u2xq0206t & 18/10/1995 & 20 \\
NGC1841 & 5897 & & F555W & u2xj0708t & 14/11/1995 & 800 & & F814W & u2xj0707t & 14/11/1995 & 800 \\
NGC1847 & 5475 & & F555W & u26m1b02p & 10/04/1994 & 20 & & F450W & u26m1b01p & 10/04/1994 & 40 \\
NGC1850 & 5475 & & F555W & u26m1e02t & 15/05/1994 & 20 & & F450W & u26m1e01t & 15/05/1994 & 40 \\
NGC1856 & 5475 & & F555W & u26m1302t & 06/02/1994 & 30 & & F450W & u26m1301t & 06/02/1994 & 60 \\
NGC1860 & 5475 & & F555W & u26m1202t & 07/02/1994 & 40 & & F450W & u26m1201t & 07/02/1994 & 80 \\
NGC1866 & 8134 & & F555W & u5ay0601r & 21/08/1999 & 350 & & F814W & u5ay0604r & 22/08/1999 & 350 \\
NGC1868 & 5475 & & F555W & u26m0z02t & 10/04/1994 & 100 & & F450W & u26m0z01t & 10/04/1994 & 200 \\
NGC1898 & 5916 & & F555W & u2xq0304t & 10/12/1995 & 500 & & F814W & u2xq0309t & 10/12/1995 & 600 \\
NGC1916 & 5916 & & F555W & u2xq0403t & 10/12/1995 & 500 & & F814W & u2xq0409t & 10/12/1995 & 600 \\
        &      & & F555W & u2xq0401t & 10/12/1995 & 20 & & F814W & u2xq0406t & 10/12/1995 & 20 \\
NGC1984 & 8134 & & F555W & u5ay0901r & 08/07/2000 & 350 & & F814W & u5ay0904r & 09/07/2000 & 350 \\
        &      & & F555W & u5ay0903r & 09/07/2000 & 10 & & F814W & u5ay0906r & 09/07/2000 & 10 \\
NGC2004 & 5475 & & F555W & u26m1d02t & 21/04/1994 & 20 & & F450W & u26m1d01t & 21/04/1994 & 40 \\
NGC2005 & 5916 & & F555W & u2xq0503t & 19/10/1995 & 500 & & F814W & u2xq0509t & 19/10/1995 & 600 \\
        &      & & F555W & u2xq0501t & 19/10/1995 & 20 & & F814W & u2xq0506t & 19/10/1995 & 20 \\
NGC2011 & 8134 & & F555W & u5ay0801r & 19/08/1999 & 350 & & F814W & u5ay0804r & 20/08/1999 & 350 \\
        &      & & F555W & u5ay0803r & 19/08/1999 & 10 & & F814W & u5ay0806r & 20/08/1999 & 10 \\
NGC2019 & 5916 & & F555W & u2xq0603t & 18/10/1995 & 500 & & F814W & u2xq0609t & 18/10/1995 & 600 \\
        &      & & F555W & u2xq0601t & 18/10/1995 & 20 & & F814W & u2xq0606t & 18/10/1995 & 20 \\
NGC2031 & 5904 & & F555W & u2y80301t & 20/10/1995 & 300 & & F814W & u2y80304p & 20/10/1995 & 300 \\
NGC2100 & 5475 & & F555W & u26m1c02n & 21/04/1994 & 20 & & F450W & u26m1c01t & 21/04/1994 & 40 \\
NGC2121 & 5475 & & F555W & u26m0x02t & 02/02/1994 & 120 & & F450W & u26m0x01t & 02/02/1994 & 230 \\
NGC2136 & 5475 & & F555W & u26m1702t & 11/02/1994 & 30 & & F450W & u26m1701t & 11/02/1994 & 60 \\
NGC2153$^{a}$ & 5475 & & F555W & u26m1502t & 19/04/1994 & 30 & & F450W & u26m1501t & 19/04/1994 & 60 \\
NGC2155 & 5475 & & F555W & u26m0k02t & 01/02/1994 & 120 & & F450W & u26m0k01t & 01/02/1994 & 230 \\
NGC2156 & 8134 & & F555W & u5ay0301r & 22/08/1999 & 350 & & F814W & u5ay0304r & 22/08/1999 & 350 \\
        &      & & F555W & u5ay0303r & 22/08/1999 & 10 & & F814W & u5ay0306r & 22/08/1999 & 10 \\
NGC2157 & 5475 & & F555W & u26m1902t & 26/03/1994 & 30 & & F450W & u26m1901t & 26/03/1994 & 60 \\
NGC2159 & 8134 & & F555W & u5ay0401r & 16/10/1999 & 350 & & F814W & u5ay0404r & 16/10/1999 & 350 \\
NGC2162 & 5475 & & F555W & u26m0t02t & 23/08/1994 & 120 & & F450W & u26m0t01t & 23/08/1994 & 230 \\
NGC2164 & 8134 & & F555W & u5ay0101r & 24/09/1999 & 350 & & F814W & u5ay0105r & 24/09/1999 & 350 \\
NGC2172 & 8134 & & F555W & u5ay0502r & 14/10/1999 & 350 & & F814W & u5ay0504r & 14/10/1999 & 350 \\
NGC2173 & 5475 & & F555W & u26m0m02t & 02/02/1994 & 120 & & F450W & u26m0m01t & 02/02/1994 & 230 \\
NGC2193 & 5475 & & F555W & u26m0p02t & 30/01/1994 & 120 & & F450W & u26m0p01t & 30/01/1994 & 230 \\
NGC2209 & 5475 & & F555W & u26m0w02t & 01/02/1994 & 120 & & F450W & u26m0w01t & 01/02/1994 & 230 \\
NGC2210 & 5897 & & F555W & u2xj0403t & 13/12/1995 & 260 & & F814W & u2xj0401t & 13/12/1995 & 260 \\
NGC2213 & 5475 & & F555W & u26m0v02t & 25/12/1994 & 120 & & F450W & u26m0v01t & 25/12/1994 & 230 \\
NGC2214 & 5475 & & F555W & u26m1802t & 16/02/1994 & 30 & & F450W & u26m1801t & 16/02/1994 & 60 \\
NGC2231 & 5475 & & F555W & u26m0s02t & 06/02/1994 & 120 & & F450W & u26m0s01t & 06/02/1994 & 230 \\
NGC2249 & 5475 & & F555W & u26m0y02t & 11/08/1994 & 120 & & F450W & u26m0y01t & 11/08/1994 & 230 \\
NGC2257 & 5475 & & F555W & u26m0d02t & 05/02/1994 & 300 & & F450W & u26m0d01t & 05/02/1994 & 600 \\
SL663$^{b}$   & 5475 & & F555W & u26m0l02t & 01/02/1994 & 120 & & F450W & u26m0l01t & 01/02/1994 & 230 \\
SL842   & 5475 & & F555W & u26m0h02t & 01/02/1994 & 120 & & F450W & u26m0h01t & 01/02/1994 & 230 \\
SL855   & 5475 & & F555W & u26m0q02t & 01/02/1994 & 120 & & F450W & u26m0q01t & 01/02/1994 & 230 \\
HODGE4  & 5475 & & F555W & u26m0o02t & 01/02/1994 & 120 & & F450W & u26m0o01t & 01/02/1994 & 230 \\
HODGE11 & 5475 & & F555W & u26m0f02t & 01/02/1994 & 300 & & F450W & u26m0f01t & 01/02/1994 & 600 \\
HODGE14 & 5475 & & F555W & u26m0n02t & 18/09/1994 & 120 & & F450W & u26m0n01t & 18/09/1994 & 230 \\
R136    & 5114 & & F555W & u2hk0304t & 25/09/1994 & 23 & & F814W & u2hk030rt & 25/09/1994 & 40 \\
        &      & & F555W & u2hk0302t & 25/09/1994 & 3 & & F814W & u2hk030qt & 25/09/1994 & 5 \\
\hline
\label{data}
\end{tabular}
\end{minipage}
\end{table*}

\subsection{Literature data}
As a supplement to the above data set, and as a reference point
for this and future publications, we have compiled nomenclature, position,
age and metallicity data from the literature for the complete cluster 
sample. This information is displayed in Table \ref{ages}. As with the 
data set construction, the emphasis is on obtaining as
homogeneous a compilation as possible, while still maintaining the 
integrity and accuracy of the data. This means using the results of large
surveys or studies for the most part, and supplementing these with the
results of high quality observations where necessary. In the cases where
several such high quality studies are available, we have chosen the
values which represent approximately the convergence of the data. For
several clusters, scarcity of information has necessitated the selection
of older or lower quality data. This compilation does not purport to be
a complete survey of the available literature; rather it is intended
to provide a consistent set of age and metallicity estimates for use in 
this and subsequent projects.

\subsubsection{Cluster names and positions}
We have taken position and nomenclature information from the Simbad
Astronomical Database ({\em http://simbad.u-strasbg.fr/}). 
While almost all recognized identifiers for a cluster are
included in this database, we have not compiled all such identifiers for
our sample, rather only the most common. In general the principal 
designation is an NGC number, with the other identifiers and 
corresponding catalogues as follows: Hodge, \cite{hodge}; SL, \cite{sl};
LW, \cite{lw}. In the Simbad database, J2000.0 positions for LMC objects 
are in general from the catalogue of Bica et al. \shortcite{bicasurvey}. 
Using these positions we calculate the projected angular 
distance $R_{opt}$ to the optical centre of the LMC at 
$\alpha = 05^{h}20^{m}56^{s}$, $\delta = -69\degr 28\arcmin 41\arcsec$
(J2000.0) \cite{bicaubv}, and the projected angular distance $R_{rot}$ 
to the H\,{\sc i} rotation curve centre at $\alpha = 05^{h}26^{m}39^{s}$,
$\delta = -69\degr 15\arcmin 36\arcsec$ (J2000.0) 
\cite{rohlfs,westerlund}. We later derive more accurate centres for each 
cluster; however, the literature values listed here are never
substantially different from our calculated positions.

\subsubsection{Cluster ages}
\label{clusterages}
Age determinations for LMC clusters are widely scattered across the
literature, and involve a large variety of techniques. The most useful
homogeneous data set we have located is the colour-magnitude diagram (CMD) 
study and age calibration of Geisler et al. \shortcite{geisler}, which 
provides age estimates for 15 clusters in our sample. Following this lead we 
have adopted where possible for the remaining clusters age determinations 
based on high resolution photometric studies resulting in CMDs. 
Pre-eminent amongst these are {\em HST} studies, and we adopt the results
of Rich, Shara \& Zurek \shortcite{rsz} for 5 intermediate age clusters 
and the results of Olsen et al. \shortcite{olsen} for 5 old clusters.
We take the age determinations from the ground-based studies of E91
for 9 clusters and those of Dirsch et al. \shortcite{dirsch} for a further 
three clusters. For NGC 1805 and NGC 1818 we adopt the ages from
the thorough discussion and review of de Grijs et al. \shortcite{richard}
and similarly for R136 we take the age as discussed by Sirianni et al.
\shortcite{sirianni}. Finally, we fill the remaining holes in our sample
with the age estimates of Elson \& Fall \shortcite{ef}, who provide
an age calibration based on a large sample of literature CMDs. We
correct these estimates to a distance modulus of 18.5 using their
prescribed method. We could not locate a reliable age for NGC 1916, but
Olsen et al. \shortcite{olsen} do present a CMD for this cluster along
with CMDs and age estimates for five old LMC clusters. The CMD for NGC
1916, though suffering from serious differential reddening, is not 
evidently different from those for the other five clusters, so we average
the ages of these clusters to obtain an estimate for NGC 1916.

\subsubsection{Cluster metallicities}
The definitive study of LMC cluster metallicities is by Olszewski et al.
\shortcite{ol} who derive metallicities for $\sim 70$ LMC clusters
from observations of the calcium triplet in the spectra of giants. From 
this study we adopt metallicities for 23 of the clusters in our sample.
We also adopt one metallicity (NGC 1841) from a subsequent paper in 
the same series \cite{suntzeff}. A further three abundances (NGC 1850, 
NGC 2004, NGC 2100) are taken from the spectroscopic study of 
Jasniewicz and Th\'{e}venin \shortcite{jt}. The authors report standard
deviations of $\pm 0.03$ dex in the averages of several observations for
each of these clusters; however we prefer their errors per single 
observation of $\sim 0.2$ dex, which are more consistent with the other 
errors in Table \ref{ages}. We adopt another two abundances from the 
infra-red spectra of Oliva \& Origlia \shortcite{oliva}, and one 
(NGC 1866) from the VLT spectroscopy of Hill et al.\shortcite{hill}. 
These complete the spectroscopic studies suitable for our data set. 
For a few clusters we therefore rely on abundances
derived from CMD fits and corresponding to the ages discussed in Section
\ref{clusterages}. Four come from the study of Dirsch et al. 
\shortcite{dirsch}, and two from the {\em HST} photometry of Rich et al. 
\shortcite{rsz}. For NGC 1805 and NGC 1818 we again take the ranges 
adopted by de Grijs et al. \shortcite{richard} after their detailed 
literature review (see also Johnson et al. \shortcite{johnson}), and 
similarly for R136 we again adopt the abundance from the discussion of 
Sirianni et al. \shortcite{sirianni}; see also Hunter et al. 
\shortcite{hunter}. For the remaining 14 clusters we were not able to 
obtain suitable abundance measurements from the literature so for each 
of these we average the literature results 
for all the clusters in our sample of comparable age. We emphasize that
these values in particular are simply rather crude estimates and that any 
use of them must account for this. In this paper, we adopt these 
abundances solely for the purpose of estimating mass to light ratios 
later in the analysis -- calculations which in any case are rather 
insensitive to the adopted metallicity. Naturally, any subsequent use 
must be carefully judged on the nature of the calculation at hand. 

\begin{table*}
\begin{minipage}{171mm}
\caption{Literature nomenclature, position, age and metallicity data for the cluster sample.}
\begin{tabular}{@{}llcccccccc}
\hline \hline
Principal & Alternative & \multicolumn{4}{c}{Position (J2000.0)} & $\log\tau$ & Age & Metallicity & Met. \vspace{0.5mm} \\
Name & Names & $\alpha$ & $\delta$ & $R_{opt}$ $(\degr)^{a}$ & $R_{rot}$ $(\degr)^{b}$ & (yr) & Ref. & $[$Fe$/$H$]$ & Ref. \\
\hline
NGC1466 & SL1, LW1 & $03^{h}44^{m}33^{s}$ & $-71\degr 40\arcmin 18\arcsec$ & $7.89$ & $8.38$ & $10.10 \pm 0.01$ & $5$ & $-2.17 \pm 0.20$ & $12$ \vspace{0mm} \\
NGC1651 & SL7, LW12 & $04^{h}37^{m}32^{s}$ & $-70\degr 35\arcmin 06\arcsec$ & $3.77$ & $4.29$ & $9.30^{+0.08}_{-0.10}$ & $5$ & $-0.37 \pm 0.20$ & $12$ \vspace{0mm} \\
NGC1711 & SL55 & $04^{h}50^{m}37^{s}$ & $-69\degr 59\arcmin 06\arcsec$ & $2.64$ & $3.17$ & $7.70 \pm 0.05$ & $2$ & $-0.57 \pm 0.17$ & $2$ \vspace{0mm} \\
NGC1718 & SL65 & $04^{h}52^{m}25^{s}$ & $-67\degr 03\arcmin 06\arcsec$ & $3.69$ & $4.00$ & $9.30 \pm 0.30$ & $4$ & $-0.42$ & $*$ \vspace{0mm} \\
NGC1754 & SL91 & $04^{h}54^{m}17^{s}$ & $-70\degr 26\arcmin 30\arcsec$ & $2.43$ & $2.96$ & $10.19^{+0.06}_{-0.07}$ & $11$ & $-1.54 \pm 0.20$ & $12$ \vspace{0mm} \\
NGC1777 & SL121, LW96 & $04^{h}55^{m}48^{s}$ & $-74\degr 17\arcmin 00\arcsec$ & $5.10$ & $5.44$ & $9.08^{+0.12}_{-0.18}$ & $5$ & $-0.35 \pm 0.20$ & $12$ \vspace{0mm} \\
NGC1786 & SL149 & $04^{h}59^{m}06^{s}$ & $-67\degr 44\arcmin 42\arcsec$ & $2.70$ & $3.02$ & $10.18 \pm 0.01$ & $5$ & $-1.87 \pm 0.20$ & $12$ \vspace{0mm} \\
NGC1805 & SL186 & $05^{h}02^{m}21^{s}$ & $-66\degr 06\arcmin 42\arcsec$ & $3.86$ & $4.00$ & $7.00^{+0.30}_{-0.10}$ & $1$ & $0.00$ to $-0.40$ & $1$,$9$ \vspace{0mm} \\
NGC1818 & SL201 & $05^{h}04^{m}14^{s}$ & $-66\degr 26\arcmin 06\arcsec$ & $3.47$ & $3.61$ & $7.40^{+0.30}_{-0.10}$ & $1$ & $0.00$ to $-0.40$ & $1$,$9$ \vspace{0mm} \\
NGC1831 & SL227, LW133 & $05^{h}06^{m}16^{s}$ & $-64\degr 55\arcmin 06\arcsec$ & $4.82$ & $4.85$ & $8.50 \pm 0.30$ & $4$ & $+0.01 \pm 0.20$ & $12$ \vspace{0mm} \\
NGC1835 & SL215 & $05^{h}05^{m}05^{s}$ & $-69\degr 24\arcmin 12\arcsec$ & $1.40$ & $1.90$ & $10.22^{+0.07}_{-0.08}$ & $11$ & $-1.79 \pm 0.20$ & $12$ \vspace{0mm} \\
NGC1841 &       & $04^{h}45^{m}23^{s}$ & $-83\degr 59\arcmin 48\arcsec$ & $14.55$ & $14.78$ & $10.09 \pm 0.01$ & $5$ & $-2.11 \pm 0.10$ & $15$ \vspace{0mm} \\
NGC1847 & SL240 & $05^{h}07^{m}08^{s}$ & $-68\degr 58\arcmin 18\arcsec$ & $1.34$ & $1.77$ & $7.42 \pm 0.30$ & $4$ & $-0.37$ & $*$ \vspace{0mm} \\
NGC1850 & SL261 & $05^{h}08^{m}44^{s}$ & $-68\degr 45\arcmin 36\arcsec$ & $1.32$ & $1.70$ & $7.50 \pm 0.20$ & $3$ & $-0.12 \pm 0.20$ & $8$ \vspace{0mm} \\
NGC1856 & SL271 & $05^{h}09^{m}29^{s}$ & $-69\degr 07\arcmin 36\arcsec$ & $1.08$ & $1.53$ & $8.12 \pm 0.30$ & $4$ & $-0.52$ & $*$ \vspace{0mm} \\
NGC1860 & SL284 & $05^{h}10^{m}39^{s}$ & $-68\degr 45\arcmin 12\arcsec$ & $1.18$ & $1.54$ & $8.28 \pm 0.30$ & $4$ & $-0.52$ & $*$ \vspace{0mm} \\
NGC1866 & SL319, LW163 & $05^{h}13^{m}39^{s}$ & $-65\degr 27\arcmin 54\arcsec$ & $4.08$ & $4.03$ & $8.12 \pm 0.30$ & $4$ & $-0.50 \pm 0.10$ & $6$ \vspace{0mm} \\
NGC1868 & SL330, LW169 & $05^{h}14^{m}36^{s}$ & $-63\degr 57\arcmin 18\arcsec$ & $5.57$ & $5.47$ & $8.74 \pm 0.30$ & $4$ & $-0.50 \pm 0.20$ & $12$ \vspace{0mm} \\
NGC1898 & SL350 & $05^{h}16^{m}42^{s}$ & $-69\degr 39\arcmin 24\arcsec$ & $0.41$ & $0.95$ & $10.15^{+0.06}_{-0.08}$ & $11$ & $-1.37 \pm 0.20$ & $12$ \vspace{0mm} \\
NGC1916 & SL361 & $05^{h}18^{m}39^{s}$ & $-69\degr 24\arcmin 24\arcsec$ & $0.21$ & $0.72$ & $10.20 \pm 0.09$ & $*$ & $-2.08 \pm 0.20$ & $12$ \vspace{0mm} \\
NGC1984 & SL488 & $05^{h}27^{m}40^{s}$ & $-69\degr 08\arcmin 06\arcsec$ & $0.69$ & $0.15$ & $7.06 \pm 0.30$ & $4$ & $-0.90 \pm 0.40$ & $10$ \vspace{0mm} \\
NGC2004 & SL523 & $05^{h}30^{m}40^{s}$ & $-67\degr 17\arcmin 12\arcsec$ & $2.38$ & $2.01$ & $7.30 \pm 0.20$ & $3$ & $-0.56 \pm 0.20$ & $8$ \vspace{0mm} \\
NGC2005 & SL518 & $05^{h}30^{m}09^{s}$ & $-69\degr 45\arcmin 06\arcsec$ & $0.84$ & $0.58$ & $10.22^{+0.12}_{-0.16}$ & $11$ & $-1.92 \pm 0.20$ & $12$ \vspace{0mm} \\
NGC2011 & SL559 & $05^{h}32^{m}19^{s}$ & $-67\degr 31\arcmin 18\arcsec$ & $2.24$ & $1.82$ & $6.99 \pm 0.30$ & $4$ & $-0.47 \pm 0.40$ & $10$ \vspace{0mm} \\
NGC2019 & SL554 & $05^{h}31^{m}56^{s}$ & $-70\degr 09\arcmin 36\arcsec$ & $1.16$ & $1.01$ & $10.25^{+0.07}_{-0.09}$ & $11$ & $-1.81 \pm 0.20$ & $12$ \vspace{0mm} \\
NGC2031 & SL577 & $05^{h}33^{m}41^{s}$ & $-70\degr 59\arcmin 12\arcsec$ & $1.83$ & $1.82$ & $8.20 \pm 0.10$ & $2$ & $-0.52 \pm 0.21$ & $2$ \vspace{0mm} \\
NGC2100 & SL662 & $05^{h}42^{m}08^{s}$ & $-69\degr 12\arcmin 42\arcsec$ & $1.90$ & $1.37$ & $7.20 \pm 0.20$ & $3$ & $-0.32 \pm 0.20$ & $8$ \vspace{0mm} \\
NGC2121 & SL725, LW303 & $05^{h}48^{m}12^{s}$ & $-71\degr 28\arcmin 48\arcsec$ & $2.95$ & $2.80$ & $9.51^{+0.06}_{-0.07}$ & $13$ & $-0.61 \pm 0.20$ & $12$ \vspace{0mm} \\
NGC2136 & SL762 & $05^{h}53^{m}17^{s}$ & $-69\degr 31\arcmin 42\arcsec$ & $2.83$ & $2.34$ & $8.00 \pm 0.10$ & $2$ & $-0.55 \pm 0.23$ & $2$ \vspace{0mm} \\
NGC2153 & SL792, LW341 & $05^{h}57^{m}51^{s}$ & $-66\degr 24\arcmin 00\arcsec$ & $4.81$ & $4.23$ & $9.11^{+0.12}_{-0.16}$ & $5$ & $-0.42$ & $*$ \vspace{0mm} \\
NGC2155 & SL803, LW347 & $05^{h}58^{m}33^{s}$ & $-65\degr 28\arcmin 36\arcsec$ & $5.59$ & $5.03$ & $9.51^{+0.06}_{-0.07}$ & $13$ & $-0.55 \pm 0.20$ & $12$ \vspace{0mm} \\
NGC2156 & SL796 & $05^{h}57^{m}45^{s}$ & $-68\degr 27\arcmin 36\arcsec$ & $3.53$ & $2.96$ & $7.60 \pm 0.20$ & $3$ & $-0.45$ & $*$ \vspace{0mm} \\
NGC2157 & SL794 & $05^{h}57^{m}34^{s}$ & $-69\degr 11\arcmin 48\arcsec$ & $3.26$ & $2.75$ & $7.60 \pm 0.20$ & $3$ & $-0.45$ & $*$ \vspace{0mm} \\
NGC2159 & SL799 & $05^{h}57^{m}57^{s}$ & $-68\degr 37\arcmin 24\arcsec$ & $3.48$ & $2.92$ & $7.60 \pm 0.20$ & $3$ & $-0.45$ & $*$ \vspace{0mm} \\
NGC2162 & SL814, LW351 & $06^{h}00^{m}31^{s}$ & $-63\degr 43\arcmin 18\arcsec$ & $7.23$ & $6.69$ & $9.11^{+0.12}_{-0.16}$ & $5$ & $-0.23 \pm 0.20$ & $12$ \vspace{0mm} \\
NGC2164 & SL808 & $05^{h}58^{m}54^{s}$ & $-68\degr 31\arcmin 06\arcsec$ & $3.61$ & $3.04$ & $7.70 \pm 0.20$ & $3$ & $-0.45$ & $*$ \vspace{0mm} \\
NGC2172 & SL812 & 06$^{h}00^{m}05^{s}$ & $-68\degr 38\arcmin 12\arcsec$ & $3.66$ & $3.11$ & $7.60 \pm 0.20$ & $3$ & $-0.44$ & $*$ \vspace{0mm} \\
NGC2173 & SL807, LW348 & $05^{h}57^{m}58^{s}$ & $-72\degr 58\arcmin 42\arcsec$ & $4.43$ & $4.37$ & $9.33^{+0.07}_{-0.09}$ & $5$ & $-0.24 \pm 0.20$ & $12$ \vspace{0mm} \\
NGC2193 & SL839, LW387 & $06^{h}06^{m}17^{s}$ & $-65\degr 05\arcmin 54\arcsec$ & $6.48$ & $5.89$ & $9.34^{+0.09}_{-0.11}$ & $13$ & $-0.60 \pm 0.20$ & $13$ \vspace{0mm} \\
NGC2209 & SL849, LW408 & $06^{h}08^{m}34^{s}$ & $-73\degr 50\arcmin 30\arcsec$ & $5.48$ & $5.43$ & $8.98^{+0.15}_{-0.24}$ & $5$ & $-0.47$ & $*$ \vspace{0mm} \\
NGC2210 & SL858, LW423 & $06^{h}11^{m}31^{s}$ & $-69\degr 07\arcmin 18\arcsec$ & $4.52$ & $4.00$ & $10.20 \pm 0.01$ & $5$ & $-1.97 \pm 0.20$ & $12$ \vspace{0mm} \\
NGC2213 & SL857, LW419 & $06^{h}10^{m}42^{s}$ & $-71\degr 31\arcmin 42\arcsec$ & $4.44$ & $4.16$ & $9.20^{+0.10}_{-0.12}$ & $5$ & $-0.01 \pm 0.20$ & $12$ \vspace{0mm} \\
NGC2214 & SL860, LW426 & $06^{h}12^{m}57^{s}$ & $-68\degr 15\arcmin 36\arcsec$ & $4.97$ & $4.40$ & $7.60 \pm 0.20$ & $3$ & $-0.45$ & $*$ \vspace{0mm} \\
NGC2231 & SL884, LW466 & $06^{h}20^{m}44^{s}$ & $-67\degr 31\arcmin 06\arcsec$ & $6.04$ & $5.46$ & $9.18^{+0.10}_{-0.13}$ & $5$ & $-0.67 \pm 0.20$ & $12$ \vspace{0mm} \\
NGC2249 & SL893, LW479 & $06^{h}25^{m}49^{s}$ & $-68\degr 55\arcmin 12\arcsec$ & $5.86$ & $5.33$ & $8.82 \pm 0.30$ & $4$ & $-0.47$ & $*$ \vspace{0mm} \\
NGC2257 & SL895, LW481 & $06^{h}30^{m}12^{s}$ & $-64\degr 19\arcmin 36\arcsec$ & $9.10$ & $8.47$ & $10.20 \pm 0.10$ & $2$ & $-1.63 \pm 0.21$ & $2$ \vspace{0mm} \\
SL663 & LW273 & $05^{h}42^{m}29^{s}$ & $-65\degr 21\arcmin 48\arcsec$ & $4.69$ & $4.23$ & $9.51^{+0.06}_{-0.07}$ & $13$ & $-0.60 \pm 0.20$ & $13$ \vspace{0mm} \\
SL842 & LW399 & $06^{h}08^{m}15^{s}$ & $-62\degr 59\arcmin 18\arcsec$ & $8.43$ & $7.85$ & $9.30^{+0.08}_{-0.10}$ & $5$ & $-0.36 \pm 0.20$ & $12$ \vspace{0mm} \\
SL855 & LW420 & $06^{h}10^{m}53^{s}$ & $-65\degr 02\arcmin 36\arcsec$ & $6.88$ & $6.29$ & $9.13 \pm 0.30$ & $4$ & $-0.42$ & $*$ \vspace{0mm} \\
HODGE4 & SL556, LW237 & $05^{h}31^{m}54^{s}$ & $-64\degr 42\arcmin 00\arcsec$ & $4.92$ & $4.59$ & $9.34^{+0.09}_{-0.11}$ & $13$ & $-0.15 \pm 0.20$ & $12$ \vspace{0mm} \\
HODGE11 & SL868, LW437 & $06^{h}14^{m}22^{s}$ & $-69\degr 50\arcmin 54\arcsec$ & $4.62$ & $4.15$ & $10.18 \pm 0.01$ & $5$ & $-2.06 \pm 0.20$ & $12$ \vspace{0mm} \\
HODGE14 & SL506, LW220 & $05^{h}28^{m}39^{s}$ & $-73\degr 37\arcmin 48\arcsec$ & $4.19$ & $4.37$ & $9.26^{+0.09}_{-0.11}$ & $5$ & $-0.66 \pm 0.20$ & $12$ \vspace{0mm} \\
R136 & NGC2070, 30Dor & $05^{h}38^{m}43^{s}$ & $-69\degr 06\arcmin 03\arcsec$ & $1.63$ & $1.09$ & $6.48^{+0.12}_{-0.18}$ & $14$ & $\sim -0.4$ & $7$,$14$ \vspace{0mm} \\
\hline
\label{ages}
\end{tabular}
\medskip
\\
Reference list: 1. de Grijs et al. \shortcite{richard}; 2. Dirsch et al. \shortcite{dirsch}; 3. Elson \shortcite{elsonbig}; 4. Elson \& Fall \shortcite{ef}; 5. Geisler et al. \shortcite{geisler}; 6. Hill et al. \shortcite{hill}; 7. Hunter et al. \shortcite{hunter}; 8. Jasniewicz \& Th\'{e}venin \shortcite{jt}; 9. Johnson et al. \shortcite{johnson}; 10. Oliva \& Origlia \shortcite{oliva}; 11. Olsen et al. \shortcite{olsen}; 12. Olszewski et al. \shortcite{ol}; 13. Rich et al. \shortcite{rsz}; 14. Sirianni et al. \shortcite{sirianni}; 15. Suntzeff et al. \shortcite{suntzeff}. \\
$^{*}$ Calculated metallicity (or age for NGC 1916), as described in the text.\\
$^{a}$ Relative to the optical centre of the LMC bar, at $\alpha = 05^{h}20^{m}56^{s}$, $\delta = -69\degr 28\arcmin 41\arcsec$ (J2000.0)\ \cite{bicaubv}. \\
$^{b}$ Relative to the H\,{\sc i} rotation centre of the LMC, at $\alpha = 05^{h}26^{m}39^{s}$, $\delta = -69\degr 15\arcmin 36\arcsec$ (J2000.0)\ \cite{rohlfs,westerlund}
\end{minipage}
\end{table*}

\section{Photometry}
\label{reduction}
As part of the archive retrieval process, all frames were reduced according 
to the standard {\em HST} pipeline, using the latest available calibrations.
This is preferable to using the original data as reduced soon after
observation because the latest calibrations have a much longer baseline for 
calculation and are therefore more accurate. Bearing in mind that our
data set covers a span of more than five years of {\em HST} observations,
we also maintain the homogeneity of our reduction process by using
calibrations from a single epoch. 

For photometric measurements we have found Dolphin's HSTphot
\cite{hstphot} to be the most suitable software package. Our large
number of observations requires the reduction process to
be run automatically where possible, without sacrificing data integrity.
HSTphot is well suited for running in batch mode and is specifically 
tailored to
the reduction of WFPC2 frames, routinely accounting for the severely 
under-sampled PC/WFC point spread functions (PSFs) and their variation
due to sub-pixel positioning. In particular, the 
{\em multiphot} routine has proven to be the most useful for our
data set. This program first aligns accurately (to fractions of
a WFC pixel) and then performs simultaneous photometry on
multiple images of the same field, regardless of the filter combination.
This makes it perfect for the current data set -- with two frames per
field, each through a different filter. Conventionally, this would make 
the removal of cosmic rays and cross identification of objects 
complicated and unreliable. By solving the two frames 
simultaneously, both these problems are accounted for. 

Before starting {\em multiphot} each frame is readied with the image 
preparation utilities included with the HSTphot package. This procedure 
includes the masking of bad data\footnote{Such as known bad pixels, saturated 
pixels, charge traps, bad columns, and the vignetted region between chips}
as flagged by the STScI data quality image which accompanies each 
observation, a first attempt at the 
removal of cosmic rays (using a routine based on the {\sc iraf} task
{\sc crrej}), the removal of hot-pixels not flagged by the data 
quality image (using a $\sigma$-clipping algorithm), and the robust
determination of a background image -- used by {\em multiphot} for 
stellar detection and measurement.

Photometric measurements are made using {\em multiphot} in PSF fitting 
mode. While aperture photometry mode is also an option, we have found that 
PSF fitting produces main sequences on our CMDs that are at least as narrow as 
those from aperture photometry using 2 WFC pixel and 3 PC pixel radii. 
This is presumably because {\em multiphot} is designed specifically for
measurements involving the unique WFPC2 PSFs. The detection algorithm
used by {\em multiphot} is similar in principle to that of DoPHOT, being
based around the recursive location and subtraction of objects at
increasingly lower threshold levels. We set a minimum threshold level for
detection of $3 \sigma$ above the background. In addition, we
enable a feature of {\em multiphot} which calculates an adjustment
to the background image before each photometry measurement using the
pixels just beyond the photometry radius \cite{hstphot}. This is designed
to account for rapidly varying backgrounds such as those expected in 
observations of the crowded central regions of globular clusters.

The object classification, sharpness and $\chi$ parameters 
produced from {\em multiphot's} PSF fitting are used to keep the 
photometry clean of non-stellar objects and spurious detections. The object 
classification parameter is a determination as to which prototype PSF the 
measured profile of an object conforms to -- the options being 
stellar (or a marginally resolved stellar pair), extended (galaxy), or single 
pixel (cosmic ray). The sharpness value for an object is a measure of 
how sharp its profile is in comparison with the stellar PSF, being 
negative if the profile is too broad, zero if the object is a perfect 
fit and positive if the profile is too sharp. Finally, the $\chi$ value 
simply measures the quality of fit of the PSF to the object profile. 
After some experimentation, we found that selecting objects with 
sharpness between $-0.6$ and $+0.6$ and $\chi \le 3.5$ provided a 
mostly complete stellar sample for the purpose of constructing surface 
brightness profiles. For the construction of clean CMDs, these values 
would be somewhat stricter.

The final photometric measurements obtained from {\em multiphot} are
from PSFs corrected for geometric distortion \cite{holtzman}, the
filter-dependent plate scale changes (determined empirically) 
\cite{hstphot}, and the 34th row error \cite{shaklan,kingrow}.
The magnitudes are also corrected for charge-transfer efficiency (CTE) 
effects using the longest baseline calibration available -- that of 
Dolphin \shortcite{wfpc}. Notably, this calibration also provides 
corrections for ``warm'' data -- that is, observations taken before the 
WFPC2 cool-down of 1994 April 23. This is especially important for our 
data set, since 25 clusters were observed in warm conditions.
The stellar magnitudes are also corrected for PSF residuals and to an 
aperture of $0 \farcs 5$ using groups of isolated stars selected 
according to a set of strict criteria \cite{hstphot}. Finally, 
{\em multiphot} uses the zero-points from the Dolphin \shortcite{wfpc} 
calibration. We do not convert from the {\em HST} instrumental system to 
the Johnson-Cousins system because this transformation is 
not important for the construction of surface brightness profiles, 
instead adding unnecessary scatter to the photometry. Dolphin 
\shortcite{hstphot} estimates the limiting photometric and astrometric 
accuracy of {\em multiphot} to be 0.011 to 0.014 mag and 0.05 pixels 
($2.5 \times 10^{-3}$ arcsec on the PC and $5.4 \times 10^{-3}$ arcsec 
on the WFCs) respectively.

\section{Surface brightness profiles}
\label{sbprofile}
Constructing a surface brightness profile for a given cluster 
is simple in concept; however in practice, our calculations were complicated
by the high resolution and peculiar chip geometry of WFPC2. The procedure is 
similar to that for ground-based data however, and we broadly follow the
techniques outlined by Djorgovski \shortcite{djorgovski}.

\subsection{Astrometry and centre determination}
\label{astro}
One problem presented by the data was due to the four chip structure of 
WFPC2. In order to maintain the accurate relative spatial positioning 
of stars from chip to chip, we overlaid a uniform coordinate system,
accounting for the geometric distortion from {\em HST's} optical system, the
changes in pixel scale from the PC to WFCs and the separations between the 
chips. This was achieved using the {\sc iraf}\ {\sc stsdas} task {\sc metric},
which converts the chip and pixel coordinates for a list of stars to
pixel coordinates relative to the WFC2 chip, making the appropriate geometric
corrections, and then converts these corrected pixel coordinates to
$(\alpha,\delta)$ (J2000.0) using the positional information in the image 
headers. This header information is not always reliable, and
can introduce errors into the $(\alpha,\delta)$ so calculated. To
avoid this, we took the corrected pixel coordinates from 
{\sc metric} for each star and derived angular separations using the WFC pixel 
scale of $0.0996$ arcsec pixel$^{-1}$, rather than taking the $(\alpha,\delta)$
for each star and calculating angular separations from these. All centering
calculations, annulus construction, and completeness and area corrections 
(see below) used these corrected pixel coordinates. The only
results which incorporate image header positional information are the calculated
centre coordinates expressed in $(\alpha,\delta)$ (see Table \ref{params}).

The next task was to locate accurately the centre of any given 
cluster\footnote{Or more correctly, the central surface brightness peak.}. 
Poor centering will tend to artificially flatten a surface 
brightness profile, and this in turn can lead to large systematic errors in 
any structural parameters derived from the profile, and may obscure important 
dynamical information such as the presence of a post-collapse core. There are 
several algorithms for locating a cluster's centre which rely on the symmetry 
properties of the cluster, one example being the mirror-autocorrelation method
described by Djorgovski \shortcite{djorgovski}. However, such algorithms were 
not completely suitable for our data. Unlike Galactic globular clusters, many
Magellanic Cloud clusters do not have smooth profiles, but instead are
clumpy, irregular and not particularly symmetric, making it difficult
to apply symmetry-algorithms robustly and uniformly across the sample. 
In addition, and compounded by the high resolution and small field of view of 
WFPC2, several clusters have very low surface brightness, adding to the problem. 
The high resolution in particular means that it is not sufficient to simply 
apply spatial smoothing to most clusters, because the magnitude of 
the smoothing required degrades the intrinsic accuracy of the centre 
determination, and the process therefore loses its value.

Instead, we employed a Monte Carlo style method to locate our cluster 
centres. In this procedure, each chip is first split into boxes of equal area 
($\sim 100$ WFC pixels on a side). For each box, the surface brightness is 
calculated by adding up the fluxes for all stars in the box (including 
completeness corrections -- see Section \ref{completenesscorr}) and dividing by the 
area of the box. The search is then narrowed to the region covered by the box with the 
highest surface brightness, and the eight boxes surrounding it. It does not matter 
if this region falls across more than one CCD -- we simply account for the vignetted
area (where no stars are detected) in the subsequent calculations. If the
box with the highest surface brightness falls right on the edge of a CCD, where
it is not joined by another CCD, then the number of surrounding boxes is
less (five if the central box is on an edge; three if it lies in a corner)
and the amalgamated region is correspondingly smaller. Within this amalgamated
region, points are randomly generated and the surface brightness calculated (as
for the boxes) in a circle of radius $r$ about each point. If part of one of
these circles falls off the edge of a chip or over a vignetted region, this 
``lost area'' is accounted for in the calculation. After $N$ tries, the point 
corresponding to the circle with the highest surface brightness is taken to be the 
cluster centre. 

This routine has the disadvantage that it is very good at finding the brightest
star in a cluster, and although this does represent a surface brightness peak, it
generally does not correspond to the overall cluster surface brightness peak.
To avoid this, we would ideally like to exclude the few brightest stars in a 
cluster from the above calculations, and instead use a sub-sample which should
closely trace the cluster's surface brightness profile (i.e., a sample which has
its greatest density coincident with the cluster's surface brightness peak). 
Main sequence stars are such an ensemble, and so for each cluster we impose
colour and magnitude limits on the cluster CMD to select only these stars for
the centre determination. In general this procedure worked well; however for
clusters with severe central crowding, it was necessary to relax the limits
and include brighter stars from the giant branch (or upper main sequence in the
case of very young clusters).

While inelegant and computationally expensive, our random sampling method is 
robust and easily adapted for clusters of different concentration and richness 
simply by variation of $r$ and $N$. As a consistency check, we run the centering 
algorithm independently on both colour frames for a field. For $N=2000$ the 
centering is repeatable to a limiting accuracy of approximately $\pm 10$ WFC pixels
-- approximately $\pm 1\arcsec$. As noted above however, the conversion from pixel
coordinates to $(\alpha,\delta)$ may introduce large errors
if the image header information is inaccurate. A comparison of our calculated 
centres (Table \ref{params}) with those in Table \ref{ages} shows good 
agreement, meaning that we have not introduced any large errors. We note however 
that this comparison is not sensitive to small (several arcsecond) errors.
Notwithstanding this, we expect that our calculated centres are at least as
accurate as the literature values.

\subsection{Annulus construction and flux corrections}
For each frame (two per cluster), four sets of circular annuli were 
constructed about the cluster centre. Two sets had narrow annulus widths 
of $1\farcs5$ and $2\arcsec$ respectively and were designed for sampling the 
central regions of a cluster. To this end, the $1\farcs5$ annuli were 
calculated to a radius of $\sim 20\arcsec$ and the $2\arcsec$ annuli to 
$\sim 30\arcsec$. The two other annulus sets had larger widths of $3\arcsec$ 
and $4\arcsec$ respectively, and were used primarily for sampling the outer 
(less dense) regions of a cluster. These two profiles were therefore extended 
to the maximum radii possible. Although many LMC clusters have slightly
elliptical isophotes at extended radii, there is little evidence to 
suggest that the mass distribution in inner regions is in general 
not spherical (cf. E91) -- hence we fit circular annuli. In addition, because 
of the small WFPC2 field of view, none of the images in the data set fully 
cover any cluster, making the use of elliptical annuli impractical.

Because of the high resolution of our images we simply counted stars to 
calculate the surface brightness for an annulus. For each set, the surface 
brightness $\mu_{i}$ of the $i$-th annulus is given by:
\begin{equation}
\mu_{i} = \frac{A_{i}}{\pi (b^{2}_{i} - a^{2}_{i})} \sum_{j=1}^{N_{s}} C_{j} F_{j}   
\label{sb}
\end{equation}
where $b_{i}$ and $a_{i}$ are the outer and inner radii of the annulus 
respectively, $N_{s}$ is the number of stars in the annulus, and $F_{j}$
is the flux of the $j$-th star. The factors $A_{i}$ and $C_{j}$ are
the area correction for an annulus and the completeness correction
for a star respectively, and must be determined before the annulus
is constructed. We outline the meaning of these factors and their
methods of calculation individually below.

\subsubsection{Area corrections}

The area correction $A_{i}$ for the $i$-th annulus is used simply to
normalize the flux in the annulus to that for a full annulus. This is
necessary because the shape of WFPC2 means that for all clusters, most annuli 
are not completely covered by the field of view. Since the flux through an 
annulus is directly related to the annulus area, variations in the fractions 
of annuli covered cause artificial fluctuations in a surface brightness 
profile, and must therefore be accounted for. 

The process of determining the area correction for an annulus is
complicated by the WFPC2 chip geometry (including the small separations
between the CCDs), the centering of the cluster on the camera and the
roll angle of {\em HST} at the time of observation. The
arbitrary nature of the second two factors from cluster to cluster
once again led us to resort to an inelegant but robust Monte Carlo scheme.
For a given cluster we determine corrected pixel coordinates for the sixteen 
CCD corner pixels which define the WFPC2 field of view, and from these we derive 
the border equations of the four CCDs. A large number of points are then randomly 
generated over the full area of each annulus, and 
using the CCD boundary equations we determine which points fall on the WFPC2 
camera, and which do not. The area correction $A_{i}$ for the $i$-th annulus 
is the total number of points generated divided by the number ``imaged''.
Again, while computationally inefficient, this method is very robust and can 
easily account for the arbitrary geometry of any particular observation, with 
an accuracy limited only by the number of random points. A small amount of 
experimentation showed ten thousand points per annulus to be sufficient.
To avoid introducing large uncertainties into the data we do not
use annuli for which the fraction covered falls below a third. This
limits the maximum radius for a surface brightness profile to be 
$\sim 75-80\arcsec$.

An example of the process is shown in Fig. \ref{areacorr} -- a pointing for
NGC 1841. Annuli of width $4\arcsec$ have been drawn about the centre to 
$100\arcsec$, and for every second annulus the random points falling in 
the field of view have been plotted. 
The accuracy with which the process handles the complicated geometry is quite 
evident -- note particularly the small separations of the chips due to the 
vignetted regions masked by HSTphot, and the significant errors these can 
cause to an annulus of the right radius which falls tangent.
At $100\arcsec$ the area correction factor is approximately $4$, and in
practice the process was halted at $r = 78\arcsec$.

\subsubsection{Completeness corrections} 
\label{completenesscorr}

Even with {\em HST} resolution, crowding and saturation can cause 
significant numbers of stars to be missed by automated detection software 
such as HSTphot. This missing flux can seriously affect a cluster's surface
brightness profile and must be accounted for. To 
quantify the correction we used the artificial star routine attached to
{\em multiphot}. For each cluster, we set this routine to generate 
$\sim 3.5\times 10^{5}$ stars per CCD (i.e., $\sim 1.4\times 10^{6}$ 
stars per cluster) on a CMD, the limits of which are set
to be two magnitudes above the saturation limit and two magnitudes below
the faint limit for an image, and 0.5 magnitudes redder than the
reddest region of the cluster's CMD and 0.5 magnitudes bluer than
the bluest region. The stars are spatially distributed according to
the flux of the image. Each star is placed on the image and solved 
(one at a time) by {\em multiphot}, using the same settings as for the 
real data. Stars are flagged if they are not recovered. The output 
photometry file is exactly similar to the {\em multiphot} output for 
real stars, and we run this artificial photometry through our 
classification, sharpness and $\chi$ selection criteria, again 
flagging stars which are removed from the sample. 

To determine the completeness function for a given chip, the fake stars
are binned according to their $x$- and $y$-position on the chip,
magnitude and colour. Bin sizes are typically 160 pixels in $x$ and $y$ in uncrowded 
regions (independent of chip), 0.2 magnitudes in brightness and 0.25 magnitudes
in colour. A large majority of our sample suffers little or no crowding,
even in the central regions of a cluster. This means that the binning resolution
described above, particularly in the positional sense, is perfectly adequate to 
account for the gradual spatial variations in completeness. This is demonstrated
below, in the example for NGC 2213, which is a typical cluster in the sample.
For severely crowded regions however, the completeness may suffer rapid 
and significant spatial variation, and in these cases the resolution of the 
positional binning was increased to account for this. In the very worst cases,
the inclusion of short exposures (see Section \ref{short} below) 
usually significantly alleviated any crowding problems.

The completeness $c$ for a given bin is the number
of successfully recovered (unflagged) stars divided by the
total number of stars generated in that bin, and the completeness correction
$C_{j}$ for the $j$-th star in an annulus the inverse of $c$ for the 
appropriate bin. To avoid large uncertainties, we eliminated any stars with 
completeness less than $c_{l} = 0.25$. Experimentation showed that variation of 
this limit (i.e., $0.05 \le c_{l} \le 0.5$) had a negligible effect on most 
profiles, reflecting the fact that incompleteness was not an issue for the large
majority of clusters.

Examples of the completeness functions for two clusters (the relatively open 
cluster NGC 2213 and the extremely compact cluster NGC 2005) are shown in 
Fig. \ref{completeness} together with the PC F555W images of the centres of 
these clusters. For ease of visualization the completeness functions have 
been integrated over position and colour to be functions of brightness 
only. For NGC 2213, crowding and saturation are not significant,
and the completeness functions reflect this. The core of the cluster is imaged
on the PC chip (solid line), and the completeness function for this chip
matches well the completeness functions for the three WFC chips, which imaged the
outer regions of the cluster. If incompleteness (or rapid spatial variation
of completeness) was a significant issue in the central regions, then the
position averaged PC function would be significantly degraded in
comparison to the three WFC functions. The second cluster in Fig. \ref{completeness},
NGC 2005, is an example of this. The effect of this cluster's extremely compact 
core on the PC completeness function is clear, and it is evident that a positional
resolution of 160 pixels is not adequate to fully describe the completeness variations.
In fact, in the very central regions, the situation is even worse than this -- the 
completeness is so low that the measurements become meaningless. To demonstrate this, 
from the three WFC functions, we expect the non-crowded regions of the PC to have 
$c\sim 0.85$ at $V_{555}\sim 24$, whereas we measure a spatially averaged value of 
$c\sim 0.45$. From the PC image, approximately half of the chip may be non-crowded, 
implying that the integrated completeness in the 
crowded region must be $c\la 0.05$ at $V_{555} \sim 24$ (cf. Fig. 
\ref{mergephot}(b)). In a case such as this, we had to use additional information from 
a short exposure (see Section \ref{short}, below), otherwise the surface brightness 
profile became useless.

We note that none of the completeness functions plotted in Fig. \ref{completeness} 
ever reaches $c=1$ -- this is due to the integration over 
colour and position, since bad pixels and bright stars prevent {\em all}
fake stars being recovered. Certainly however, individual bins in 
position-brightness-colour space can (and do) have full completeness. 
Completeness functions can also tell us a small amount 
about the background field near a cluster. The degradation evident in the WFC 
completeness functions for NGC 2005 is primarily due to the dense field 
population in the region of this cluster (NGC 2005 lies very near the LMC bar). 
By comparison, the field population near NGC 2213 is quite sparse.

\subsubsection{Short exposures}
\label{short}

As discussed in Section \ref{completenesscorr}, in nine cases 
where severe crowding and/or saturation significantly degraded the completeness
in parts of an image, we used information from short 
exposure images to complete the surface brightness profiles. This occurred
specifically for the very old, compact clusters NGC 1754, NGC 1835, NGC 1916, 
NGC 2005, and NGC 2019, and the very young clusters NGC 1984, NGC 2011, NGC 2156, and
R136. NGC 1786 required correction but no additional data was available.

To add the short exposure data is a relatively straightforward process. For a given 
cluster, the short exposure image is reduced in exactly the same manner as the original 
image, and a surface brightness profile constructed. When plotted 
over the original surface brightness profile, the short exposure profile is found 
to have a shape matching that of the original at large radii (where neither
crowding nor saturation are an issue), but with a larger scatter and brighter
by some tenths of a magnitude. This offset is a product of the shorter 
image having a brighter saturation limit and therefore including
more bright stars in its profile than the longer image, whereas the larger 
scatter is due to both the inclusion of many of the brightest stars in the
cluster (see also Section \ref{saturated}) and the exclusion of large numbers of 
faint stars, which do not get measured on the short image. To eliminate the offset,
we impose a brightness limit $V_{lim}$ on the short exposure photometry 
to cut out the brightest stars, and vary $V_{lim}$ until the profiles 
overlap in their outer regions. This is equivalent to removing the brightest stars
in the cluster from the calculations, as discussed below in Section \ref{saturated},
and so also reduces the noise in the short exposure profile. It is then simple to 
measure the radius $r_{dev}$ at which incompleteness becomes an issue and the two 
profiles begin to deviate, and form a composite profile using photometry 
for stars with $r \le r_{dev}$ from the short exposure (with 
$V_{lim}$ imposed) and the remainder (with $r > r_{dev}$) from the long 
exposure. This keeps scatter to a minimum while alleviating 
the incompleteness at small radii.  

Fig. \ref{mergephot} shows an example of this process for NGC 2005. In the previous
Section (\ref{completenesscorr}) we showed that the long exposure photometry for this 
cluster suffered serious incompleteness in the central region. Further evidence of this
can be seen in Fig. \ref{mergephot}(a). The two surface brightness profiles show a 
similar shape up to $\log r_{dev} \sim 0.8$, or $r_{dev} \sim 6\farcs3$. At this point 
incompleteness in the long exposure photometry becomes significant and the
surface brightness profile eventually turns over because of the 
amount of missing flux. Not only is the completeness very low in this
region, but it must vary on a scale shorter than $\sim 6\arcsec$, or
$\sim 130$ pixels. To account for the crowding, measurements are taken from the short
photometry within $6\farcs3$. The advantage gained from this is evident in Fig.
\ref{mergephot}(b), which shows the completeness functions for the long and short
exposure photometry, for $r < r_{dev}$. As before, the completeness functions have 
been integrated over position and colour to be a function of brightness only. The terrible
incompleteness in the long exposure photometry is very evident. It is also clear
that the short exposure photometry within the central region accounts very well for this
incompleteness, and we are therefore justified in using only the short exposure
photometry for the surface brightness profile in this region.

\subsubsection{Saturated stars}
\label{saturated}
\begin{table}
\caption{Clusters with more than five saturated stars within two core radii.}
\centering
\begin{tabular}{@{}lccc}
\hline \hline
Cluster \hspace{5mm} & \hspace{2.5mm} $\log\tau$ \hspace{2.5mm} & \hspace{2.5mm} $N_{s}$ $^{a}$\hspace{2.5mm} & \hspace{2.5mm} $f_{s}$ $^{b}$ \hspace{2.5mm} \vspace{0.5mm} \\
 & (yr) & & \\
\hline
NGC1466	& $10.10$ & $15$ & $0.002$ \\
NGC1711	& $7.70$  & $45$ & $0.018$ \\
NGC1786	& $10.18$ & $\sim20^{c}$ & $\sim0.008$ \\
NGC1805	& $7.00$  & $30$ & $0.034$ \\
NGC1818	& $7.40$  & $35$ & $0.016$ \\
NGC1841	& $10.09$ & $\sim50^{d}$ & $0.003$ \\
NGC1866	& $8.12$  & $30$ & $0.009$ \\
NGC1898 & $10.15$ & $35$ & $0.006$ \\
NGC2031 & $8.20$  & $35$ & $0.012$ \\
NGC2159 & $7.60$  & $25$ & $0.022$ \\
NGC2164 & $7.70$  & $30$ & $0.016$ \\
NGC2172 & $7.60$  & $30$ & $0.033$ \\
NGC2210 & $10.20$ & $30$ & $0.005$ \\
NGC2257 & $10.20$ & $10$ & $0.001$ \\
\hline
\end{tabular}
\medskip
\\
\begin{flushleft}
$^{a}$ Approximate number of saturated stars within two core radii. \\
$^{b}$ $N_s$ divided by the total number of stars within two core radii. \\
$^{c}$ This figure is an estimate since NGC 1786 suffers from extreme crowding and no short exposure was available. \\
$^{d}$ NGC 1841 has an extremely large core radius.\\
\end{flushleft}
\label{satclus}
\end{table}

At this stage we comment briefly on the treatment of saturated
stars in our images. Such stars were not measured by {\em multiphot} and were
therefore not included in the construction of surface brightness profiles.
For most clusters, saturated stars are not a significant factor. Of the 53 clusters
in the sample, thirty had less than five saturated stars within approximately
two core radii of the centre. When data from short exposures were included to
circumvent extreme crowding and saturation (see Section \ref{short}), this number
rose to 39. For all of these clusters, the presence of so few saturated stars
meant that their neglect did not significantly alter the surface brightness profiles
derived for these clusters, especially given that on average several thousand stars
were measured in the central regions of each.

The remaining fourteen clusters are listed in Table \ref{satclus}, along with an
estimate of the total number of saturated stars within two core radii for each cluster.
We felt comfortable leaving these stars out of the calculations, for the following
reasons. Firstly, their presence did not degrade the quality of the images or the
measurements made from these images (otherwise short exposure data would have been
obtained -- see Section \ref{short}). Secondly, it is clear that for each cluster, the 
fraction listed in Table \ref{satclus} is tiny (although larger than for the other
39 clusters). The fourteen clusters in Table \ref{satclus} are either
young ($\tau \la 100$ Myr) or very old ($\tau > 10$ Gyr). In such clusters, the 
brightest stars are either giants or upper
main-sequence stars and are in very brief (but very luminous) phases of
evolution, especially compared to the dynamical timescales (e.g., the median relaxation
time) of the clusters they are in. They therefore may not in general represent the 
spatial distribution of the underlying stellar population. It does not make sense to 
measure a surface brightness profile which is dominated by the output of an essentially
randomly distributed tiny fraction of the stars in a cluster -- in practice, this
simply adds large amounts of noise to the profile. 

Removing these stars from the calculations provides less noisy profiles 
without compromising the measurement of structural parameters. Elson and 
collaborators for example (Elson et al. \shortcite{efl}; E91; Elson 
\shortcite{elsonfive}) routinely remove the brightest stars when calculating their 
surface brightness profiles, in order to make sure the profile reflects the 
underlying stellar distribution. Specifically, Elson et al. \shortcite{efl}
found that the removal of these stars (in the construction of their ``cleaned'' 
profiles) reduced the scatter in their parameter measurements by $\sim 40$ 
per cent on average without altering the parameters by more than 5 per cent.
Elson \shortcite{elsonfive} removed the brightest 120 stars per cluster in the 
construction of her cleaned profiles - again, this provided much smoother profiles 
without altering the derived structural parameters by a significant amount.
Some experimentation on our part showed similar results. While we do not 
find it necessary to remove quite so many stars, in the cases 
listed in Table \ref{satclus}, it is evident from the final results presented 
in Fig. \ref{plots} that leaving the brightest (saturated) stars out of the
construction process has resulted in clean, low noise profiles, which should reflect
the underlying stellar distribution of each cluster.

\subsubsection{Error estimation}
\label{errors}
To estimate the internal error $\sigma_{i}$ for an annulus we divided it 
into eight sectors of equal area, and calculated the surface 
brightness in each using Eq. \ref{sb} with the factors $A_{i}$ set to unity.
The internal error for the annulus is the standard deviation of the 
surface brightness values for these eight sectors \cite{djorgovski}. This 
technique ensures that the internal error of an annulus reflects the error 
due to the discreteness of the light distribution - that is, clumps of stars
or single bright stars will increase the surface brightness
of an annulus, but also its error by an appropriate amount. This is essential
in the case of LMC clusters, many of which are patchy and irregular. 
Similarly, the error for an annulus reflects the width 
of that annulus -- important given our four different annulus sets.
  
In the outer regions of most clusters however, we found that the
error bars so calculated were considerably smaller than the random scatter in the
points. As an example see Fig. \ref{backgroundfit}, which demonstrates
the background subtraction technique described in Section \ref{background}
but also well illustrates the present issue. The under-estimation is 
caused because $A_{i} \sim 3$ for the outer annuli, so that 
while the flux for such an annulus is scaled to that for the full area, 
the errors are calculated over only a third of the full area and 
therefore do not reflect any large scale variations for that annulus. 
This effect is coupled with field star contamination for many clusters.
This background is negligible in the inner regions of a cluster, but
it can become significant in the outer regions. The background tends to be
much more smoothly distributed than the cluster itself, and so artificially 
shrinks the error bars determined using sectors. To solve this problem, for 
every annulus we also calculated the error from Poisson statistics. When 
this became significantly larger than the sector error, we took the Poisson error 
instead. This condition only ever became true in the outer regions of clusters,
meaning that in the central regions of a cluster the errors are sector errors, while
in the outer regions they are Poisson errors, with a smooth transition across some 
intermediate region (usually $\log r \sim 1.6$). The Poisson errors tend to 
over-estimate the scatter between points -- this is due to the relatively low numbers
of stars in the large outer annuli -- but we find this preferable to an under-estimate,
especially given our process of background subtraction (see Section \ref{background}).

Finally, we note that for some clusters, the (inner) sector errors are also larger than 
the RMS point-to-point scatter. In considering this, it is 
important that only the points in a single annulus set are compared -- consecutive 
points from different annulus sets use at least some of the same information, so one
would expect their scatter to be small. One should also examine the brightness range
of a profile -- some cover up to 9 magnitudes while others only 2 or 3 magnitudes. This
can give the impression of enormous error bars on some profiles and tiny ones on others.
Nonetheless, particularly in low surface brightness clusters such as NGC 1841, NGC 2121, 
NGC 2209, SL 663, and SL855, the effect persists.
It is due to the low density of these clusters -- the stars (especially the
bright stars) are sparsely distributed. Therefore, while the surface brightness
for each annulus is the azimuthal average over a comparatively large area, the eight sectors 
used to calculate the error for an annulus may have a considerable standard deviation in 
their surface brightnesses and correspondingly result in errors which are larger
than the RMS scatter between points. In the richer clusters (e.g., NGC 1466, NGC 1805,
NGC 1831, NGC 2210, etc) the errors match the RMS point-to-point scatter well.

\subsection{Profile fitting and background subtraction}
\subsubsection{King versus EFF profiles}

Traditionally, the surface brightness profiles of old globular 
clusters are described by families of King-type models -- for example the 
empirical King \shortcite{empking} models:
\begin{equation}
\mu(r) = k \left\{ \frac{1}{\left[1+(r/r_{c})^{2}\right]^{\frac{1}{2}}} - \frac{1}{\left[1+(r_{t}/r_{c})^{2}\right]^{\frac{1}{2}}} \right\}^{2}
\label{kp}
\end{equation}
where $r_{t}$ is the tidal radius of the cluster, and $r_{c}$ the core
radius. Provided $r_{t} >> r_{c}$, the core radius may be taken as the
radius at which the surface brightness has dropped to half its central
value. It is therefore useful to define the concentration parameter
$c = \log(r_{t}/r_{c})$. For LMC clusters, concentration 
parameters of $1.0 \la c \la 2.0$ are measured (cf. EFF87) implying
ratios $10 \la r_{t} / r_{c} \la 100$.

However, studies of young LMC clusters by Elson and collaborators
(EFF87; Elson et al. \shortcite{efl}; E91) have shown that these clusters
do not appear to be tidally truncated, even at radii of several hundred 
arcseconds. EFF87 argue that the lack of tidal truncation in a young cluster 
is due to the cluster expanding over its Roche limit as a result of mass loss 
or violent relaxation. Given that a young cluster has only completed a 
fraction of its orbit about the LMC, any stars outside the Roche limit will 
not have been tidally stripped and instead surround the cluster in an unbound 
halo. In this case, a more suitable profile is:
\begin{equation}
\mu(r) = \mu_{0} \left(1 + \frac{r^{2}}{a^{2}} \right)^{-\frac{\gamma}{2}}
\label{ep}
\end{equation}
where $\mu_{0}$ is the central surface brightness, $a$ is a measure of the
core radius and $\gamma$ is the power-law slope at large radii. For 
$\gamma \sim 2$ this is essentially Eq. \ref{kp} with 
$r_{t} \rightarrow \infty$. We hereafter refer to profiles of the form of 
Eq. \ref{ep} as EFF models, after Elson, Fall \& Freeman \shortcite{eff}, 
who first introduced them into star cluster research. Fig. \ref{profiles} 
shows an empirical King model compared with an EFF model, both having been 
calculated with a core radius of $r_{c} = 10\arcsec$, typical of the LMC 
clusters measured by EFF87 and Mateo \shortcite{mateo}. For the King profile 
we take $c=1.5$. It is evident that there is little deviation between the two 
profiles in the inner regions. The difference only becomes large at radii 
approaching $r_{t}$.

The presence of both very young and very old clusters in our sample 
raises a potential dilemma over the fitting of one type
of profile or the other, particularly for intermediate age clusters.
Fortunately this is solved for us by the small field of view of WFPC2,
which limits our surface brightness profiles to have maximum radii
of $r \sim 80\arcsec$ or $\log r \sim 1.9$. From Fig. \ref{profiles}
we see that at such radii the two profiles are essentially 
indistinguishable, and so we may fit either. Because of the virtual
impossibility of obtaining any accurate information about $r_{t}$ without
measuring to at least this radius, we chose to fit EFF profiles to
our data.

The parameter $a$ in an EFF profile may be linked to the King core 
radius by setting $\mu = \mu_{0} / 2$ in Eq. \ref{ep} and rearranging to
obtain
\begin{equation}
r_{c} = a(2^{2 / \gamma} - 1)^{1 / 2}
\label{rc}
\end{equation}
We note that this assumes $r_{t} >> r_{c}$, which is valid for most
of the clusters in our sample. For several however (e.g., NGC 1841 and
NGC 2257) $r_{c}$ is very large and $\gamma$ quite steep. In such cases, 
Eq. \ref{rc} may underestimate the core radius by 3 to 10 per cent 
\cite{elsonfive}.

\subsubsection{Fitting procedure and background subtraction}
\label{background}
When fitting the profiles described in the previous section, in most 
cases it was necessary to correct for field star contamination.
Especially for clusters near the LMC bar, the stellar background 
contribution is not negligible and tends to flatten out the cluster
profiles in their outer regions (see e.g., Fig. \ref{mergephot}). Because 
$\gamma$ is largely determined from the slope of a profile well away from the 
core and is therefore very sensitive to field star contamination, we had to
subtract the background level in order to avoid large systematic errors.
Traditionally one would use an observation well away from the
cluster to obtain the local background, and subtract this level from each
annulus. However, we had no access to such background fields for our
entire sample, nor were any of our fields large enough in area to
allow background determinations away from a cluster. We therefore
had to try and determine the background levels individually from each 
surface brightness profile.

One way of doing this is to fit Eq. \ref{ep} but with an extra parameter 
$\phi(r)$ added -- we set $\phi$ as a constant to represent a
simple background. However, these four-parameter fits did not work as well as 
desired -- they tended to over-estimate the background, causing $\gamma$ to 
be too steep by factors of up to three. The reason for this is not clear; it 
would seem that the creation of an additional degree of freedom degraded the 
quality of the fit for the original three parameters. We therefore implemented
a recursive method using the fact that for $r >> a$, Eq. \ref{ep}
with a background contribution $\phi$ becomes:
\[
\mu(r) \,\,\approx\,\, \mu_{0}\left(\frac{r}{a}\right)^{-\gamma} + \,\, \phi
\]
\begin{equation}
\hspace{7mm} \,\,\approx\,\, \mu_{0}\left(\frac{r}{r_{c}}\right)^{-\gamma} (2^{2/\gamma}-1)^{-\gamma/2} \,\, + \,\,\phi
\label{bgrfit}
\end{equation}
with Eq. \ref{rc} substituted for $a$.

We first fit an EFF model to the inner region of a cluster profile using 
the two narrow annulus sets, to estimate $\mu_{0}$ and $r_{c}$. In this region 
($r \la 30\arcsec$) the background contribution is negligible, in all cases
being at least 5 magnitudes below the central surface brightness. The 
background therefore does not affect a measurement of the central surface
brightness $\mu_{0}$ nor the core radius $r_{c}$, which is essentially the
shape of the profile in the inner region. Hence it is a good approximation
to determine these without subtracting any background. We 
next use these estimated values and fit a two-parameter model $(\gamma,\phi)$
of the form of Eq. \ref{bgrfit} to the outer region of the profile, using
the two wider annulus sets and ensuring $r >> a$. This allowed us to 
determine $\phi$ and subtract it from every annulus. After 
the background subtraction, we calculate the Poisson errors for each point 
and substitute as appropriate (see Section \ref{errors}). Finally, we fit 
another EFF model to the new subtracted profiles, using all four annulus 
sets, and thereby determine the best-fitting parameters. This method allows 
us to treat severely contaminated clusters (e.g., NGC 1898 
and NGC 1916) equally with clusters that have no evident background 
contribution (e.g., Hodge 11 and NGC 1841). 

To test the approximation described above, we examined the final 
parameters $(\mu_{0},r_{c})$ to see how well they corresponded to the
determinations from the central fit. In no cases did we observe significant
deviation, suggesting that the approximation is good. In addition, we checked 
for any systematic errors introduced by this method by comparing our
determinations of $\gamma$, which is very sensitive to the background 
level, with previously published work. This is presented in Section \ref{prevstud},
and suggests that we introduced no significant systematic errors. A final
point concerns the effect of non-constant backgrounds such as might
be encountered near the bar -- in this case, $\phi = \phi(r)$. Such a background
would have two effects given the procedure described above. The first would
be to cause a very small inaccuracy in our determined centre. The second would
be to increase the noise in the outer region of a profile, because of the
fact that we image only a fraction of any outer annulus. This in turn would
cause a larger random error in the determination of $\gamma$, but this
would be accounted for by our bootstrapping method of determining errors in
the derived parameters (see below).


Fig. \ref{backgroundfit} shows an example of the background fitting,
again for NGC 2005. In this case, the estimated parameters
from the initial fit were $a\sim 4\arcsec$ and $\mu_{0}\sim16.9$, 
and the background measured as $V_{bg} = 21.83$. 
As mentioned in Section \ref{completenesscorr}, NGC 2005
is very near the LMC bar, and has a densely populated background field. The 
derived background intensities for this cluster and the others in the sample are 
consistent with the LMC isophote maps of de Vaucouleurs \shortcite{dev}. 
Equivalently, a plot of background intensity vs. $R_{opt}$ shows the expected
decrease with increasing $R$. 

In implementing each fit, we develop a grid in parameter space and
choose the combination of parameters which minimizes the weighted sum:
\begin{equation}
\chi^{2} = \sum_{i=1}^{N_{a}} \left ( \frac{\mu_{i} - \mu(r_{i},\mu_{0},a,\gamma)}{\sigma_{i}} \right )^{2}
\label{chisq}
\end{equation}
where $\mu_{i}$ is the surface brightness of the $i$-th annulus, 
$\sigma_{i}$ is the error in this value, $N_{a}$ is the total number of
annuli in the set in question, and the other parameters are as defined 
in Eq. \ref{ep}. A refined mesh is then expanded about this parameter 
combination and the iteration continued until convergence. Given that our
parameter space is topologically well behaved, this method was accurate 
and efficient, with convergence typically within ten iterations. 
Because our technique of using four annulus widths effectively counts
each star four times, to maintain independence between data points we
fit each annulus set individually as appropriate. To obtain the best
fit curve for an iteration we average the individual fits, weighted
by their $\chi^{2}$ values. We determined errors in the three parameters 
obtained from the final fits to the background-subtracted profiles using 
a bootstrap method (Press et al. 1992, p691) with 1000 recursions per fit.
Again, this was done separately for each annulus set to maintain the
independence of the points. The errors so determined represent
random errors in the best-fit parameters. Systematic errors such as those
potentially introduced by the background fitting are not included in the 
estimation.

\section{Results}
\label{results}
\subsection{Profiles and structural parameters}
The background-subtracted F555W surface brightness profiles for each
of the 53 clusters are presented in Fig. \ref{plots}.
We plot each of the four annulus sets on the same axes, to demonstrate the
high degree of consistency between them. For each cluster, the best fit 
EFF profile is also plotted, the core radius indicated, and the best 
fit parameters listed. These results are summarized in Table 
\ref{params} along with their corresponding errors, the calculated 
centre of each cluster and the maximum radial extent $r_{m}$ of each 
profile. Three of the profiles (NGC 1754, NGC 1786, NGC 1916) are 
incomplete in their inner regions -- this is due to the effects of 
crowding, even including supplementary photometry from short exposures
(a short exposure was unavailable for NGC 1786, however). The derived
core radii for these clusters, and for the clusters NGC 1835, NGC 2005,
NGC 2019 and R136 which also required short exposure photometry, should
be considered upper limits. 

This is one of the largest published studies of LMC cluster 
surface brightness profiles, and as far as we are aware, the only one 
to use {\em HST} data. In addition, approximately one quarter of the 
clusters in the sample do not have previously published surface 
brightness profiles. There is much to be learned from such detailed 
data, covering many aspects of globular cluster and LMC astronomy. We 
can however split the sample into sub-groups and obtain some immediate 
and interesting results, which we discuss below. We first 
compare our results with those from the several other 
published large-scale studies.











\begin{table*}
\begin{minipage}{164mm}
\caption{Structural parameters for the cluster sample derived from the best-fitting F555W EFF profiles.}
\begin{tabular}{@{}lcccccccc}
\hline \hline
Cluster & \multicolumn{2}{c}{Centre (J2000.0)$^{a}$} & $\mu_{555}(0)^{b}$ & $a$ & $\gamma$ & $r_{c}$ & $r_{c}$ & $r_{m}$ \vspace{0.5mm} \\
Name & $\alpha$ & $\delta$ & & $(\arcsec)$ & & $(\arcsec)$ & (pc)$^{c}$ & $(\arcsec)$ \\
\hline
NGC1466 & $03^{h}44^{m}32\fs9$ & $-71\degr 40\arcmin 13\arcsec$ & $19.08 \pm 0.05$ & $14.89 \pm 0.79$ & $3.31 \pm 0.12$ & $10.73 \pm 0.36$ & $2.61 \pm 0.09$ & 76 \\
NGC1651 & $04^{h}37^{m}31\fs1$ & $-70\degr 35\arcmin 02\arcsec$ & $20.38 \pm 0.04$ & $15.82 \pm 1.26$ & $2.21 \pm 0.12$ & $14.75 \pm 0.70$ & $3.58 \pm 0.17$ & 76 \\
NGC1711 & $04^{h}50^{m}37\fs3$ & $-69\degr 59\arcmin 04\arcsec$ & $18.28 \pm 0.05$ & $10.93 \pm 0.83$ & $2.78 \pm 0.13$ & $8.78 \pm 0.44$ & $2.13 \pm 0.11$ & 72 \\
NGC1718 & $04^{h}52^{m}25\fs6$ & $-67\degr 03\arcmin 06\arcsec$ & $19.18 \pm 0.07$ & $9.38 \pm 0.76$ & $2.31 \pm 0.10$ & $8.52 \pm 0.49$ & $2.07 \pm 0.12$ & 76 \\
NGC1754 & $04^{h}54^{m}18\fs9$ & $-70\degr 26\arcmin 31\arcsec$ & $17.48 \pm 0.26$ & $4.12 \pm 0.65$ & $2.43 \pm 0.09$ & $3.61 \pm 0.49$ & $0.88 \pm 0.12$ & 76 \\
NGC1777 & $04^{h}55^{m}48\fs9$ & $-74\degr 17\arcmin 03\arcsec$ & $20.01 \pm 0.02$ & $16.62 \pm 0.89$ & $2.60 \pm 0.10$ & $13.96 \pm 0.44$ & $3.39 \pm 0.11$ & 76 \\
NGC1786 & $04^{h}59^{m}07\fs9$ & $-67\degr 44\arcmin 45\arcsec$ & $17.46 \pm 0.11$ & $6.79 \pm 0.67$ & $2.78 \pm 0.13$ & $5.45 \pm 0.38$ & $1.33 \pm 0.09$ & 76 \\
NGC1805 & $05^{h}02^{m}21\fs8$ & $-66\degr 06\arcmin 42\arcsec$ & $18.01 \pm 0.06$ & $6.84 \pm 0.42$ & $2.81 \pm 0.10$ & $5.47 \pm 0.23$ & $1.33 \pm 0.06$ & 69 \\
NGC1818 & $05^{h}04^{m}13\fs8$ & $-66\degr 26\arcmin 02\arcsec$ & $18.36 \pm 0.05$ & $12.50 \pm 0.78$ & $2.76 \pm 0.12$ & $10.10 \pm 0.39$ & $2.45 \pm 0.09$ & 76 \\
NGC1831 & $05^{h}06^{m}17\fs4$ & $-64\degr 55\arcmin 11\arcsec$ & $19.08 \pm 0.04$ & $25.81 \pm 1.59$ & $3.41 \pm 0.19$ & $18.28 \pm 0.59$ & $4.44 \pm 0.14$ & 76 \\
NGC1835 & $05^{h}05^{m}06\fs7$ & $-69\degr 24\arcmin 15\arcsec$ & $16.37 \pm 0.08$ & $6.35 \pm 0.46$ & $3.11 \pm 0.10$ & $4.76 \pm 0.26$ & $1.16 \pm 0.06$ & 76 \\
NGC1841 & $04^{h}45^{m}23\fs9$ & $-83\degr 59\arcmin 56\arcsec$ & $21.59 \pm 0.02$ & $53.62 \pm 5.07$ & $4.55 \pm 0.61$ & $32.00 \pm 0.72$ & $7.77 \pm 0.17$ & 78 \\
NGC1847 & $05^{h}07^{m}07\fs7$ & $-68\degr 58\arcmin 17\arcsec$ & $18.54 \pm 0.06$ & $8.58 \pm 0.44$ & $2.05 \pm 0.13$ & $8.44 \pm 0.34$ & $2.05 \pm 0.08$ & 42 \\
NGC1850 & $05^{h}08^{m}41\fs2$ & $-68\degr 45\arcmin 31\arcsec$ & $16.70 \pm 0.07$ & $11.11 \pm 1.03$ & $2.18 \pm 0.11$ & $10.48 \pm 0.64$ & $2.55 \pm 0.16$ & 69 \\
NGC1856 & $05^{h}09^{m}31\fs5$ & $-69\degr 07\arcmin 46\arcsec$ & $16.68 \pm 0.07$ & $6.81 \pm 0.46$ & $2.01 \pm 0.05$ & $6.80 \pm 0.36$ & $1.65 \pm 0.09$ & 76 \\
NGC1860 & $05^{h}10^{m}38\fs9$ & $-68\degr 45\arcmin 12\arcsec$ & $19.72 \pm 0.11$ & $9.74 \pm 0.53$ & $2.25 \pm 0.13$ & $9.00 \pm 0.50$ & $2.19 \pm 0.12$ & 76 \\
NGC1866 & $05^{h}13^{m}38\fs9$ & $-65\degr 27\arcmin 52\arcsec$ & $18.52 \pm 0.02$ & $18.60 \pm 0.72$ & $3.04 \pm 0.09$ & $14.15 \pm 0.32$ & $3.44 \pm 0.08$ & 76 \\
NGC1868 & $05^{h}14^{m}36\fs2$ & $-63\degr 57\arcmin 14\arcsec$ & $17.99 \pm 0.04$ & $8.83 \pm 0.41$ & $3.07 \pm 0.10$ & $6.67 \pm 0.19$ & $1.62 \pm 0.05$ & 76 \\
NGC1898 & $05^{h}16^{m}42\fs4$ & $-69\degr 39\arcmin 25\arcsec$ & $19.05 \pm 0.06$ & $8.79 \pm 0.91$ & $2.14 \pm 0.14$ & $8.40 \pm 0.55$ & $2.04 \pm 0.13$ & 76 \\
NGC1916 & $05^{h}18^{m}37\fs5$ & $-69\degr 24\arcmin 25\arcsec$ & $15.93 \pm 0.09$ & $4.08 \pm 0.28$ & $2.70 \pm 0.08$ & $3.35 \pm 0.18$ & $0.81 \pm 0.04$ & 76 \\
NGC1984 & $05^{h}27^{m}40\fs8$ & $-69\degr 08\arcmin 05\arcsec$ & $18.62 \pm 0.11$ & $4.44 \pm 0.41$ & $2.27 \pm 0.11$ & $4.07 \pm 0.28$ & $0.99 \pm 0.07$ & 76 \\
NGC2004 & $05^{h}30^{m}40\fs9$ & $-67\degr 17\arcmin 09\arcsec$ & $16.93 \pm 0.08$ & $7.56 \pm 0.86$ & $2.53 \pm 0.12$ & $6.47 \pm 0.53$ & $1.57 \pm 0.13$ & 76 \\
NGC2005 & $05^{h}30^{m}10\fs3$ & $-69\degr 45\arcmin 09\arcsec$ & $17.04 \pm 0.13$ & $4.33 \pm 0.54$ & $2.60 \pm 0.14$ & $3.63 \pm 0.33$ & $0.88 \pm 0.08$ & 69 \\
NGC2011 & $05^{h}32^{m}19\fs6$ & $-67\degr 31\arcmin 14\arcsec$ & $18.93 \pm 0.15$ & $5.17 \pm 0.60$ & $2.22 \pm 0.09$ & $4.81 \pm 0.51$ & $1.17 \pm 0.12$ & 76 \\
NGC2019 & $05^{h}31^{m}56\fs6$ & $-70\degr 09\arcmin 33\arcsec$ & $16.73 \pm 0.10$ & $4.21 \pm 0.37$ & $2.52 \pm 0.09$ & $3.61 \pm 0.24$ & $0.88 \pm 0.06$ & 76 \\
NGC2031 & $05^{h}33^{m}41\fs1$ & $-70\degr 59\arcmin 13\arcsec$ & $18.93 \pm 0.06$ & $11.13 \pm 0.98$ & $2.09 \pm 0.11$ & $10.81 \pm 0.61$ & $2.63 \pm 0.15$ & 76 \\
NGC2100 & $05^{h}42^{m}08\fs6$ & $-69\degr 12\arcmin 44\arcsec$ & $16.26 \pm 0.14$ & $5.73 \pm 0.75$ & $2.44 \pm 0.14$ & $5.02 \pm 0.50$ & $1.22 \pm 0.12$ & 76 \\
NGC2121 & $05^{h}48^{m}11\fs6$ & $-71\degr 28\arcmin 51\arcsec$ & $20.90 \pm 0.02$ & $33.83 \pm 3.83$ & $2.25 \pm 0.28$ & $31.22 \pm 1.14$ & $7.59 \pm 0.28$ & 76 \\
NGC2136 & $05^{h}52^{m}58\fs0$ & $-69\degr 29\arcmin 36\arcsec$ & $17.37 \pm 0.09$ & $12.33 \pm 1.06$ & $3.79 \pm 0.23$ & $8.19 \pm 0.43$ & $1.99 \pm 0.10$ & 69 \\
NGC2153 & $05^{h}57^{m}51\fs2$ & $-66\degr 23\arcmin 58\arcsec$ & $19.17 \pm 0.09$ & $4.71 \pm 0.79$ & $2.64 \pm 0.22$ & $3.91 \pm 0.44$ & $0.95 \pm 0.11$ & 69 \\
NGC2155 & $05^{h}58^{m}33\fs3$ & $-65\degr 28\arcmin 35\arcsec$ & $20.44 \pm 0.05$ & $20.30 \pm 2.20$ & $2.85 \pm 0.25$ & $16.06 \pm 0.98$ & $3.90 \pm 0.24$ & 76 \\
NGC2156 & $05^{h}57^{m}49\fs7$ & $-68\degr 27\arcmin 41\arcsec$ & $18.98 \pm 0.07$ & $8.15 \pm 0.51$ & $2.79 \pm 0.09$ & $6.53 \pm 0.31$ & $1.59 \pm 0.08$ & 76 \\
NGC2157 & $05^{h}57^{m}32\fs4$ & $-69\degr 11\arcmin 49\arcsec$ & $17.78 \pm 0.05$ & $13.76 \pm 0.91$ & $3.45 \pm 0.18$ & $9.68 \pm 0.36$ & $2.35 \pm 0.09$ & 76 \\
NGC2159 & $05^{h}58^{m}03\fs0$ & $-68\degr 37\arcmin 27\arcsec$ & $19.66 \pm 0.05$ & $12.73 \pm 0.94$ & $3.00 \pm 0.15$ & $9.77 \pm 0.47$ & $2.37 \pm 0.11$ & 69 \\
NGC2162 & $06^{h}00^{m}30\fs4$ & $-63\degr 43\arcmin 19\arcsec$ & $19.92 \pm 0.08$ & $12.97 \pm 1.40$ & $2.91 \pm 0.23$ & $10.13 \pm 0.66$ & $2.46 \pm 0.16$ & 76 \\
NGC2164 & $05^{h}58^{m}55\fs9$ & $-68\degr 31\arcmin 00\arcsec$ & $18.22 \pm 0.05$ & $10.16 \pm 0.43$ & $2.96 \pm 0.07$ & $7.85 \pm 0.24$ & $1.91 \pm 0.06$ & 72 \\
NGC2172 & $06^{h}00^{m}06\fs4$ & $-68\degr 38\arcmin 15\arcsec$ & $20.21 \pm 0.07$ & $13.77 \pm 0.97$ & $2.94 \pm 0.14$ & $10.68 \pm 0.49$ & $2.60 \pm 0.12$ & 76 \\
NGC2173 & $05^{h}57^{m}58\fs5$ & $-72\degr 58\arcmin 40\arcsec$ & $19.86 \pm 0.05$ & $13.13 \pm 1.35$ & $2.37 \pm 0.16$ & $11.70 \pm 0.74$ & $2.84 \pm 0.18$ & 76 \\
NGC2193 & $06^{h}06^{m}17\fs0$ & $-65\degr 05\arcmin 54\arcsec$ & $19.77 \pm 0.08$ & $7.65 \pm 0.71$ & $2.50 \pm 0.11$ & $6.59 \pm 0.43$ & $1.60 \pm 0.10$ & 72 \\
NGC2209 & $06^{h}08^{m}34\fs8$ & $-73\degr 50\arcmin 12\arcsec$ & $22.09 \pm 0.06$ & $22.97 \pm 2.39$ & $2.08 \pm 0.16$ & $22.34 \pm 1.34$ & $5.43 \pm 0.33$ & 76 \\
NGC2210 & $06^{h}11^{m}31\fs5$ & $-69\degr 07\arcmin 17\arcsec$ & $18.18 \pm 0.05$ & $11.75 \pm 0.56$ & $3.51 \pm 0.12$ & $8.18 \pm 0.24$ & $1.99 \pm 0.06$ & 76 \\
NGC2213 & $06^{h}10^{m}42\fs2$ & $-71\degr 31\arcmin 46\arcsec$ & $19.16 \pm 0.05$ & $9.36 \pm 0.69$ & $2.84 \pm 0.13$ & $7.43 \pm 0.36$ & $1.80 \pm 0.09$ & 76 \\
NGC2214 & $06^{h}12^{m}55\fs8$ & $-68\degr 15\arcmin 38\arcsec$ & $18.36 \pm 0.07$ & $9.55 \pm 0.99$ & $2.26 \pm 0.14$ & $8.79 \pm 0.59$ & $2.14 \pm 0.14$ & 76 \\
NGC2231 & $06^{h}20^{m}42\fs7$ & $-67\degr 31\arcmin 10\arcsec$ & $20.48 \pm 0.12$ & $11.42 \pm 2.79$ & $2.07 \pm 0.26$ & $11.14 \pm 1.55$ & $2.71 \pm 0.38$ & 76 \\
NGC2249 & $06^{h}25^{m}49\fs8$ & $-68\degr 55\arcmin 13\arcsec$ & $19.05 \pm 0.04$ & $12.13 \pm 0.83$ & $3.40 \pm 0.17$ & $8.61 \pm 0.35$ & $2.09 \pm 0.09$ & 76 \\
NGC2257 & $06^{h}30^{m}12\fs1$ & $-64\degr 19\arcmin 42\arcsec$ & $20.93 \pm 0.03$ & $38.61 \pm 2.71$ & $3.54 \pm 0.27$ & $26.75 \pm 0.73$ & $6.50 \pm 0.18$ & 76 \\
SL663 & $05^{h}42^{m}28\fs8$ & $-65\degr 21\arcmin 44\arcsec$ & $22.40 \pm 0.05$ & $28.83 \pm 2.24$ & $2.13 \pm 0.15$ & $27.60 \pm 1.27$ & $6.71 \pm 0.31$ & 76 \\
SL842 & $06^{h}08^{m}14\fs9$ & $-62\degr 59\arcmin 15\arcsec$ & $21.07 \pm 0.11$ & $14.55 \pm 1.75$ & $3.37 \pm 0.29$ & $10.37 \pm 0.76$ & $2.52 \pm 0.18$ & 69 \\
SL855 & $06^{h}10^{m}53\fs7$ & $-65\degr 02\arcmin 30\arcsec$ & $22.93 \pm 0.03$ & $23.10 \pm 2.91$ & $2.12 \pm 0.23$ & $22.18 \pm 1.35$ & $5.39 \pm 0.33$ & 76 \\
HODGE4 & $05^{h}32^{m}25\fs2$ & $-64\degr 44\arcmin 11\arcsec$ & $20.59 \pm 0.06$ & $15.84 \pm 1.65$ & $2.13 \pm 0.15$ & $15.20 \pm 1.08$ & $3.69 \pm 0.26$ & 76 \\
HODGE11 & $06^{h}14^{m}22\fs3$ & $-69\degr 50\arcmin 50\arcsec$ & $19.60 \pm 0.05$ & $13.66 \pm 1.17$ & $2.38 \pm 0.12$ & $12.14 \pm 0.66$ & $2.95 \pm 0.16$ & 76 \\
HODGE14 & $05^{h}28^{m}39\fs3$ & $-73\degr 37\arcmin 49\arcsec$ & $20.28 \pm 0.09$ & $8.38 \pm 0.93$ & $2.41 \pm 0.16$ & $7.39 \pm 0.56$ & $1.80 \pm 0.14$ & 76 \\
R136 & $05^{h}38^{m}42\fs5$ & $-69\degr 06\arcmin 03\arcsec$ & $12.66 \pm 0.09$ & $1.48 \pm 0.12$ & $2.43 \pm 0.09$ & $1.30 \pm 0.08$ & $0.32 \pm 0.02$ & 10 \\
\hline
\end{tabular}
\medskip
\\
$^{a}$ We find our centering algorithm to be repeatable to approximately $\pm 1\arcsec$, notwithstanding image header inaccuracies (see Section \ref{astro}). Given this precision, coordinates in $\delta$ are provided to the nearest arcsecond. Those in $\alpha$ are reported to the nearest tenth of a second, but the reader should bear in mind that at $\delta = -69\degr$, one second of RA corresponds to approximately five seconds of arc -- in other words, the uncertainty in $\alpha$ is approximately $\pm 0\fs2$.\\
$^{b}$ The $V_{555}$ magnitude of one square arcsecond at the centre of a given cluster.\\
$^{c}$ When converting to parsecs we assume an LMC distance modulus of $18.5$ which equates to a scale of $4.116$ arcsec pc$^{-1}$.
\label{params}
\end{minipage}
\end{table*}

\subsubsection{Comparison with previous work}
\label{prevstud}
At least three-quarters of the clusters studied in this paper have 
published surface brightness profiles, concentrated mainly in four large 
studies -- those of EFF87, Mateo \shortcite{mateo}, E91 (see also 
Elson et al. \shortcite{efl}), and Kontizas and collaborators
\cite{kck,khk,mkk,ckk}. It is useful to compare our results to those
from these papers and thereby establish the validity of our
reduction procedure.


In particular, our background subtraction procedure might introduce
systematic errors into the measured structural parameters. We checked this 
by comparing our derived
values for $\gamma$ with those from EFF87 for the 10 young clusters 
($\tau \leq 3\times 10^{8}$ yr) in 
their sample, since $\gamma$ is very sensitive to the subtracted
background. The comparison is plotted in Fig. \ref{gammacomp}. 
The profiles from EFF87 extend to typically $r_{m}\sim 250\arcsec$ and 
their background measurements are made at these large radii. The 
agreement with our fitted values of $\gamma$ is generally good, and 
within the errors for most points. If anything there is a slight tendency
for our $\gamma$ measurements to be slightly larger than those from EFF87
(i.e., we slightly over-estimated the background -- as would be expected),
but this discrepancy is not significant. EFF87 find a range of 
$2.2\la \gamma \la 3.2$ and a median value $\gamma \approx 2.6$. For
the same ten clusters we find a range of $2.26\la \gamma \la 3.45$
but a median value $\gamma \approx 2.95$. We note however that the
sample size is barely large enough for meaningful statistics. Extending
the average to cover the 22 clusters in our sample with
$\tau \leq 3\times 10^{8}$ yr we obtain a range $2.01\la \gamma \la 3.79$
and a median $\gamma \approx 2.59$. We are therefore confident that
our background subtraction method has not introduced any significant 
systematic errors into the structural parameter measurements.

We have also taken the core radii from the first three studies listed 
above for all the clusters in common with our sample, and plotted these 
against our measured values (Fig. \ref{prevstudplot}). Again, the 
agreement is relatively good; however there is some considerable scatter 
in the plot. Nonetheless, there are good explanations for many of the
errant points. Our core radii are generally somewhat larger than those
measured by EFF87 (filled triangles) -- however the inner regions of the 
profiles from EFF87 are a combination of literature data and 
photomultiplier aperture measurements and the errors quoted for $a$ are 
up to $50$ per cent and more, which would bring most if not all of our 
results well into the error margins. More encouraging is the good agreement
of our measurements with those from E91 (filled circles), again for relatively
young clusters ($10^{7} < \tau < 10^{9}$ yr). The profiles from this study 
were measured from CCD images, with backgrounds determined at 
$r\sim 200\arcsec$. The most errant point (falling well below the 
equality line) is for the young cluster NGC 2100. We 
speculate that the different saturation limits between the two studies 
($12 \la V \la 19$ from E91 as compared with $15 \la V_{555} \la 22$ from the 
present study) might be the root of this difference. 
Furthermore, we have adopted the core radius measurements from E91 not 
corrected for seeing, because there is some evidence of over-correction in the 
quoted values. The seeing corrections move the errant point 
considerably to the left on the plot, but also move most of the 
other points to the left and away from the equality line.

Both the studies by EFF87 and E91 describe the profiles of many young LMC 
clusters as being very irregular in comparison to the smooth profiles typical 
of old globular clusters. We also observe such irregularities, in the same
forms as those noted by E91 -- that is, bumps (e.g., NGC 2004), 
steps and wiggles (e.g., NGC 1711, NGC 1856), sharp shoulders or dips near 
the core radius (e.g., NGC 1850, NGC 1860, NGC 1984) and central dips (e.g., 
NGC 2031).

The measurements from the study of Mateo \shortcite{mateo} for old LMC
clusters (filled squares), are from $B$-band images and so are not 
strictly comparable with ours; however we proceed for the sake of completeness.
Again, there are no significant systematic differences, and most errant 
points can be explained. The four smallest core radii from Mateo all 
correspond to upper limit measurements in our study -- so the points likely 
lie closer to the equality line than indicated. The two errant points to the 
right of the plot are estimates taken from the literature -- the exact method 
of estimation is not clear, but we conjecture that it might be a significant 
cause of the observed differences. We also note good agreement with the 
core radius for NGC 2257 ($25\farcs1$ as compared with our measurement of 
$26\farcs75\pm 0\farcs73$), which falls outside the plot. 

It is more difficult to reconcile our results with those from the four
studies of Kontizas et al. \shortcite{kck,khk}, Metaxa et al. 
\shortcite{mkk}, and Chrysovergis et al. \shortcite{ckk}, which use
density profiles from number counts. While this again means the derived 
parameters are not strictly comparable with ours, the two sets 
should be roughly similar. However, we observe a large, apparently random 
scatter between them. The authors report that their core radius
estimates are derived from King model fits to the density profiles, which yield
$r_{t}$ and the concentration $c$, and that the errors are
$\sim 15$ per cent in $r_{t}$ and $\pm 0.25$ in $c$, which is $\sim 20$ per
cent in the typical value $c\approx 1.5$. We estimate that these errors
could cause uncertainties of the order of $\pm 60$ per cent or more in 
derived values of $r_{c}$, consistent with the difficulties associated with 
counting stars in crowded regions on photographic plates and the resolution 
of the profiles (most do not extend within $\sim 25\arcsec$). Such
uncertainties can account for most of the observed scatter. Nonetheless, 
several measurements are discrepant by factors of up to 10 (e.g., 
$r_{c} \approx 55\arcsec$ for NGC 1805 as compared with our measurement of 
$r_{c} = 5\farcs5$) and we are unable offer a plausible explanation to 
fully reconcile these.

\subsubsection{Double clusters and profile bumps}
\label{binarycl}

There is a small amount of evidence for double clusters in our sample. In particular,
NGC 1850 is a well known double cluster, with a much smaller and younger
companion -- NGC 1850B -- separated by approximately $32\arcsec$ (see 
e.g., Gilmozzi et al. \shortcite{gilmozzi}). It is not clear that NGC 1850B is
physically associated with NGC 1850 -- it may be a chance superposition of two
clusters along the same line of sight. Measurements of the mass of each 
cluster, and their three dimensional separation and relative velocity are required to 
settle this issue. 

Nonetheless, NGC 1850B appears in our 
images, and its presence can be seen in the surface brightness profile 
for NGC 1850 as a bump at $\log r \sim 1.55$. We also show our WFC2 $V$-band
image of NGC1850 in Fig. \ref{doubleclus} -- in this image, the
centre of NGC1850 is to the lower left, and NGC1850B is evident to the upper right
of the main cluster.
The over-sampling of our data has the effect of smearing out sharp
features in the surface brightness profiles. For example, a very bright 
star will appear once in each annulus set, resulting in four points at 
slightly different radii and giving the impression of a bump rather than 
a spike. This makes the bump in the profile of NGC 1850 almost 
indistinguishable from that which would be caused by a bright star, 
although upon examination of the actual image, the nature of NGC 1850B 
is immediately apparent. Our unpublished CMD for NGC 1850 also shows evidence 
of this young sub-cluster in the form of a few main sequence stars above the 
turn-off for the main cluster. We also note that because each annulus is
essentially an azimuthal average, the amplitude of the bump due to NGC 1850B
is reduced - that is, the peak surface brightness of NGC 1850B is certainly
much greater than indicated in the surface brightness profile for NGC 1850.

The above result prompted us to search the other cluster images and profiles 
(correlating the two against each other) for similar evidence of double clusters, 
even though only two of our sample, NGC 2011 and NGC 2136, appear in the catalogue 
of LMC double clusters published by Bhatia et al. \shortcite{bhatia}. 
None of the images showed any evident double clusters. Surprisingly, 
we did however locate two very prominent bumps in the profiles
of NGC 2153 and NGC 2213, at $\log r \sim 1.15$ ($14\arcsec$) and 
$\log r \sim 1.25$ ($18\arcsec$) respectively. We show DSS2 $R$-band images of 
each of these clusters in Fig.\ref{doubleclus}. It is clear from these images,
and from our WFPC2 images, that secondary clusters are not the cause of these bumps.
Furthermore, although it is evident from the 
images that each cluster contains bright stars, neither of the bumps can be
due to a single bright star, because each bump appears in at
least two consecutive points from single annulus sets (the $1\farcs5$,
$2\arcsec$ and $3\arcsec$ sets for NGC 2156, and the $1\farcs5$ and $2\arcsec$
sets for NGC 2213). It is therefore not clear what 
the cause of either bump is. It is however clear that the binarity of a cluster
cannot be judged from its surface brightness profile, because these examples show
that even single clusters can have significant bumps in their profiles.

One possibility which is evident from the DSS images (but not necessarily so
clear on the high resolution WFPC2 images) is that the chance positioning 
of several bright stars at exactly the right distance from a given cluster's centre 
but with different position angles could result in a bump. For NGC 2153, there is 
an arc of stars to the north of the main cluster and a fainter arc to the south
which are at approximately the right distance to match the bump in the profile. There
is also a bright star to the south, but this is saturated on our WFPC2 images. 
Whether these arcs are simply random placement
of several stars, or a physical association is not clear. For NGC 2213, it would
appear that there are several bright stars at approximately the right distance
from the centre but at different angles about the cluster, which might cause the 
bump in its profile. Again, whether any physical significance can be attached to 
this bump is uncertain.

In addition, three other clusters -- NGC 2004, NGC 2155, and NGC 2159 -- have 
less statistically significant bumps in their profiles, at $\log r \sim 1.25$ 
($18\arcsec$), $\log r \sim 1.3$ ($20\arcsec$), and $\log r \sim 1.3$ 
($20\arcsec$) respectively. We show DSS2 $R$-band images of these clusters,
also in Fig. \ref{doubleclus}. Just as for NGC 2153 and NGC 2213, it is not 
immediately clear what the causes of these bumps are. Certainly they are not due 
to single bright stars, and again there are no clear sub-clusters like NGC 1850B.
NGC 2004 shows a ring of bright stars about its centre, but these are at
too great a radius to cause the observed bump. We do note a slight extension of
the main cluster towards the upper left -- this is also apparent in the WFPC2
image and might be the source of the bump in the profile. Again, the physical
significance of this extension is not clear. For NGC 2155, there is no evident
source for the bump, although this cluster does show a clumpy and
dispersed structure. Finally, for NGC 2159, again the bump may be caused by
several bright stars at the right distance from the centre but at different
position angles. There is no other structure evident which might cause the
bump in its profile.


The profiles for NGC 2011 and NGC 2136, the only clusters in our sample apart 
from NGC 1850 to appear in the catalogue by Bhatia et al. \shortcite{bhatia} 
do not show any structure that might be due to their binary nature. However, 
both companions lie outside the respective fields of view for these clusters. 
NGC 1818 also likely has a small companion (see e.g., de Grijs et al. 
\shortcite{rdg}) but the separation is $\sim 90\arcsec$ and it again lies 
outside our field of view. NGC 2214 is regarded by some authors as a binary 
cluster in the late stages of a merger \cite{bhatiaa}, and has been noted to 
have an anomalous core \cite{meylan}. We do not observe any significant bumps 
in our surface brightness profile for this cluster, or any interesting 
structure in our image, but we do see a similar 
effect to that described by Meylan \& Djorgovski where the surface brightness 
profile in the core appears similar to that for a post core-collapse cluster. 
That NGC 2214 is such a cluster is unlikely given its age ($\sim 60$ Myr).
Finally, NGC 2156 is rendered interesting because of the results of de 
Oliveira et al. \shortcite{deoliveira} who measure this cluster to have 
somewhat elliptical isophotes. Upon comparison with the results of $N$-body 
simulations of star cluster encounters, they conclude that this ellipticity 
could result from a binary cluster interaction, given a favourable viewing
angle. Moreover, the young age of NGC 2156 combined with the behaviour of the 
simulation suggests that the merging cluster will not yet have been disrupted.
However, we observe no evidence for such a distinct sub-cluster in our
images, or any structure in the surface brightness profile. 

\subsubsection{Post core-collapse clusters}
\label{pcc}
We find strong evidence for several post core-collapse (PCC) clusters 
amongst the old clusters in our sample. Theoretical studies, as well as observations
of the Galactic globular cluster system have shown that the surface brightness 
profiles of PCC clusters are characterized by power-law cusps in their central regions,
rather than ordinary King-type models. We have identified two clusters with profiles
which clearly show such cusps as well as breaks to the power-law regions, and another
two clusters which show cusps and breaks at lower signficance. A further three clusters 
have incomplete profiles in their central regions (a result of severe crowding in these 
regions -- see Section \ref{completenesscorr} and the beginning of Section \ref{results})
but do appear to show breaks similar to the other four PCC candidates. The three groups
of clusters are, respectively: NGC 2005 and NGC 2019; NGC 1835 and NGC 1898; and NGC 1754,
NGC 1786, and NGC 1916.

We first address NGC 2005 and NGC 2019, which have profiles clearly showing PCC-like 
cusps. This has previously been noted by Mateo \shortcite{mateo} for both clusters, 
and by Meylan \& Djorgovski \shortcite{meylan} for NGC 2019. Mateo estimates power-law 
slopes of $\beta \sim 1$ for both clusters, where we assume the profile goes as
$r^{-\beta}$ in the central regions. He also notes sharp breaks in both profiles at the 
transition to the power-law region, which occur at $4\farcs4\pm 0\farcs4$ for 
NGC 2005 and $3\farcs6\pm 0\farcs4$ for NGC 2019. We observe well defined power-law regions 
for both clusters, with sharp breaks at $5\farcs6\pm 0\farcs3$ for NGC 2005 and 
$5\farcs0\pm 0\farcs6$ for NGC 2019, both slightly larger than the measurements of Mateo.
If we fit power-law profiles to these inner sections, we obtain slopes of $\beta = 0.74$ 
for NGC 2005 and $\beta = 0.71$ for NGC 2019. The profiles and power-law models are shown 
in Fig. \ref{corecollapse}(a) and (b), together with the best fitting EFF models. 
Clearly, for each cluster the power-law is a better fit in the centre than the EFF model.
Further, it is interesting that studies of Galactic PCC clusters have measured 
slopes in the range $0.6$ to $0.8$ (Lugger, Cohn \& Grindlay 1995), or with a median 
value $\beta \approx 0.9$ \cite{djki}, and many of the PCC Galactic clusters also show 
breaks in their profiles (see Fig. 2 in Djorgovski \& King, and Fig. 2(a-r) in Lugger et al.).

NGC 1835 and NGC 1898 also appear to show power-law cusps and breaks, but at lower 
significance than NGC 2005 and NGC 2019. The profiles for these two clusters are shown in 
Fig. \ref{corecollapse}(c) and (d) respectively, along with power-law fits to the central
profile regions, and the best fitting EFF models. For each cluster, a power-law seems to 
model the central data better than an EFF profile, although for NGC 1898 the difference 
is small. The power law slopes are $\beta = 0.45$ for NGC 1835, and $\beta = 0.30$ for 
NGC 1898, and the break radii are $4\farcs2\pm 0\farcs4$ and $6\farcs3\pm 1\farcs2$
respectively. Profiles for NGC 1835 have been published by both Elson \& Freeman 
\shortcite{efree} and Mateo \shortcite{mateo}; however, while both show NGC 1835 to have a 
small core, neither find a power law region or a break.

The three remaining clusters have incomplete profiles in their inner regions 
because of severe crowding. Even using short exposure photometry (Section \ref{short})
for NGC 1754 and NGC 1916, the incompleteness could not be overcome, while NGC 1786 had
no supplementary images available. The profiles for these three clusters are 
shown in Fig. \ref{corecollapse}(e-g). Given the lack of data, we cannot assert even that
these three clusters are PCC candidates; nonetheless, they warrant further attention 
because they all appear to show breaks in their profiles at very similar radii to 
those observed for NGC 2005 and NGC 2019. Clusters with ordinary King-type profiles should 
not show such breaks. Each of the three profiles is complete to several points further 
in than its break, and we attempted to fit power-law models to these points even though 
there are very few of them in each case -- the idea being to explore the matter further
by comparing potential power-law models to the best fitting EFF models. We obtained 
slopes of $\beta = 0.74$ for NGC 1754, $\beta = 0.90$ for NGC 1786, and 
$\beta = 1.17$ for NGC 1916. Interestingly, these values are very close to those 
found for NGC 2005 and NGC 2019, especially given that the uncertainties for 
fits to only a few points (just three in the case of NGC 1754) are large. 

Profiles for both NGC 1754 and NGC 1786 are given by Mateo \shortcite{mateo};
however he finds that neither show power-law regions or breaks. Similarly,
Meylan \& Djorgovski \shortcite{meylan} observe NGC 1786, and they classify it as 
having a normal King-type profile. There is no published profile for NGC 1916.
We find that for each of the three clusters, both the power-law profile and the 
EFF profile fit the data equally well around the break radius -- hence we 
do not assert that these three clusters are PCC candidates. However, we believe that all
three clusters warrant further high resolution observations to obtain complete profiles.
It is suggestive that not only do these clusters seem to show breaks in their profiles, 
but each appears at least as compact as both NGC 2005 and NGC 2019, which are our leading 
PCC candidates, but which do not suffer as badly from crowding as do NGC 1754, NGC 1786
and NGC 1916.

\begin{table}
\caption{Power-law slopes and break radii for seven potential PCC clusters.}
\centering
\begin{tabular}{@{}lccccc}
\hline \hline
Cluster & $r_{c}$ & $\beta$ & $\log r_{break}$ & $r_{break}$ \vspace{0.5mm} \\
 & $(\arcsec)$ & & $(\arcsec)$ & $(\arcsec)$ \vspace{0.5mm} \\
\hline
NGC2005	& $3.63 \pm 0.33$ & $0.74$ & $0.75 \pm 0.02$ & $5.6 \pm 0.3$ \\
NGC2019	& $3.61 \pm 0.24$ & $0.71$ & $0.70 \pm 0.05$ & $5.0 \pm 0.6$ \\
\hline
NGC1835 & $4.76 \pm 0.26$ & $0.45$ & $0.62 \pm 0.04$ & $4.2 \pm 0.4$ \\
NGC1898 & $8.40 \pm 0.55$ & $0.30$ & $0.80 \pm 0.08$ & $6.3 \pm 1.2$ \\
\hline
NGC1754 & $3.61 \pm 0.41$ & $0.74$ & $0.62 \pm 0.04$ & $4.2 \pm 0.4$ \\
NGC1786 & $5.45 \pm 0.38$ & $0.90$ & $0.75 \pm 0.04$ & $5.6 \pm 0.5$ \\
NGC1916 & $3.35 \pm 0.18$ & $1.17$ & $0.82 \pm 0.07$ & $6.6 \pm 1.1$ \\
\hline
\end{tabular}
\medskip
\\
\begin{flushleft}
Note: $r_c$ is the core radius of the best fitting EFF profile, from Table \ref{params}; $\beta$ is the slope of the best fitting power-law model; $r_{break}$ is the break radius observed from Fig. \ref{corecollapse} with errors estimated by eye.
\end{flushleft}
\label{betafits}
\end{table}

The results of the power-law fits for the seven clusters are summarized in 
Table \ref{betafits}, including the break radius for each cluster. On the basis of the 
above analysis, we conclude that NGC 2005 and NGC 2019 are strong PCC candidates,
while NGC 1754, NGC 1786, and NGC 1916 deserve further detailed study. 
The status of NGC 1835 and NGC 1898 is unclear. While they both have central regions which
are fit better by power-law models than EFF models, the power-law slopes are shallower 
than those typically associated with PCC clusters. However, both show breaks at similar 
radii to NGC 2005 and NGC 2019. Physically, NGC 1835 is very similar in appearance 
to these two clusters, and we therefore consider it also a PCC candidate. 
NGC 1898 is a much more open cluster, and we speculate that it is perhaps on the verge of 
core-collapse. It is interesting that Olsen et al. \shortcite{olsen} find NGC 1898 to be 
$\sim 3$ Gyr younger than NGC 1835, NGC 2005 and NGC 2019.

Given these results, we briefly consider the LMC old cluster population as a whole. In 
addition to the twelve clusters in the present sample (adding NGC 1466, NGC 1841, NGC 2210, 
NGC 2257, and Hodge 11 to the seven discussed above) there are three more suspected 
old LMC clusters -- Reticulum, which is definitely old and likely a member of the LMC based 
on its radial velocity \cite{suntzeff}; and NGC 1928 and NGC 1939, which are definitely LMC 
members and likely to be old based on the spectroscopic study of Dutra et al. 
\shortcite{dutra}. Of the twelve measured in the present study, we suggest that $3 \pm 1$ 
are good PCC candidates. Of the three not measured in the present study, Reticulum is 
definitely not (Suntzeff et al. \shortcite{suntzeff} estimate a core radius of $\sim 13$ pc), 
and NGC 1928 and NGC 1939 are unknown as far as we are aware. Therefore we estimate
that $20 \pm 7$ per cent of old LMC clusters are PCC clusters, a value which matches well
the $\sim 20$ per cent estimated by Djorgovski \& King \shortcite{djki} for 
Galactic clusters. In addition, all the candidate LMC PCC clusters, as well as NGC 1754,
NGC 1786, NGC 1916, NGC 1928 and NGC 1939 (which all require further study) are in the bar 
region, at projected radii less than $\sim 2.5\degr$ from the optical centre of the LMC. 
Similarly, Djorgovski \& King find the Galactic PCC clusters to be centrally concentrated.

\subsubsection{R136}

R136 is the extremely compact, very young ($\tau = 3$ Myr) central cluster of 
the 30 Doradus H\,{\sc ii} region, which is the most luminous H\,{\sc ii} 
region in the local group. Because of its unique characteristics, we briefly 
describe it separately here. 

The surface brightness profile for R136 and the surrounding cluster NGC 2070 
is clearly separable into two regions -- an inner region which is well fit by 
an EFF profile with $r_{c} = 1\farcs3$ and $\gamma = 2.43$ to a radius 
$r\sim 10\arcsec$ (see Fig. \ref{plots} and Table \ref{params}), and beyond this, 
a break to a shallower profile. This is consistent with the review of Meylan 
\shortcite{meylanr} in which he finds a two-component King model necessary to fit 
his observed surface brightness profile. The first component in his model has core 
radius $r_{c} = 1\farcs3$, equivalent with our measurement, and the 
second component a core radius $r_{c} = 15\arcsec$. Following in this vein, we
attempted a two-component fit to our profile. Fitting two EFF profiles, we found
the best fitting core component to match our previously measured core profile, 
with the outer profile best fit by $\mu \sim 18.7$, $\gamma \sim 2.6$, and 
$a \sim 39\farcs4$, meaning a core radius $r_{c} \approx 33\arcsec$. This composite
profile is plotted in Fig. \ref{R136bits} (dashed line). It is clear that it provides
a good fit to the entire profile, as opposed to our original one-component profile,
which only fit the core. 

In the very outer part of our measured profile, we noticed a turn-down at low 
significance -- this may be evidence for a tidal cut-off. If so, it would be more 
suitable to fit a King profile (Eq. \ref{kp}) as the second component. Doing so,
we found that once again the core was well fit by our original EFF model, with 
the King component having $k \sim 18.2$, $r_{t} \sim 130\arcsec$, and concentration 
$c \sim 0.6$, meaning a core radius $r_{c} \approx 33\arcsec$, which matches well the
value from the two component EFF profile. The composite EFF-King profile is
the solid line in Fig. \ref{R136bits}. It is clear that it differs significantly from
the two component EFF profile only in the very outer regions.

Fitting two-component models to our overall profile did not alter our original core
radius meanurement for R136, which suggests a degree of robustness to our result. There
is considerable variation in the previously published values 
of this core radius. Meylan \shortcite{meylanr} finds $r_{c} = 1\farcs3$ from ground
based data, while Elson et al. \shortcite{elsondor} find 
$r_{c} = 0\farcs5$ from {\em HST} observations. Malumuth \& Heap 
\shortcite{malumuth} obtain $r_{c} = 0\farcs96$ from a mass-density profile 
based on {\em HST} data, and Hunter et al. \shortcite{hunter} find $r_{c} < 0\farcs08$,
also using {\em HST} observations. Campbell et al. \shortcite{campbell} measure 
$r_{c} = 0\farcs25$ based on a pure power-law surface brightness fit, again 
derived from {\em HST} observations. Our measured core radius of 
$r_{c} = 1\farcs3$ is consistent with that of Meylan, but considerably larger 
than the other three. We reiterate however, that our measurement is an upper 
limit, due to the crowding in our images and also because our selected annulus 
widths mean that we are not equipped to measure core radii smaller than 
$r_{c} \sim 1\farcs5$ accurately. Brandl et al. \shortcite{brandl} find considerable
evidence for mass segregation in R136, and show that the derived core radius is
therefore sensitive to the lower luminosity/mass cut-off for stars used in
its calculation. For example, they find a core radius of 
$r_{c} = 0\farcs48$ using stars with $m > 15 M_{\odot}$, but increasing to 
$r_{c} = 0\farcs97$ using stars with $m > 4 M_{\odot}$. This may go some way towards
explaining the above variations.

Similarly, there is some argument in the literature as to whether R136 might have 
undergone core collapse (see e.g., Campbell et al. \shortcite{campbell};
Malumuth \& Heap \shortcite{malumuth}) -- mostly the arguments are dependent on the 
core radius assumed for the cluster, and the relaxation time chosen to characterize
the system. Campbell et al. calculate the relaxation time at the centre (their Eq. 2),
and using their measured core radius ($0.06$ pc) and central density 
($5\times 10^{4} M_{\odot}$pc$^{-3}$) obtain an estimate of several$\times 10^{4}$ 
yr -- much shorter than the age of the cluster. They therefore suggest that R136 is
likely in an advanced state of dynamical evolution and has undergone core collapse.
In contrast, Malumuth \& Heap derive a core radius four times larger, and a 
correspondingly larger central relaxation time -- comparable to the age of the cluster.
They also calculate the relaxation time at the median (half-mass) radius (their Eq. 5) 
and obtain a similar estimate, therefore concluding that R136 is probably 
mass-segregated but not post core collapse (according to Binney \& Tremaine 
\shortcite{galdyn} p527, core collapse occurs at $12$ to $19$ median relaxation times). 
Similarly, Brandl et al. \shortcite{brandl} show that dynamical mass segregation is
likely to have occurred, at least for the most massive stars in the system, but
that the median relaxation time is too long for core collapse to have taken place.


Even though our resolution is not quite sufficient to provide more than an upper limit
for the core radius of R136, our surface brightness profile can help shed some light
on the issue. If R136 is a PCC cluster then its profile should show a cusp similar to 
those observed for the old clusters discussed in Section \ref{pcc}.  Both Campbell
et al. \shortcite{campbell} and Brandl et al. \shortcite{brandl} show that the profile
of R136 can be fit by a power law (although in the case of Brandl et al., this is only
for stars with $m > 40 M_{\odot}$). We observe a small
amount of evidence for such a power-law cusp and a break. The central region of the 
profile is plotted in Fig. \ref{R136core}, along with the best fitting 
power-law model and EFF model. We measure a slope $\beta = 1.17$ with the break at 
$r \sim 2\farcs6$. This is only marginally consistent with the values of $\beta$ 
measured for the old PCC candidates. From Fig. \ref{R136core} it
is also clear that the EFF profile fits equally well, unlike for the best old
PCC candidates. It is likely then that in the case of R136 the apparent power-law 
structure in the central region of the profile is due to the brightest stars in
the cluster residing in this space, consistent with the argument of Brandl et al.
We can calculate the median relaxation time (in the form of Malumuth \& Heap 
\shortcite{malumuth}, but see also e.g., Spitzer \& Hart \shortcite{spitzerh}; 
Spitzer \shortcite{spitzerb}; Binney \& Tremaine \shortcite{galdyn}; 
Meylan \shortcite{mey}):
\begin{equation}
t_{rh} = \frac{6.5\times10^{8}}{\ln(0.4N)}\left(\frac{r_{h}^{3/2}}{m}\right)\left(\frac{M_{tot}}{10^{5}}\right)^{1/2} {\rmn yr}
\label{relax}
\end{equation}
where $N$ is the total number of stars and $m$ is the typical stellar mass, $M_{tot}$ is
the total mass of the cluster and $r_{h}$ is the median (or half mass) radius. Eq.
\ref{relax} is essentially equivalent to Eq. 4.5 from Brandl et al. In the following
Section (\ref{lumsec}) we derive estimates for the total masses of all the clusters
studied in this paper. For R136 we obtain $M_{tot} \sim 2.5\times10^{4} M_{\odot}$, and
using this and Eq. \ref{lum} we can solve for the half-mass radius, obtaining
$r_{h} \sim 1.25$ pc. Both this radius and the total mass are entirely consistent with
the values obtained by Brandl et al. ($3\times10^{4} M_{\odot}$ and $1.1$ pc 
respectively). Following Brandl et al., we adopt $m\sim 0.5 M_{\odot}$ and therefore
estimate $N\sim5\times10^{4}$. Substituting these values into Eq. \ref{relax} 
provides us with $t_{rh}\sim 2.1\times10^{8}$ yr, very similar to the estimate of 
Brandl et al. ($t_{rh}\sim 2.5\times10^{8}$ yr). For the most massive stars (i.e.,
$m\sim 40M_{\odot}$) we obtain $t_{rh}\sim 2.1\times10^{6}$ yr, 
which is comparable to the age of the cluster. This suggests that for these 
stars, dynamical mass segregation has indeed had time to occur, but both calculated
values of $t_{rh}$ argue strongly against core collapse having happened. In 
confirmation of this, we can estimate the central relaxation time (in the form of 
Campbell et al. \shortcite{campbell}, but see also e.g., Spitzer \shortcite{spitzerb}; 
Binney \& Tremaine \shortcite{galdyn}; Meylan \shortcite{mey}):
\[
t_{r0}\approx 3\times10^{4} \left(\frac{10M_{\odot}}{m}\right)\left(\frac{r_{c}}{0.06 {\rmn pc}}\right)^{3}
\]
\begin{equation}
\hspace{10mm}
\times \left(\frac{\rho_{0}}{5\times10^{4}M_{\odot}{\rmn pc}^{-3}}\right)^{1/2} \left(\frac{1.4}{\ln(0.4N)}\right) {\rmn yr}
\label{crelax}
\end{equation}
where $r_c$ is the core radius and $\rho_{0}$ the central mass density. In the following
Section (\ref{lumsec}) we also estimate the central mass densities for all the clusters
studied in this paper. Using our derived value of 
$\rho_{0} \sim 3\times10^4 M_{\odot}{\rmn pc}^{-3}$ for R136, and our upper limit
for the core radius of $r_{c} \sim 0.32$ pc, we obtain $t_{r0} \sim 1\times10^{7}$ yr for
 $m\sim 0.5 M_{\odot}$, again considerably longer than
the estimated age of the cluster. Given that $t_{r0} \propto r_c^3$, the true core
radius would have to be at least five times smaller than our upper limit in order to 
bring the central relaxation timescale down to the size required to make R136 
dynamically old enough to be in the epoch of core collapse 
($t_{r0} \sim 1\times 10^{5}$ yr).

\subsection{Luminosity and mass estimates}
\label{lumsec}
We can use our measured structural parameters to obtain luminosity and mass estimates
for each cluster. Eq. \ref{ep} must first be deprojected, which is
done by means of an Abel integral equation (EFF87; see also Binney \& Tremaine 
\shortcite{galdyn}, Section 4.2 and Appendix 1.B.4) for the luminosity 
density $j(r)$ of the cluster in question. We obtain:
\begin{equation}
j(r) = j_{0} \left ( 1 + \frac{r^{2}}{a^{2}} \right )^{-\frac{(\gamma+1)}{2}}
\label{lumdensi}
\end{equation}
which has the same functional form as the surface brightness profile 
$\mu(r)$ but with index $\gamma + 1$. In Eq. \ref{lumdensi},
$j_{0}$ represents the central luminosity density and is given by
\begin{equation}
j_{0} = \frac{\mu_{0} \, \Gamma \left( \frac{\gamma + 1}{2} \right) }{a \sqrt{\pi} \,\, \Gamma \left( \frac{\gamma}{2} \right) }
\label{lumdensii}
\end{equation}
where $\Gamma$ is a standard gamma function. To obtain the measured enclosed 
luminosity $L$ as a function of radius, we integrate Eq. \ref{lumdensi}
within a cylinder of radius $r$ along the line of sight, since this is the 
relevant observational quantity. This gives:
\begin{equation}
L(r) = 4 \pi j_0 \int_{0}^{\infty} \int_{0}^{r} \ell \left( 1 + \frac{x^{2}+\ell^{2}}{a^{2}} \right )^{-\frac{(\gamma+1)}{2}} d\ell \,\, dx
\label{lum}
\end{equation}
where $\ell$ is the radial variable and $x$ the line-of-sight variable. Evaluating 
the integral for $r=r_m$, the maximum radial extent of the measured
surface brightness profile, gives:
\begin{equation}
L_m = \frac{2 \pi \mu_{0}}{\gamma - 2} \left( a^2 - a^\gamma (a^2 + r_m^2 )^{-\frac{(\gamma - 2)}{2}} \right)
\label{lrmax}
\end{equation}
By taking the limit $r_m \rightarrow \infty$, we can also obtain an estimate for the 
asymptotic cluster luminosity $L_{\infty}$:
\begin{equation}
L_{\infty} = \frac{2 \pi \mu_{0} a^{2}}{\gamma - 2}
\label{linf}
\end{equation}
provided $\gamma > 2$, otherwise the limit is divergent. 

The calculated values for $j_{0}$, $L_{\infty}$ and $L_{m}$ are listed
in Table \ref{luminmass}. When calculating $L_{m}$ we take $r_{m}$ from
Table \ref{params}. In order to use Eq. \ref{lumdensii}, \ref{lrmax} and 
\ref{linf}, we must have $\mu_{0}$, $a$ and $r_{m}$ in physical units. As 
always, to convert to parsecs we take the LMC distance modulus to be 
$DM = 18.5$, which implies a scale of $4.116$ arcsec pc$^{-1}$. This
takes care of $a$ and $r_{m}$, but to convert $\mu_{0}$ to 
$L_{\odot}\,{\rmn pc}^{-2}$ is more complicated. We first need to know the 
$V_{555}$ magnitude of the sun. We assume an absolute standard magnitude of 
$V = +4.82$ and a standard colour $B-V = +0.65$ and combine Eq. 7 and 8 from
Holtzman et al. \shortcite{holtzmanb}:
\[
{\rmn WFPC2} = {\rmn SMAG} - T_{1,FS} \times {\rmn SCOL}
\]
\begin{equation}
\hspace{20mm}
- T_{2,FS} \times {\rmn SCOL}^{2} + Z_{FG} - Z_{FS}
\label{wfpcmag}
\end{equation}
where the notation is as in Holtzman et al. \shortcite{holtzmanb}. Using our 
assumed standard solar magnitude and colour, $Z_{FG}$ from Table 6 of Holtzman
et al., and $T_{1,FS}$, $T_{2,FS}$ and $Z_{FS}$ from Table 7 of Holtzman et 
al., we obtain $V_{555}^{\odot} = +4.85$. Alternatively, using Eq. 11 and 12 
from Dolphin \shortcite{wfpc} provides an identical result. The central 
surface brightness for each cluster must be corrected for absorption, and we use 
$E(B-V) = 0.10$, which is a reasonable average in the direction of the LMC 
(cf. EFF87). R136 however, has significantly higher reddening, with 
$E(B-V) = 0.38$ \cite{hunter}. Armed with these numbers, we use the relation
\[
\log \mu_{0} = 0.4(V_{555}^{\odot} - \mu_{555}(0) + DM
\]
\begin{equation}
\hspace{15mm} + 3.1E(B-V)) + \log(4.116^{2})\,\, L_{\odot} \,{\rmn pc}^{-2}
\label{convlsun}
\end{equation}
to convert $\mu_{0}$ to physical units. 

\begin{table*}
\begin{minipage}{175mm}
\caption{Luminosity and mass estimates calculated using the structural parameters from the best fitting EFF profiles.}
\begin{tabular}{@{}lccccccccc}
\hline \hline
Cluster & $\log \mu_{0}$$^{a}$ & Adopted & Adopted & $\log j_{0}$ & $\log L_{\infty}$ & $\log L_{m}$ & $\log \rho_{0}$ & $\log M_{\infty}$ & $\log M_{m}$ \vspace{0.5mm} \\
 & $(L_{\odot}\,{\rmn pc}^{-2})$ & [Fe/H] & $M/L_V$ & $(L_{\odot}\,{\rmn pc}^{-3})$ & $(L_{\odot})$ & $(L_{\odot})$ & $(M_{\odot}\,{\rmn pc}^{-3})$ & $(M_{\odot})$ & $(M_{\odot})$ \\
\hline
NGC1466 & $3.06 \pm 0.02$ & $-2.25$ & $2.80$ & $2.33 \pm 0.05$ & $4.86 \pm 0.11$ & $4.80 \pm 0.09$ & $2.78 \pm 0.05$ & $5.31 \pm 0.11$ & $5.25 \pm 0.09$ \\
NGC1651 & $2.54 \pm 0.02$ & $-0.33$ & $1.14$ & $1.68^{+0.07}_{-0.06}$ & $5.19^{+0.45}_{-0.28}$ & $4.64^{+0.10}_{-0.11}$ & $1.74^{+0.07}_{-0.06}$ & $5.24^{+0.45}_{-0.28}$ & $4.70^{+0.10}_{-0.11}$ \\
NGC1711 & $3.38 \pm 0.02$ & $-0.64$ & $0.12$ & $2.74^{+0.07}_{-0.06}$ & $5.14 \pm 0.16$ & $5.02^{+0.11}_{-0.12}$ & $1.82^{+0.07}_{-0.06}$ & $4.21 \pm 0.16$ & $4.10^{+0.11}_{-0.12}$ \\
NGC1718 & $3.02 \pm 0.03$ & $-0.33$ & $1.14$ & $2.40^{+0.08}_{-0.07}$ & $5.04^{+0.26}_{-0.22}$ & $4.72 \pm 0.13$ & $2.46^{+0.08}_{-0.07}$ & $5.10^{+0.26}_{-0.22}$ & $4.78 \pm 0.13$ \\
NGC1754 & $3.70 \pm 0.10$ & $-1.65$ & $3.36$ & $3.45^{+0.19}_{-0.18}$ & $4.87^{+0.33}_{-0.34}$ & $4.72^{+0.26}_{-0.28}$ & $3.98^{+0.19}_{-0.18}$ & $5.39^{+0.33}_{-0.34}$ & $5.25^{+0.26}_{-0.28}$ \\
NGC1777 & $2.69 \pm 0.01$ & $-0.33$ & $0.67$ & $1.85 \pm 0.04$ & $4.92^{+0.13}_{-0.12}$ & $4.70^{+0.07}_{-0.08}$ & $1.68 \pm 0.04$ & $4.75^{+0.13}_{-0.12}$ & $4.53^{+0.07}_{-0.08}$ \\
NGC1786 & $3.71 \pm 0.04$ & $-1.65$ & $3.29$ & $3.28 \pm 0.10$ & $5.05 \pm 0.20$ & $4.98^{+0.16}_{-0.17}$ & $3.79 \pm 0.10$ & $5.57 \pm 0.20$ & $5.50^{+0.16}_{-0.17}$ \\
NGC1805 & $3.49 \pm 0.02$ & $-0.33$ & $0.05$ & $3.06 \pm 0.06$ & $4.82 \pm 0.13$ & $4.75^{+0.10}_{-0.11}$ & $1.75 \pm 0.06$ & $3.52 \pm 0.13$ & $3.45^{+0.10}_{-0.11}$ \\
NGC1818 & $3.35 \pm 0.02$ & $-0.33$ & $0.08$ & $2.65 \pm 0.06$ & $5.23^{+0.15}_{-0.14}$ & $5.11 \pm 0.10$ & $1.55 \pm 0.06$ & $4.13^{+0.15}_{-0.14}$ & $4.01 \pm 0.10$ \\
NGC1831 & $3.06 \pm 0.02$ & $-0.33$ & $0.32$ & $2.10 \pm 0.06$ & $5.30 \pm 0.13$ & $5.21^{+0.09}_{-0.10}$ & $1.60 \pm 0.06$ & $4.81 \pm 0.13$ & $4.71^{+0.09}_{-0.10}$ \\
NGC1835 & $4.14 \pm 0.03$ & $-1.65$ & $3.56$ & $3.77 \pm 0.07$ & $5.27 \pm 0.13$ & $5.24 \pm 0.12$ & $4.32 \pm 0.07$ & $5.83 \pm 0.13$ & $5.79 \pm 0.12$ \\
NGC1841 & $2.06 \pm 0.01$ & $-2.25$ & $2.75$ & $0.85 \pm 0.08$ & $4.68^{+0.21}_{-0.19}$ & $4.57^{+0.12}_{-0.14}$ & $1.29 \pm 0.08$ & $5.12^{+0.21}_{-0.19}$ & $5.00^{+0.12}_{-0.14}$ \\
NGC1847 & $3.28 \pm 0.02$ & $-0.33$ & $0.09$ & $2.66 \pm 0.06$ & $6.01^{+0.48}_{-0.63}$ & $4.91^{+0.09}_{-0.10}$ & $1.62 \pm 0.06$ & $4.97^{+0.48}_{-0.63}$ & $3.86^{+0.09}_{-0.10}$ \\
NGC1850 & $4.01 \pm 0.03$ & $-0.33$ & $0.10$ & $3.30 \pm 0.08$ & $6.42^{+0.52}_{-0.32}$ & $5.87^{+0.13}_{-0.14}$ & $2.30 \pm 0.08$ & $5.42^{+0.52}_{-0.32}$ & $4.87^{+0.13}_{-0.14}$ \\
NGC1856 & $4.02 \pm 0.03$ & $-0.64$ & $0.18$ & $3.50 \pm 0.06$ & $7.26^{+0.69}_{-0.87}$ & $5.63^{+0.10}_{-0.11}$ & $2.76 \pm 0.06$ & $6.51^{+0.69}_{-0.87}$ & $4.89^{+0.10}_{-0.11}$ \\
NGC1860 & $2.80 \pm 0.04$ & $-0.64$ & $0.22$ & $2.16 \pm 0.08$ & $4.95^{+0.41}_{-0.27}$ & $4.55^{+0.12}_{-0.13}$ & $1.50 \pm 0.08$ & $4.30^{+0.41}_{-0.27}$ & $3.90^{+0.12}_{-0.13}$ \\
NGC1866 & $3.28 \pm 0.01$ & $-0.64$ & $0.18$ & $2.44 \pm 0.03$ & $5.38 \pm 0.08$ & $5.26 \pm 0.06$ & $1.69 \pm 0.03$ & $4.63 \pm 0.08$ & $4.52 \pm 0.06$ \\
NGC1868 & $3.50 \pm 0.02$ & $-0.64$ & $0.40$ & $2.97 \pm 0.04$ & $4.93 \pm 0.10$ & $4.89 \pm 0.09$ & $2.58 \pm 0.04$ & $4.53 \pm 0.10$ & $4.49 \pm 0.09$ \\
NGC1898 & $3.07 \pm 0.02$ & $-1.65$ & $3.13$ & $2.46^{+0.09}_{-0.08}$ & $5.38^{+0.19}_{-0.42}$ & $4.80^{+0.15}_{-0.16}$ & $2.96^{+0.09}_{-0.08}$ & $5.88^{+0.19}_{-0.42}$ & $5.29^{+0.15}_{-0.16}$ \\
NGC1916 & $4.32 \pm 0.04$ & $-2.25$ & $3.37$ & $4.10 \pm 0.07$ & $5.27^{+0.15}_{-0.14}$ & $5.21^{+0.12}_{-0.13}$ & $4.63 \pm 0.07$ & $5.79^{+0.15}_{-0.14}$ & $5.73^{+0.12}_{-0.13}$ \\
NGC1984 & $3.24 \pm 0.04$ & $-0.64$ & $0.05$ & $2.94 \pm 0.10$ & $4.68^{+0.35}_{-0.28}$ & $4.40^{+0.16}_{-0.17}$ & $1.64 \pm 0.10$ & $3.38^{+0.35}_{-0.28}$ & $3.10^{+0.16}_{-0.17}$ \\
NGC2004 & $3.92 \pm 0.03$ & $-0.64$ & $0.08$ & $3.42^{+0.10}_{-0.09}$ & $5.52^{+0.24}_{-0.23}$ & $5.37^{+0.16}_{-0.17}$ & $2.32^{+0.10}_{-0.09}$ & $4.43^{+0.24}_{-0.23}$ & $4.27^{+0.16}_{-0.17}$ \\
NGC2005 & $3.88 \pm 0.05$ & $-1.65$ & $3.56$ & $3.62 \pm 0.12$ & $4.94^{+0.27}_{-0.26}$ & $4.85^{+0.21}_{-0.22}$ & $4.17 \pm 0.12$ & $5.49^{+0.27}_{-0.26}$ & $5.40^{+0.21}_{-0.22}$ \\
NGC2011 & $3.12 \pm 0.06$ & $-0.33$ & $0.05$ & $2.75 \pm 0.12$ & $4.77^{+0.38}_{-0.32}$ & $4.42^{+0.19}_{-0.20}$ & $1.45 \pm 0.12$ & $3.47^{+0.38}_{-0.32}$ & $3.12^{+0.19}_{-0.20}$ \\
NGC2019 & $4.00 \pm 0.04$ & $-1.65$ & $3.80$ & $3.75 \pm 0.09$ & $5.10^{+0.20}_{-0.19}$ & $4.99 \pm 0.15$ & $4.33 \pm 0.09$ & $5.68^{+0.20}_{-0.19}$ & $5.57 \pm 0.15$ \\
NGC2031 & $3.12 \pm 0.02$ & $-0.64$ & $0.20$ & $2.40 \pm 0.08$ & $5.83^{+0.24}_{-0.45}$ & $5.03^{+0.12}_{-0.13}$ & $1.70 \pm 0.08$ & $5.13^{+0.24}_{-0.45}$ & $4.33^{+0.12}_{-0.13}$ \\
NGC2100 & $4.19 \pm 0.06$ & $-0.33$ & $0.07$ & $3.80^{+0.13}_{-0.12}$ & $5.63^{+0.33}_{-0.30}$ & $5.46^{+0.22}_{-0.23}$ & $2.64^{+0.13}_{-0.12}$ & $4.48^{+0.33}_{-0.30}$ & $4.31^{+0.22}_{-0.23}$ \\
NGC2121 & $2.33 \pm 0.01$ & $-0.64$ & $1.33$ & $1.15 \pm 0.09$ & $5.56^{+0.22}_{-0.44}$ & $4.86^{+0.11}_{-0.13}$ & $1.27 \pm 0.09$ & $5.69^{+0.22}_{-0.44}$ & $4.99^{+0.11}_{-0.13}$ \\
NGC2136 & $3.74 \pm 0.04$ & $-0.64$ & $0.16$ & $3.13 \pm 0.09$ & $5.24 \pm 0.17$ & $5.22 \pm 0.16$ & $2.33 \pm 0.09$ & $4.45 \pm 0.17$ & $4.42 \pm 0.16$ \\
NGC2153 & $3.02 \pm 0.04$ & $-0.33$ & $0.69$ & $2.74^{+0.14}_{-0.13}$ & $4.13^{+0.35}_{-0.32}$ & $4.04^{+0.26}_{-0.28}$ & $2.58^{+0.14}_{-0.13}$ & $3.97^{+0.35}_{-0.32}$ & $3.88^{+0.26}_{-0.28}$ \\
NGC2155 & $2.52 \pm 0.02$ & $-0.64$ & $1.33$ & $1.61 \pm 0.09$ & $4.77^{+0.26}_{-0.23}$ & $4.61^{+0.15}_{-0.16}$ & $1.74 \pm 0.09$ & $4.90^{+0.26}_{-0.23}$ & $4.73^{+0.15}_{-0.16}$ \\
NGC2156 & $3.10 \pm 0.03$ & $-0.33$ & $0.11$ & $2.59^{+0.07}_{-0.06}$ & $4.59^{+0.14}_{-0.13}$ & $4.51^{+0.10}_{-0.11}$ & $1.63^{+0.07}_{-0.06}$ & $3.63^{+0.14}_{-0.13}$ & $3.55^{+0.10}_{-0.11}$ \\
NGC2157 & $3.58 \pm 0.02$ & $-0.33$ & $0.11$ & $2.90 \pm 0.06$ & $5.27 \pm 0.13$ & $5.23 \pm 0.11$ & $1.94 \pm 0.06$ & $4.31 \pm 0.13$ & $4.27 \pm 0.11$ \\
NGC2159 & $2.83 \pm 0.02$ & $-0.33$ & $0.11$ & $2.14^{+0.07}_{-0.06}$ & $4.61 \pm 0.15$ & $4.52 \pm 0.12$ & $1.18^{+0.07}_{-0.06}$ & $3.65 \pm 0.15$ & $3.56 \pm 0.12$ \\
NGC2162 & $2.72 \pm 0.03$ & $-0.33$ & $0.69$ & $2.02 \pm 0.10$ & $4.56^{+0.25}_{-0.23}$ & $4.46^{+0.17}_{-0.18}$ & $1.86 \pm 0.10$ & $4.40^{+0.25}_{-0.23}$ & $4.30^{+0.17}_{-0.18}$ \\
NGC2164 & $3.40 \pm 0.02$ & $-0.33$ & $0.13$ & $2.81 \pm 0.04$ & $5.01 \pm 0.09$ & $4.93 \pm 0.07$ & $1.93 \pm 0.04$ & $4.12 \pm 0.09$ & $4.04 \pm 0.07$ \\
NGC2172 & $2.61 \pm 0.03$ & $-0.33$ & $0.11$ & $1.88 \pm 0.07$ & $4.48^{+0.16}_{-0.15}$ & $4.39^{+0.12}_{-0.13}$ & $0.92 \pm 0.07$ & $3.52^{+0.16}_{-0.15}$ & $3.43^{+0.12}_{-0.13}$ \\
NGC2173 & $2.75 \pm 0.02$ & $-0.33$ & $1.19$ & $1.99 \pm 0.08$ & $4.99^{+0.35}_{-0.27}$ & $4.67^{+0.14}_{-0.15}$ & $2.06 \pm 0.08$ & $5.06^{+0.35}_{-0.27}$ & $4.75^{+0.14}_{-0.15}$ \\
NGC2193 & $2.78 \pm 0.03$ & $-0.64$ & $1.00$ & $2.27^{+0.09}_{-0.08}$ & $4.42^{+0.22}_{-0.20}$ & $4.25 \pm 0.14$ & $2.27^{+0.09}_{-0.08}$ & $4.42^{+0.22}_{-0.20}$ & $4.25 \pm 0.14$ \\
NGC2209 & $1.86 \pm 0.02$ & $-0.33$ & $0.61$ & $0.82 \pm 0.09$ & $5.25^{+0.36}_{-0.60}$ & $4.22^{+0.11}_{-0.13}$ & $0.60 \pm 0.09$ & $5.03^{+0.36}_{-0.6}$ & $4.01^{+0.11}_{-0.13}$ \\
NGC2210 & $3.42 \pm 0.02$ & $-2.25$ & $3.37$ & $2.81 \pm 0.05$ & $4.95 \pm 0.10$ & $4.92^{+0.08}_{-0.09}$ & $3.34 \pm 0.05$ & $5.48 \pm 0.10$ & $5.45^{+0.08}_{-0.09}$ \\
NGC2213 & $3.03 \pm 0.02$ & $-0.33$ & $0.87$ & $2.46 \pm 0.06$ & $4.62 \pm 0.15$ & $4.54^{+0.11}_{-0.12}$ & $2.40 \pm 0.06$ & $4.56 \pm 0.15$ & $4.48^{+0.11}_{-0.12}$ \\
NGC2214 & $3.35 \pm 0.03$ & $-0.33$ & $0.11$ & $2.71 \pm 0.09$ & $5.46^{+0.45}_{-0.31}$ & $5.09^{+0.16}_{-0.17}$ & $1.76 \pm 0.09$ & $4.50^{+0.45}_{-0.31}$ & $4.13^{+0.16}_{-0.17}$ \\
NGC2231 & $2.50 \pm 0.05$ & $-0.64$ & $0.69$ & $1.77^{+0.20}_{-0.18}$ & $5.34^{+0.43}_{-0.96}$ & $4.44^{+0.29}_{-0.35}$ & $1.60^{+0.20}_{-0.18}$ & $5.18^{+0.43}_{-0.96}$ & $4.28^{+0.29}_{-0.35}$ \\
NGC2249 & $3.07 \pm 0.02$ & $-0.33$ & $0.48$ & $2.44 \pm 0.06$ & $4.66 \pm 0.13$ & $4.63 \pm 0.12$ & $2.12 \pm 0.06$ & $4.35 \pm 0.13$ & $4.31 \pm 0.12$ \\
NGC2257 & $2.32 \pm 0.01$ & $-1.65$ & $3.43$ & $1.19 \pm 0.06$ & $4.88 \pm 0.15$ & $4.72^{+0.09}_{-0.10}$ & $1.73 \pm 0.06$ & $5.41 \pm 0.15$ & $5.26^{+0.09}_{-0.10}$ \\
SL663 & $1.73 \pm 0.02$ & $-0.64$ & $1.33$ & $0.60 \pm 0.07$ & $5.11^{+0.24}_{-0.42}$ & $4.21^{+0.09}_{-0.10}$ & $0.73 \pm 0.07$ & $5.23^{+0.24}_{-0.42}$ & $4.33^{+0.09}_{-0.10}$ \\
SL842 & $2.26 \pm 0.04$ & $-0.33$ & $1.14$ & $1.55 \pm 0.12$ & $4.02^{+0.25}_{-0.24}$ & $3.97^{+0.20}_{-0.21}$ & $1.61 \pm 0.12$ & $4.08^{+0.25}_{-0.24}$ & $4.02^{+0.20}_{-0.21}$ \\
SL855 & $1.52 \pm 0.01$ & $-0.33$ & $0.70$ & $0.49^{+0.10}_{-0.09}$ & $4.74^{+0.34}_{-0.59}$ & $3.88^{+0.13}_{-0.15}$ & $0.33^{+0.10}_{-0.09}$ & $4.58^{+0.34}_{-0.59}$ & $3.72^{+0.13}_{-0.15}$ \\
HODGE4 & $2.46 \pm 0.02$ & $-0.33$ & $1.21$ & $1.59 \pm 0.09$ & $5.31^{+0.22}_{-0.45}$ & $4.59^{+0.13}_{-0.14}$ & $1.67 \pm 0.09$ & $5.39^{+0.22}_{-0.45}$ & $4.67^{+0.13}_{-0.14}$ \\
HODGE11 & $2.85 \pm 0.02$ & $-2.25$ & $3.25$ & $2.08 \pm 0.07$ & $5.11^{+0.26}_{-0.22}$ & $4.79^{+0.11}_{-0.12}$ & $2.59 \pm 0.07$ & $5.63^{+0.26}_{-0.22}$ & $5.31^{+0.11}_{-0.12}$ \\
HODGE14 & $2.58 \pm 0.04$ & $-0.64$ & $0.88$ & $2.02 \pm 0.10$ & $4.38^{+0.34}_{-0.28}$ & $4.16^{+0.18}_{-0.19}$ & $1.96 \pm 0.10$ & $4.33^{+0.34}_{-0.28}$ & $4.10^{+0.18}_{-0.19}$ \\
R136$^{b}$ & $5.98 \pm 0.04$$^{c}$ & $-0.33$ & $0.02$ & $6.17 \pm 0.08$ & $6.25^{+0.21}_{-0.19}$ & $6.01 \pm 0.13$ & $4.47 \pm 0.08$ & $4.55^{+0.21}_{-0.19}$ & $4.31 \pm 0.13$ \\
\hline
\end{tabular}
\medskip
\\
$^{a}$ Corrected for reddening using $E(B-V) = 0.10$. \\
$^{b}$ Luminosity and mass calculated using the structural parameters derived for the inner regions only -- that is, those parameters listed in Table \ref{params}. Given the break to a shallower profile at $r \sim 15\arcsec$, we likely under-estimate $L_{\infty}$ and $M_{\infty}$ -- the asymptotic luminosity and mass including NGC 2070.\\
$^{c}$ Corrected for reddening using $E(B-V) = 0.38$ (see text). \\
\label{luminmass}
\end{minipage}
\end{table*}

We calculate mass and density estimates for each cluster by multiplying 
the appropriate luminosity equations by the average mass to light ratio 
$M/L_V$ for the cluster in question. The density $\rho(r)$ then corresponds 
to Eq. \ref{lumdensi}, with the central density $\rho_{0}$ obtained from 
Eq. \ref{lumdensii}. Similarly, the mass $M_{m}$ inside $r_{m}$ is given by Eq.
\ref{lrmax} multiplied by $M/L_V$ and the asymptotic mass $M_{\infty}$
is given by Eq. \ref{linf} times this ratio. 

To estimate $M/L_{V}$ for each cluster, we use the evolutionary
synthesis code of Fioc \& Rocca-Volmerange \shortcite{pegase} (PEGASE v2.0, 1999). 
This code determines the integrated properties of a synthetic 
stellar population as a function of time, using libraries of isochrones and stellar 
spectra to make the detailed calculations, and accounting for factors such as the IMF 
of the system, a range of metallicities, and any ongoing star-formation. We proceed 
using the simplest possible model -- a population of stars formed simultaneously in one
initial burst and with the same metallicity -- presumably a fairly good approximation
to the formation of a rich stellar cluster. For the initial burst, we assume
the IMF of Kroupa, Tout \& Gilmore \shortcite{ktg} over the mass range 
$0.1$ to $120 M_{\odot}$. There are four available initial abundances which cover 
the metallicity range of the cluster sample -- $Z = 0.0001$ ([Fe/H] 
$\approx -2.25$); $Z = 0.0004$ ([Fe/H] $\approx -1.65$); $Z = 0.004$ ([Fe/H] 
$\approx -0.64$); and $Z = 0.008$ ([Fe/H] $\approx -0.33$). The $M/L_V$ 
values for each metallicity are plotted as a function of cluster age in Fig. 
\ref{pegaseml}. Although the mass-to-light ratios obtained from 
these models are purely theoretical, they do agree well with observations 
(see e.g., Parmentier \& Gilmore \shortcite{pg}). From Fig. \ref{pegaseml},
it is clear that for most of the evolution, the $M/L_V$ ratio is relatively
insensitive to the chosen abundance. For completeness, we adopt the metallicity for 
each cluster based on the estimates in Table \ref{ages}. Because of the insensitivity
of the calculation to the selected abundance, we are confident in using even those
abundances which were calculated as averages rather than being taken directly from
the literature -- it is sufficient to be able to differentiate between ``metal poor''
([Fe/H] $\approx -2.25$) and ``metal rich'' ([Fe/H] $\approx -0.33$). We then use the 
age estimates from Table \ref{ages} to obtain a mass to light ratio from the 
appropriate evolutionary synthesis model. The adopted abundances, and values for 
$M/L_V$, $\rho_{0}$, $M_{\infty}$ and $M_{m}$ are listed in Table \ref{luminmass}.

The values of $L_{\infty}$ and $M_{\infty}$ are intended to provide 
reasonable estimates of total cluster luminosities and masses in the 
absence of tidal limit information. There is very little difference in
the result obtained from using Eq. \ref{lrmax} with $r_{m}$ several 
hundred arcseconds (a reasonable tidal cut-off), and the result obtained
from Eq. \ref{linf}. We provide estimates of $L_{m}$ and $M_{m}$ to
show the luminosity and mass over the radial range we actually measure,
and as such, these provide reliable lower limits for total cluster 
luminosities and masses. For most clusters there is not a great difference
between these values and the asymptotic values; however, for clusters with 
$\gamma \approx 2$, the value of $L_{\infty}$ (and $M_{\infty}$) can become 
unreasonably large (e.g, NGC 1856), and the extrapolation
$r \rightarrow \infty$ may not be justified. 

The presence of saturated stars on some images, and the imposition of 
brightness limits on others (see Sections \ref{short} and \ref{saturated}) causes some 
luminosity (and mass) to be unaccounted for in several surface brightness profiles
(that is, the zero-point $\mu_{555}(0)$ is fainter than would be expected were the
brightest stars included in the profiles). However, this effect is not particularly 
significant. It is most pronounced in the youngest clusters, where bright stars
are relatively massive. At worst we miss $\sim 30$ stars from any young cluster, 
which results in $\la 600 M_{\odot}$ being neglected. Addition of this missing mass
gives totals still within the quoted uncertainties for such clusters. The errors 
listed in Table \ref{luminmass} represent the uncertainties in the calculated 
parameters due to the random errors in the values of $\mu_{0}$, $\gamma$ and $a$. We 
do not account for any systematic errors such as those introduced by any saturated
stars or those which might be present in the calculated values for $M/L_{V}$.

\section{The core radius vs. age relationship}
\label{discussion}
With a large sample such as this, there is the opportunity for a full
statistical analysis of the parameters presented in Tables \ref{ages},
\ref{params} and \ref{luminmass}, in search of correlations
and physical insight into the overall properties of the LMC cluster
system. Such an analysis is beyond the scope of the present paper and will be 
presented in the future. We do however observe one particularly noteworthy 
trend. 

When core radius is plotted against age for all clusters in the sample (Fig. 
\ref{coreage}), a clear relationship exists between these two parameters -- 
namely that the spread in core radius increases significantly with increasing 
cluster age. Without exception, all of the young clusters have compact cores 
with $r_{c} \la 2.5$ pc, while for clusters older than $\log\tau \sim 9$ the 
full range $r_{c} \approx 0-8$ pc is covered. This trend has previously been 
discovered and discussed by Elson et al. \shortcite{efl}, E91 and Elson 
\shortcite{elsonfive}, using a combination of literature data and 
ground-based measurements. 

The key question is whether this observed relationship is indicative
of true structural evolution in LMC clusters as they grow older, or
is merely the result of a secondary correlation, a product of the reduction
process or a selection effect (i.e., we systematically missed all the young, 
low surface brightness clusters from our sample). The fact that we confirm 
exactly the result of Elson, using a larger (and distinct) sample, 
space-based observations, and uniformly selected and reduced data is argument 
against the relationship being a reduction artifact, or due to a selection 
effect. In addition, we did not choose our young clusters in a systematic 
fashion, but rather simply used all those available in the {\em HST}
archive -- in effect, a random sample. This adds weight to the argument
against a selection effect being responsible for the observed upper envelope;
however we cannot guarantee that there are not very low surface brightness
young clusters with large cores in the LMC system. It is likely however,
that such diffuse clusters would not remain bound for long, and so their 
presence (or otherwise) does not affect the discussion and conclusions 
below.

Given that the radius-age relationship does not seem to be
the product of data reduction or a selection effect, E91 discusses the 
possibility that a correlation between mass and core radius and between 
mass and age might be responsible -- that is, if both more massive clusters 
(with larger cores) and less massive clusters (with smaller cores) were formed 
in the past, but only less massive clusters (with smaller cores) formed recently, 
it might seem that core radius evolves with age. However, Elson observed no 
such correlations in her sample of 10 young clusters. 

Similarly, we plot $M_{\infty}$ against age, and $M_{\infty}$ against
$r_{c}$ for our (much larger) sample, to look for correlations. These 
plots are Fig. \ref{agemass} and Fig. \ref{coremass} respectively. 
While $M_{\infty}$ is not always a reliable extrapolation
to the true mass of a cluster, we use it here as a convenient parameter for 
eliminating the age bias in $L_{m}$ and $L_{\infty}$ (i.e., young clusters 
are intrinsically more luminous than old clusters), as well as any 
observational bias present in $M_{m}$ (i.e., a cluster with $\gamma \sim 2$ 
will have more mass at large radii, where we did not observe, than
a cluster with $\gamma \sim 4$, and consequently will likely have a 
correspondingly small $M_{m}$). In Fig. \ref{agemass}, we see no significant 
correlation of mass with age. It is true that there are a few young clusters 
with low masses, and that all the old clusters seem to have larger masses; 
however we attribute this to the fact that the low-mass young clusters will 
have faded and probably dispersed by age $10^{10}$ yr -- explaining why there 
are no low-mass old clusters (a detailed discussion of this effect is 
presented in Meylan \shortcite{meylanr}). There is no evidence for any 
significant intrinsic difference between the masses of the largest newly 
formed clusters and the old clusters in the sample. In Fig. \ref{coremass} we 
also see no evidence for a strong correlation. While it is true that 
those clusters with large core radii all seem to be relatively massive, 
they are not more massive than most of the clusters with smaller core radii.

We therefore conclude, like Elson, that the trend observed in Fig.
\ref{coreage} represents real evolution in the structure of clusters as they 
grow older. Given this, it is useful to return to
Fig. \ref{coreage} and examine closely the distribution of clusters, which
seems to exhibit a bifurcation at around several$\times10^{8}$ yr.
This bifurcation grows into a large separation by 10 Gyr. It is possible
that the apparent dearth of older clusters with core radii in the range
$3-6$ pc could be a small-sample effect. A simple Kolmogorov-Smirnov test
on the distribution of core radii for all clusters older than 10 Gyr, shows that
they are not drawn from a uniform distribution, at about the 99 per cent level.
If we increase the sample size by including all clusters older than 1 Gyr,
the significance increases to better than 99.5 per cent. The bifurcation
is therefore statistically significant.

The lower branch in the radius-age diagram represents the standard picture of
(isolated) cluster evolution, where a newly formed cluster suffers significant
and rapid mass-loss due to stellar evolution, causing the cluster to expand.
The rate of expansion is regulated by the initial mass spectrum present in the 
cluster (i.e., the slope of the mass function) (see e.g., Chernoff \& 
Weinberg \shortcite{cw}). The more heavily weighted the high mass end of a 
cluster's mass function is, the more severe the early mass loss. As the cluster
grows older and evolves, the expansion slows and is eventually reversed by core 
collapse. E91 present Fokker-Planck models of evolving clusters with different 
IMF slopes, plotted over the equivalent of Fig. \ref{coreage}. Clusters with
IMF slopes similar to that for the Salpeter IMF ($x=1.35$, where the 
IMF is given by $\Phi \propto m^{-(1+x)}$) and steeper follow tracks which match
well the shape of the lower branch in Fig. \ref{coreage}. 

There is a noticeable
scatter about this sequence, but its cause is not clear. Because the IMF
slope governs the rate of expansion, IMF variations could be responsible. However,
more and more evidence is pointing towards the universality of the IMF --
Gilmore \shortcite{gilimf} provides a detailed review, and more recently
de Grijs et al. \shortcite{grijsimf} have shown that a sample of six LMC clusters
(widely scattered on Fig. \ref{coreage}) must have had very similar IMFs. It 
therefore seems likely that a combination of other factors is instead responsible.
Even the youngest clusters in the present sample have a spread in core radii --
this probably reflects the different formation conditions of each cluster (E91), 
as well as the spread in initial mass (Fig. \ref{agemass}). Variations in initial 
mass would cause a spread in the lower sequence - it may be that this alone could
reproduce the required scatter. A second possibility is that the scatter is a
result of the cluster sample being spread over a considerable range of
distances from the LMC centre (Table \ref{ages}). Gilmore \shortcite{gilimf} 
shows that tidal forces are at least as significant in the comparative evolution 
of Galactic globular clusters as IMF variations between clusters. The spread in 
the lower sequence of Fig. \ref{coreage} might simply reflect the spread in orbital
radii between clusters. We are currently employing $N$-body simulations
to explore these two mechanisms in detail, to measure their relative 
effects, and to see whether either (or a combination of both) is able to
reproduce the lower region of Fig. \ref{coreage} for a suitable cluster sample.

The fact that the models of isolated evolving clusters presented by E91 match
the lower sequence of Fig. \ref{coreage} so well adds plausibility to our argument
that the radius-age plot tracks the physical evolution of LMC clusters.
This renders the clusters which follow the upper branch especially intriguing.
What makes these clusters different -- why do they diverge from the standard
sequence beyond a few$\times 10^{8}$ yr, to finish in the upper right of the plot?
Possibly, these clusters are on the verge of dissolution -- after a certain
time, mass loss can be so significant that rather than undergoing core collapse, 
a cluster simply dissipates. However, the upper sequence covers the timescale 
$10^{8}$ to $10^{10}$ yr, which is far too long for dissolution. In addition, 
Fig. \ref{coremass} shows that clusters with large core radii are all relatively
massive, whereas one would expect a cluster dissolving due to mass loss
to be less massive. Upper sequence clusters are therefore probably not undergoing 
dissolution, but are expanding for another reason. Given this, we suggest two ways 
in which upper sequence clusters might be distinguished from the 
``standard'' clusters -- either they have (or have had) radically different 
stellar populations, or they have been influenced by some external process (i.e.,
they are not isolated). 

Examples falling into the first category include very flat IMF slopes, and very
large binary star fractions. It seems unlikely that either could cause
the degree of core expansion observed. As discussed previously, the IMF slope
for a cluster regulates its early expansion, but the rapid expansion phase
does not last the entire lifetime of the cluster (as would be required for the
cluster to evolve along the upper sequence). In addition, as shown by E91, 
a cluster with a very flat IMF suffers such extreme early mass loss that it becomes 
unbound after only a very short time ($\sim 10^7$ yrs). Our earlier argument with 
respect to the universality of the IMF also holds here. Similarly, although the binary
star fraction in a cluster is important in the context of halting core collapse, 
binary stars alone cannot drive the large scale core expansion we have observed. 
To extract energy from a cluster's binary stars requires close stellar encounters,
and an expanding core radius implies lower stellar density and therefore fewer
encounters in a given period. It is unlikely therefore that significant cluster 
expansion can be driven by binary stars over the required timescale, even given
an exceptional binary fraction. 

Two examples which fall into the second category are the merger of double clusters
and the effects of strongly variable, perturbative tidal fields. Again, from simple
physical arguments, it is difficult to see how either mechanism could produce the
scale of core expansion which is observed without destroying the clusters in 
question. Double cluster pairs are relatively common in the LMC 
\cite{bhatia,deoliv,deoliveira}, and Bhatia \& Hatzidimitriou \shortcite{bhati} 
have shown that it is statistically unlikely that they are {\em all} coincidental
alignments -- hence some are likely gravitationally bound and interacting. 
An interacting pair will merge after a relatively brief period \cite{bhat}, and 
$N$-body studies by de Oliveira et al. \shortcite{deoliv,deoliveira} have shown that
a binary cluster merger can result in a single cluster with a
structure which is stable after $\sim 200$ Myr and distinct from the 
structure of the original clusters. However, it is unclear whether large scale 
core expansion is a result -- small scale expansion implies many mergers would
be required, and this is statistically very unlikely. In the case of 
tidal forces, it is uncertain as to what effects a tidal field can have on a cluster
core. Tidal forces increase as the cube of the distance from the centre of a cluster
and therefore affect mainly the outer parts. A tidal force strong enough to
significantly alter a cluster's core would probably destroy the cluster. However,
as with the merger of double clusters, no models have been explored in the
context of core expansion, and so neither mechanism can yet be quantitatively 
eliminated. We are currently carrying out detailed $N$-body simulations 
(e.g., Wilkinson et al., in prep.) to explore this further.

\section{Summary and Conclusions}
We have compiled a pseudo-snapshot data set of two-colour observations
from the {\em HST} archive for a sample of 53 rich LMC clusters
spanning the full age range $10^{6}$ to $10^{10}$ yr. The emphasis
has been on trying to make this compilation and the subsequent reduction
process as homogeneous as possible without sacrificing data integrity.
We have also compiled literature estimates for the ages and metallicities
of these clusters, again trying to maintain consistency as far as 
possible. From the {\em HST} observations, we have constructed 
surface brightness profiles for the entire sample and obtained structural
parameters for each cluster, including the core radius and power-law
slope at large radii. Using these parameters we have also estimated the
total luminosity and mass for each cluster. These data, along with the surface
brightness profiles, are available on-line at 
{\em http://www.ast.cam.ac.uk/STELLARPOPS/LMC\_clusters/}.

The surface brightness profiles show a rich amount of detail, with young
clusters in particular exhibiting bumps, shoulders and dips in their
profiles. We see evidence for double clusters in our sample, as well as
post core-collapse clusters. The PCC candidates are especially interesting
-- the two best examples show clear power-law profiles at small radii,
with slopes $\beta \sim 0.7$ in the $\log\mu - \log r$ plane. Our sample
covers twelve of the fifteen definite old LMC globular clusters, and
we are able to estimate that $20 \pm 7$ per cent of these
clusters are PCC objects, matching the $20$ per cent
estimated for the Galactic globular cluster system. We have also shown that
R136 requires a two component fit to its profile, and that it is
likely not yet in a PCC state.

If core radius is plotted against age for the entire sample, we see
that while all the young clusters have compact cores, the spread in
core radius increases with age, with the oldest clusters covering the
full range of core radii measured. We have argued that this trend 
reflects real evolution in cluster structure with age. The distribution 
of clusters on the plot suggests a bifurcation at several hundred Myr, 
with most clusters maintaining small cores consistent with standard
isolated globular cluster evolution, but with several moving to
the upper right of the diagram and evolving large diffuse cores. 
We suggest that these clusters must be different to the ``standard''
clusters, either by having exceptional stellar populations, or by 
being subjected to an external influence. We are currently employing 
$N$-body simulations to explore several physical processes which fall
into these categories.

\section*{Acknowledgments}
We would like to thank Mark Wilkinson, Richard de Grijs and Jarrod
Hurley for useful discussions and suggestions. ADM is supported by
a Trinity College ERS grant and a British government ORS award.
This paper is based on observations made with the NASA/ESA 
{\em Hubble Space Telescope}, obtained from the data archive at the 
Space Telescope Institute. STScI is operated by the association of 
Universities for Research in Astronomy, Inc. under the NASA contract 
NAS 5-26555.


\clearpage


\begin{figure*}
\begin{minipage}{175mm}
\section*{Figure Captions}
\caption{Observation geometry and area correction points for cluster NGC 1841. The centre of the cluster falls on WFC3 at pixel coordinates $(417,401)$ and is marked with a small cross. Annuli of width $4\arcsec$ have been drawn to radius $100\arcsec$ and the random points falling in the field of view are plotted for every second annulus.}
\label{areacorr}
\end{minipage}
\end{figure*}

\begin{figure*}
\begin{minipage}{175mm}
\caption{Completeness functions for NGC 2213 (left) and NGC 2005 (right). The function for the PC is denoted by the solid line, that for WFC2 by the long-dashed line, that for WFC3 by the short-dashed line, and that for WFC4 by the dotted line. The functions have been integrated over position and colour to be a function of $V_{555}$ magnitude only. Shown above the completeness functions are the PC F555W images of the core of each cluster, with exposure times of 120 s (NGC 2213) and 500 s (NGC 2005). The images are slightly cropped to be approximately 730 pixels ($33\farcs2$) on a side. It is clear from these images that the core of NGC 2005 is extremely compact, on a scale less than $\sim 160$ pixels, while NGC 2213 is well resolved even in its very centre.}
\label{completeness}
\end{minipage}
\end{figure*}

\begin{figure*}
\begin{minipage}{175mm}
\caption{An example of the inclusion of information from a short exposure to alleviate significant incompleteness due to crowding, in this case for NGC 2005. Our long exposure image of the core of this cluster is shown in Fig. \ref{completeness}. {\em (a)} Surface brightness profiles for all four annulus widths, calculated from the original exposure (solid circles) and short exposure (open squares). The dotted line shows the approximate radius $r_{dev}$ at which the profiles start to deviate. {\em (b)} Completeness functions for the short exposure (solid line) and original exposure (dashed line) within $r_{dev}$. The completeness functions have been integrated over colour and position within $r_{dev}$ to be a function of $V_{555}$ only. The vertical dotted line marks the brightness limit $V_{lim}$ imposed on the short exposure photometry to remove the offset between the two surface brightness profiles.}
\label{mergephot}
\end{minipage}
\end{figure*}

\begin{figure*}
\begin{minipage}{175mm}
\caption{Surface brightness profiles for an EFF model (solid line) and empirical King model (dashed line). The profiles both have core radii of $r_{c} = 10\arcsec$ and central surface brightnesses of $\mu(0) = 17$. For the EFF profile we have chosen $\gamma = 2.5$ which is approximately the median value measured by EFF87 for young LMC clusters, and this implies $a = 11\farcs62$ to obtain the correct $r_{c}$. For the King profile we have chosen a concentration $c = 1.5$, implying $r_{t} = 316\arcsec$ and $k = 16.93$. The core and tidal radii are marked. The vertical dotted line shows the typical maximum radius for surface brightness profiles constructed in the present paper.}
\label{profiles}
\end{minipage}
\end{figure*}

\begin{figure*}
\begin{minipage}{175mm}
\caption{Background fit for NGC 2005. Models of the form of Eq. 
\ref{bgrfit} were fit to the profiles with annulus widths $3\arcsec$ (filled triangles) and $4\arcsec$ (filled circles) beyond $r=20\arcsec$. The two estimated parameters were $a\sim 4\arcsec$ and $\mu_{0}\sim 16.9$. The solid line shows the average of the two best-fitting models, used to determine the background level $V_{bg} = -2.5\log\phi = 21.83$, marked with an arrow. Errors shown are those determined from the light distribution (Section \ref{errors}) rather than the Poisson errors, which are necessarily calculated after the background subtraction but before the final fit (see text). Note the apparent under-estimation of the scatter at large radii.}
\label{backgroundfit}
\end{minipage}
\end{figure*}

\begin{figure*}
\begin{minipage}{175mm}
\caption{Background-subtracted F555W surface brightness profiles for each of the 53 clusters in the sample. The four different annulus widths are marked with different point types: $1\farcs5$ width are crosses, $2\arcsec$ width are open squares, $3\arcsec$ width are filled triangles, and $4\arcsec$ width are filled circles. Error bars marked with down-pointing arrows fall below the bottom of their plot. The solid lines show the best-fit EFF profiles. For each cluster the core radius $r_{c}$ is indicated and the best-fit parameters listed. When converting to parsecs, we assume an LMC distance modulus of 18.5.}
\label{plots}
\end{minipage}
\end{figure*}

\begin{figure*}
\begin{minipage}{175mm}
\caption{Measured values of $\gamma$ plotted against those determined by EFF87 from measurements extending to $r\sim 250\arcsec$ for ten young clusters. Errors in the $y$-direction represent the random errors determined as described in Section \ref{errors} and do not include any estimation of the systematic errors which may be introduced by our background fitting algorithm. The dashed line is plotted for reference and indicates equality between the two sets of measurements.}
\label{gammacomp}
\end{minipage}
\end{figure*}

\begin{figure*}
\begin{minipage}{175mm}
\caption{Measured values of $r_{c}$ plotted against those determined from the following studies: EFF87 (filled triangles); Mateo (1987) (filled squares); Elson (1991) (not seeing corrected) (filled circles). The cross marks the value measured for NGC 1856 by Elson (1992). Errors in the $x$-direction are those quoted by the various authors, and in the $y$-direction are those determined as described in Section \ref{errors}. The dashed line is plotted for reference and indicates equality between the measurements. Radii are in arcseconds to account for the different distance moduli adopted in each study.}
\label{prevstudplot}
\end{minipage}
\end{figure*}

\begin{figure*}
\begin{minipage}{175mm}
\caption{Images of the six clusters described in the text of Section \ref{binarycl}. {\em Top left}: Our WFC2 image of the centre of NGC 1850 (lower left) and NGC 1850B (centre). The image is approximately $70\arcsec$ (730 WFC pixels) on a side. {\em Top centre}: NGC 2153. {\em Top right}: NGC 2213. {\em Bottom row, left to right}: NGC 2004, NGC 2155, and NGC 2159. Each of the latter five is an $R$-band DSS2 image, approximately $195\arcsec$ on a side.}
\label{doubleclus}
\end{minipage}
\end{figure*}

\begin{figure*}
\begin{minipage}{175mm}
\caption{Central surface brightness profiles and power-law fits for the seven PCC cluster candidates in the present sample. On the left are shown the four clusters with complete profiles; on the right those three with incomplete profiles, but which seem to exhibit breaks. The four different point styles represent the four annulus sets, as in Fig. \ref{plots}. The best fitting power-law models are shown, and the radii of the breaks to the power-law regions are indicated by arrows, with errors indicated. The slopes $\beta$ of the power-law models are listed. Note that because of the magnitude scale the slopes of the models as plotted are $2.5\beta$.}
\label{corecollapse}
\end{minipage}
\end{figure*}

\begin{figure*}
\begin{minipage}{175mm}
\caption{Two-component fits for the complete surface brightness profile of R136/NGC 2070. The four different point styles represent the four annulus sets, as in Fig. \ref{plots}. We show two different two-component profiles. The solid line represents an EFF profile for the core (R136) and a King profile for the outer region, while the dashed line represents an EFF profile in the core and an EFF profile in the outer region. Best-fit parameters for each of the two models are as discussed in the text. Note that the two only differ significantly at large radii.}
\label{R136bits}
\end{minipage}
\end{figure*}

\begin{figure*}
\begin{minipage}{175mm}
\caption{Power-law fit to the inner core of R136. The four different point styles represent the four annulus sets, as in Fig. \ref{plots}. The best-fit EFF and power-law profiles are shown as is the power-law slope ($\beta$) and break radius (arrow), just as for the PCC candidate clusters in Fig. \ref{corecollapse}. Again, because of the magnitude scale, the slopes of the power-law model as plotted is $2.5\beta$.}
\label{R136core}
\end{minipage}
\end{figure*}

\begin{figure*}
\begin{minipage}{175mm}
\caption{Mass to light ratios predicted from the single-burst stellar population models of Fioc \& Rocca-Volmerange (1997) (PEGASE v2.0, 1999), using the IMF of Kroupa, Tout \& Gilmore (1993) over the mass range $0.1$ to $120 M_{\odot}$, and calculated for the four available abundances which cover the range spanned by the cluster sample: [Fe/H] $= -2.25$ (long dashes); [Fe/H] $= -1.65$ (solid line); [Fe/H] $= -0.64$ (short dashes); and [Fe/H] $= -0.33$ (dotted line).}
\label{pegaseml}
\end{minipage}
\end{figure*}

\begin{figure*}
\begin{minipage}{175mm}
\caption{Core radius vs. age for all clusters in the sample. Ages are the literature estimates from Table \ref{ages}, and the core radii as listed in Table \ref{params}. The points marking clusters with radii smaller than $\sim 1$ pc should be considered upper limits, as discussed in Section \ref{results}.}
\label{coreage}
\end{minipage}
\end{figure*}

\begin{figure*}
\begin{minipage}{175mm}
\caption{Asymptotic mass vs. age for all clusters in the sample. Again, ages are the literature estimates from Table \ref{ages}, while the masses are from Table \ref{luminmass}.}
\label{agemass}
\end{minipage}
\end{figure*}

\begin{figure*}
\begin{minipage}{175mm}
\caption{Asymptotic mass vs. core radius for all clusters in the sample. Core radii are from Table \ref{params}, and masses from Table \ref{luminmass}.}
\label{coremass}
\end{minipage}
\end{figure*}

\begin{figure*}
\begin{minipage}{175mm}
\end{minipage}
\end{figure*}



\bsp 

\label{lastpage}

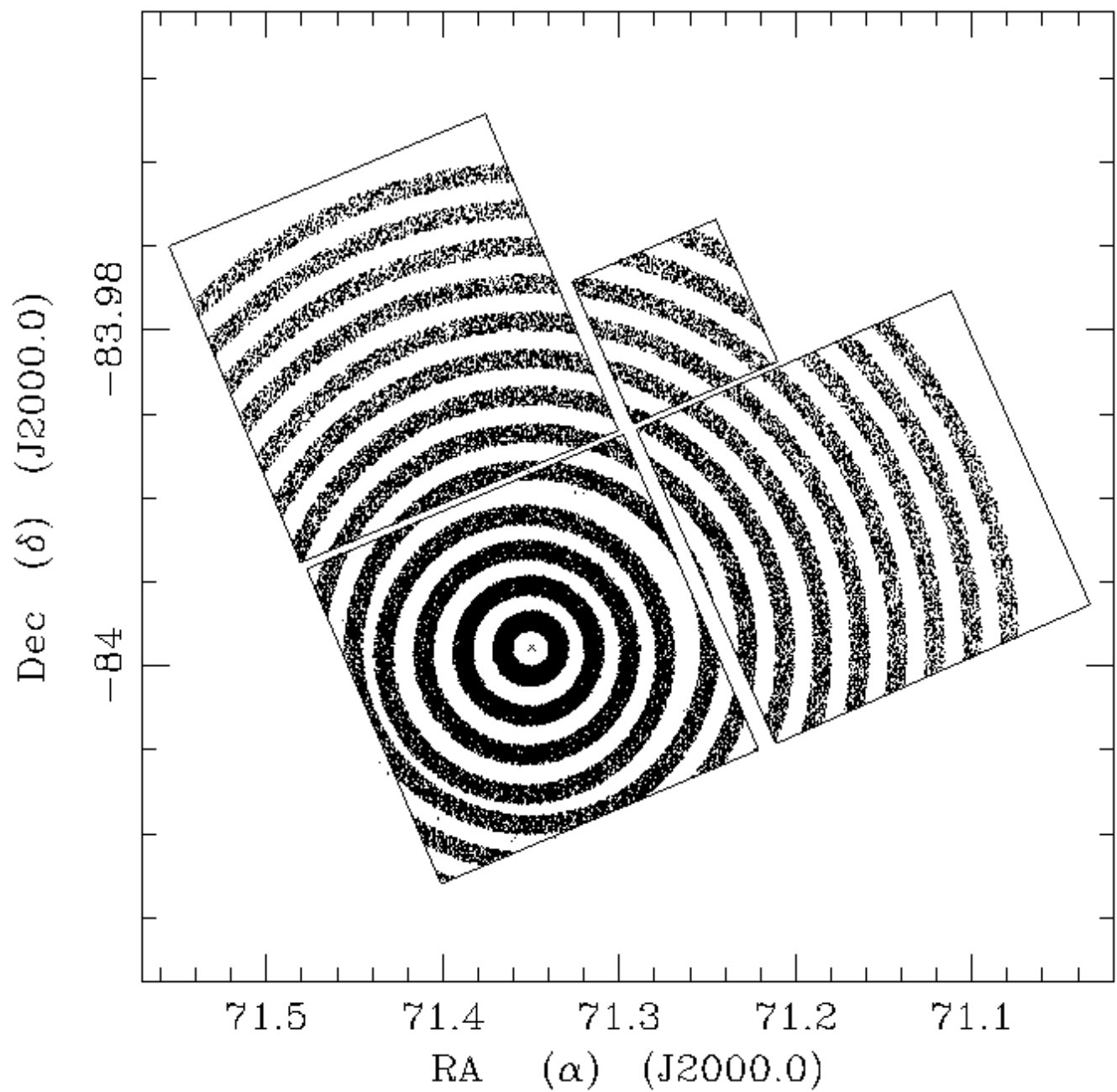
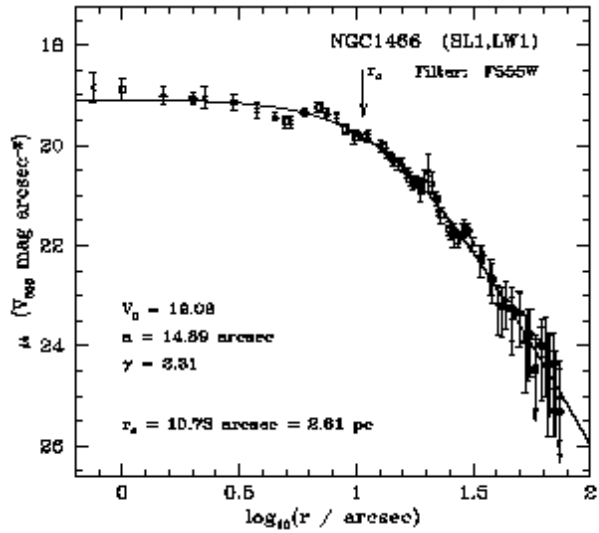
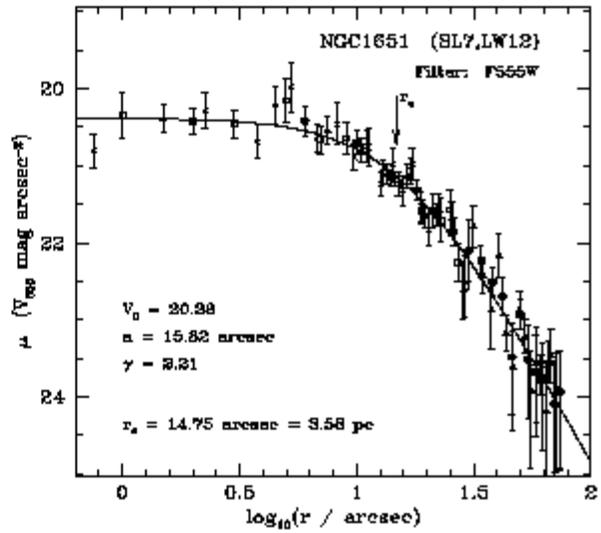
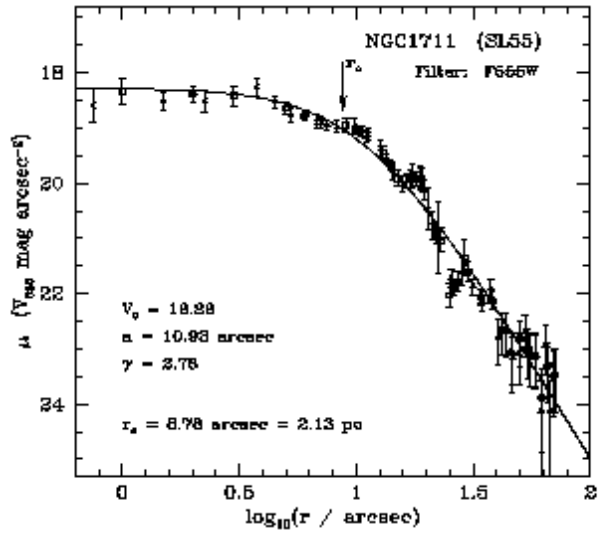
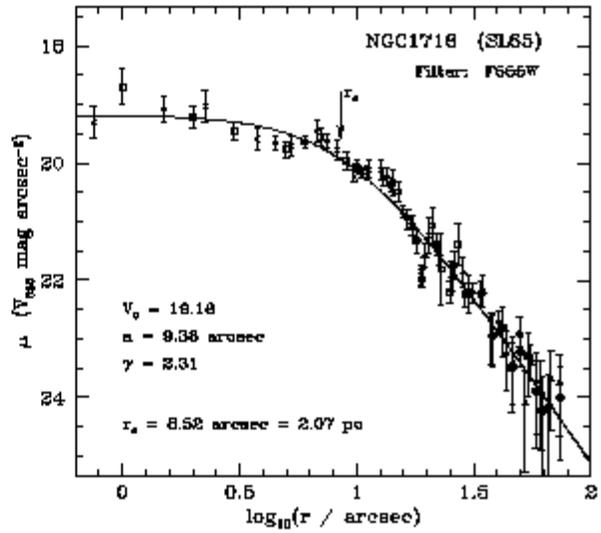
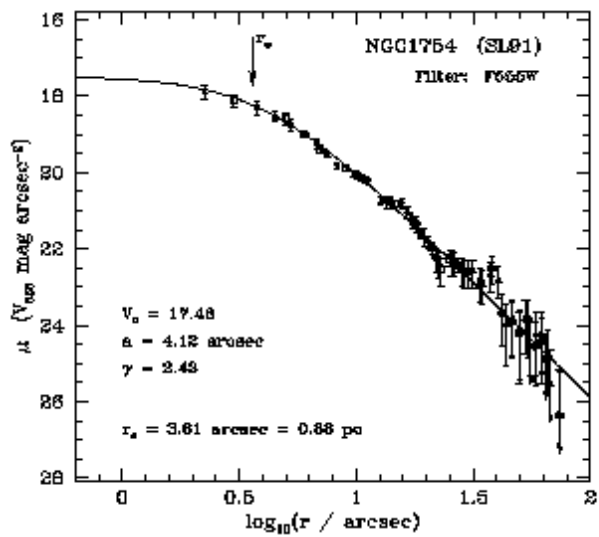
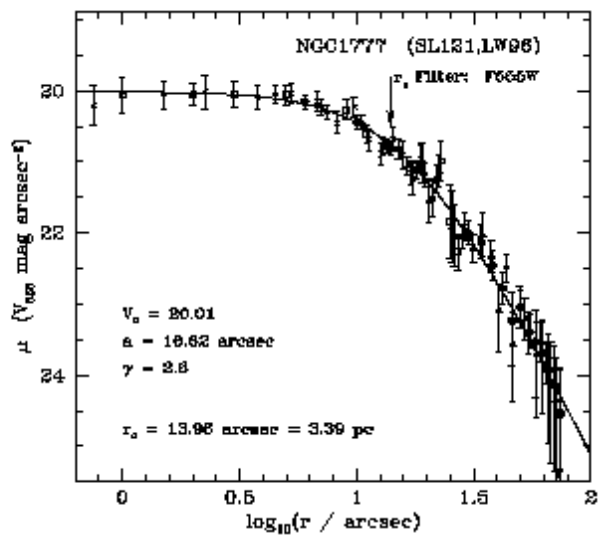
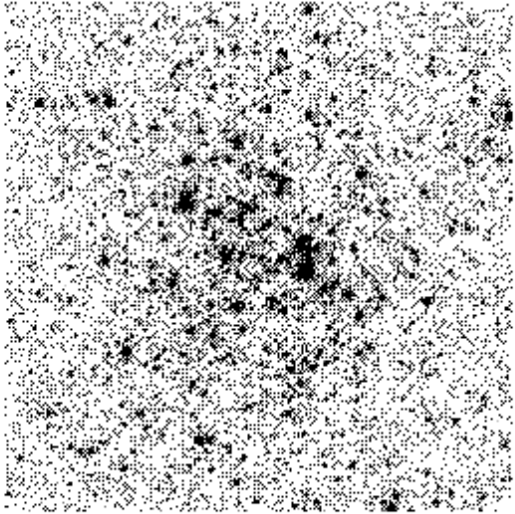
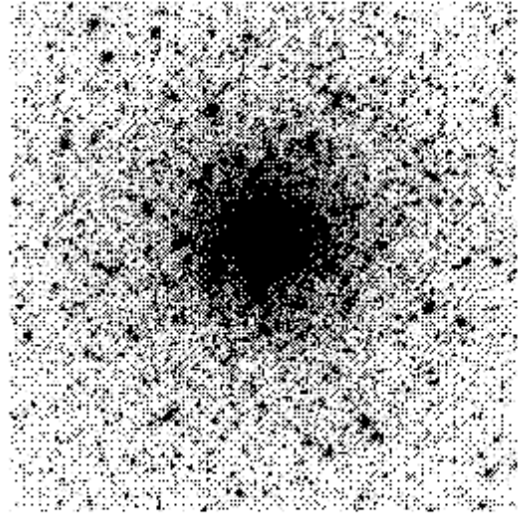
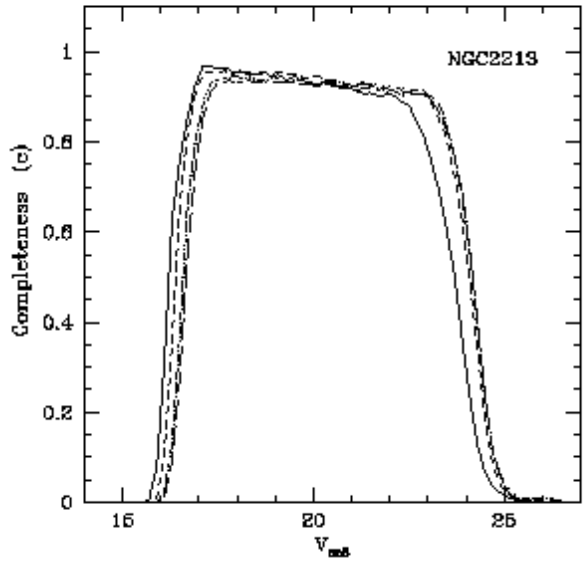
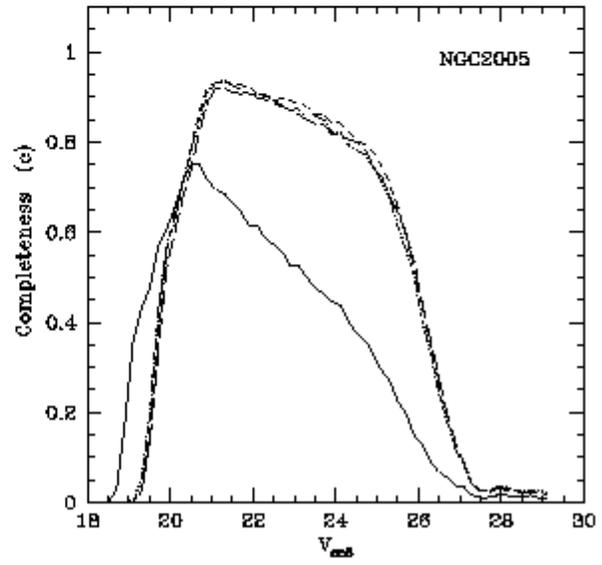
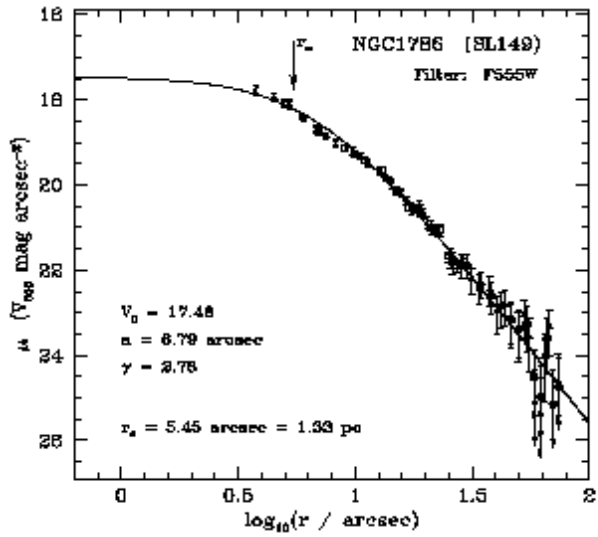
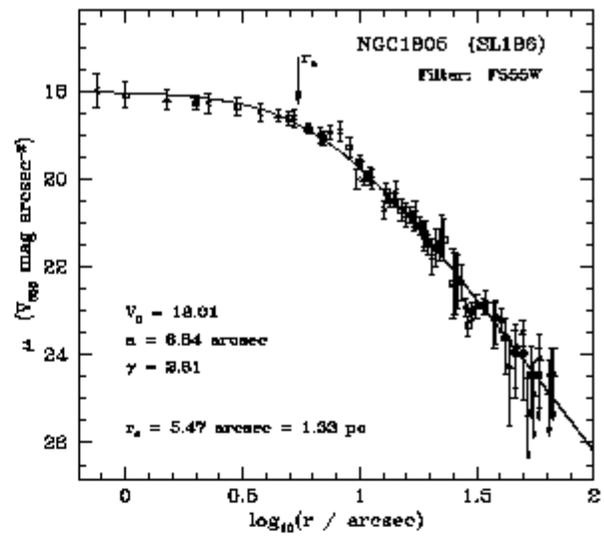
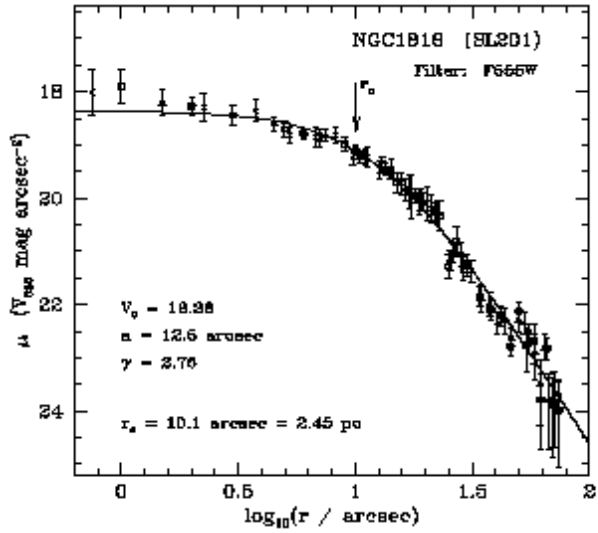
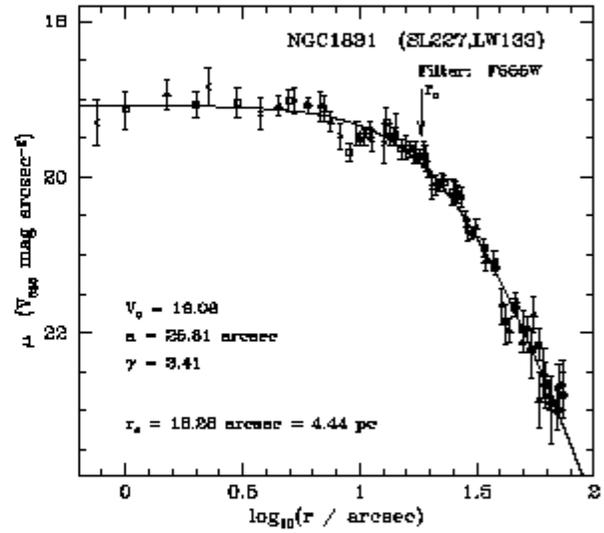
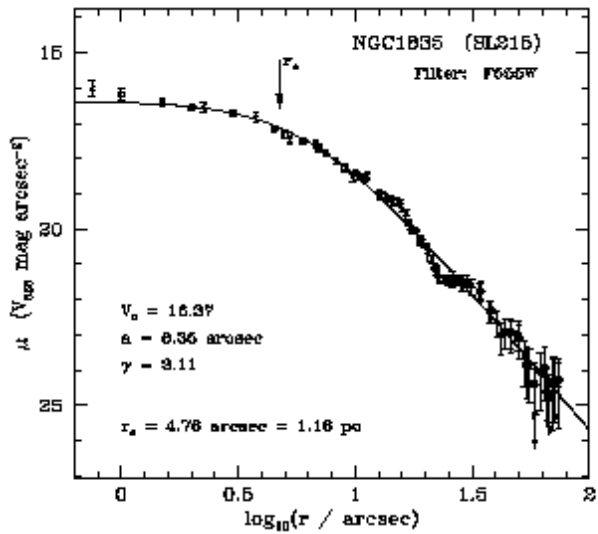
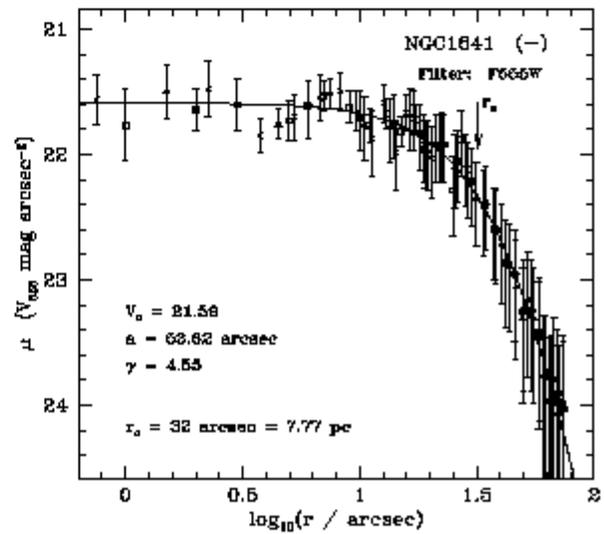
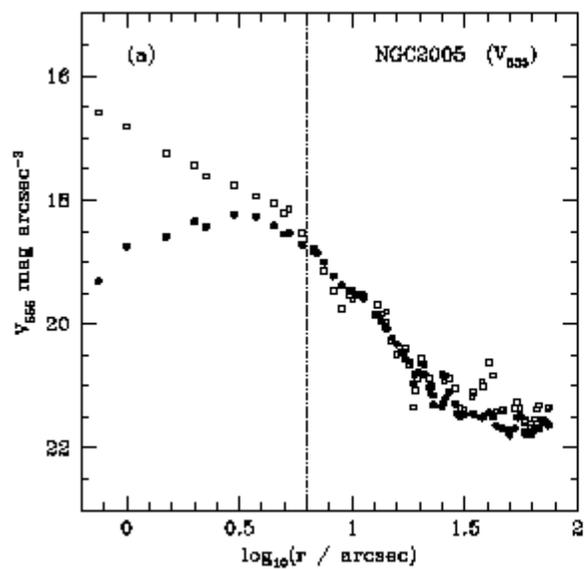
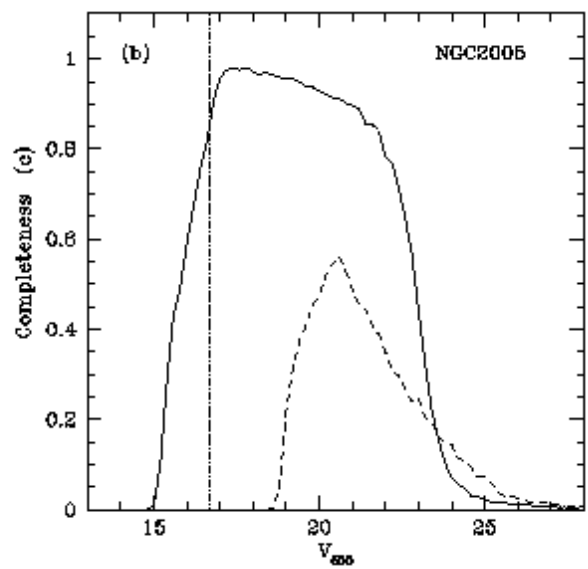
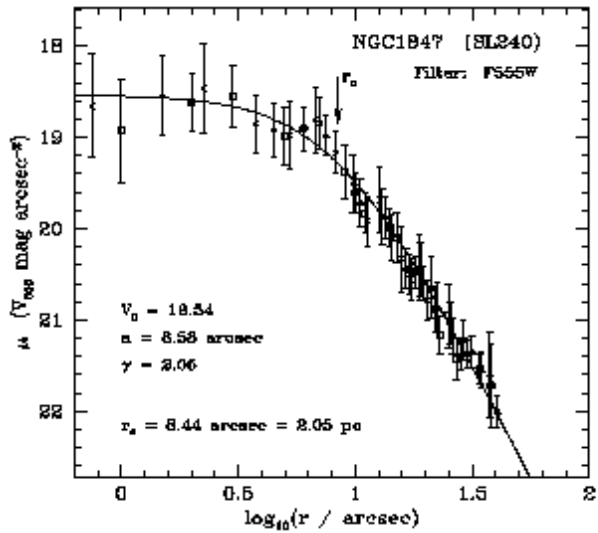
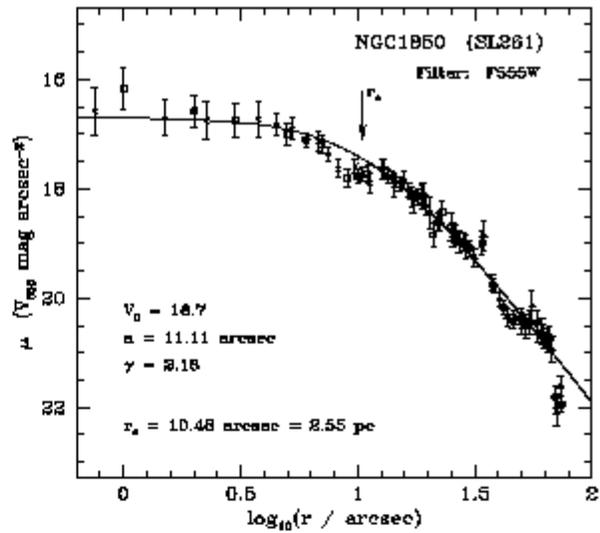
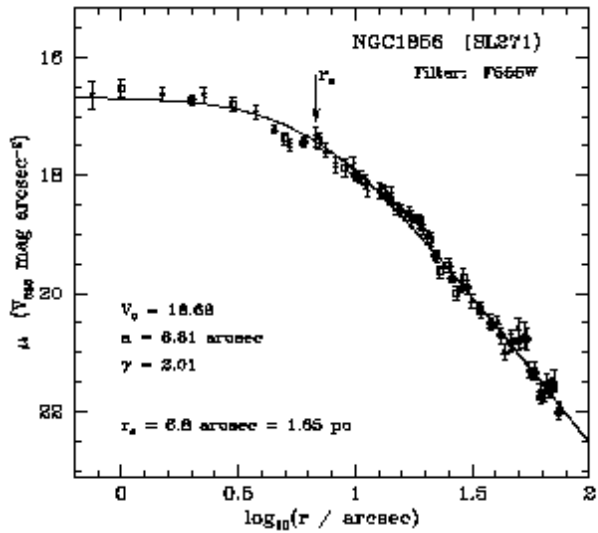
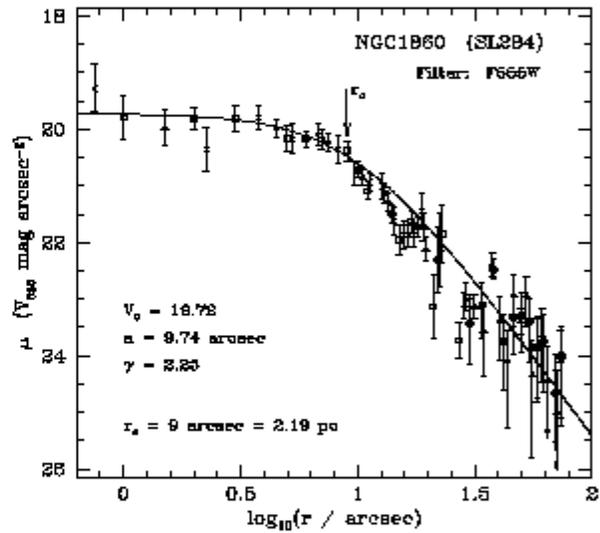
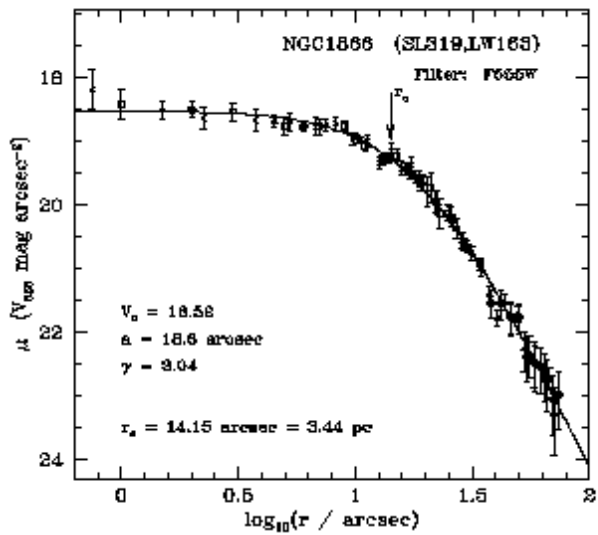
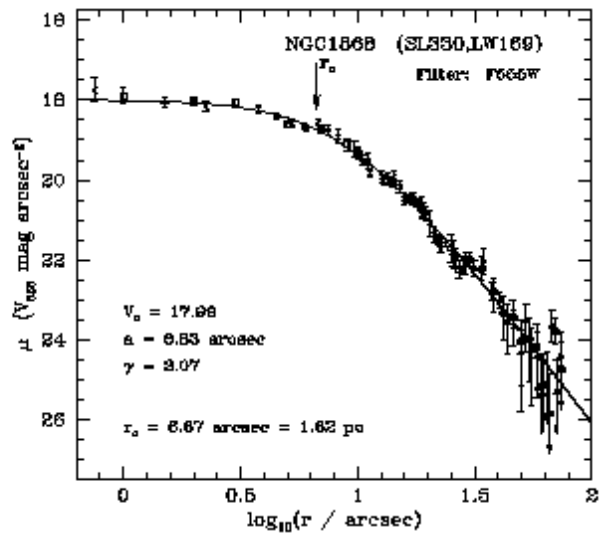
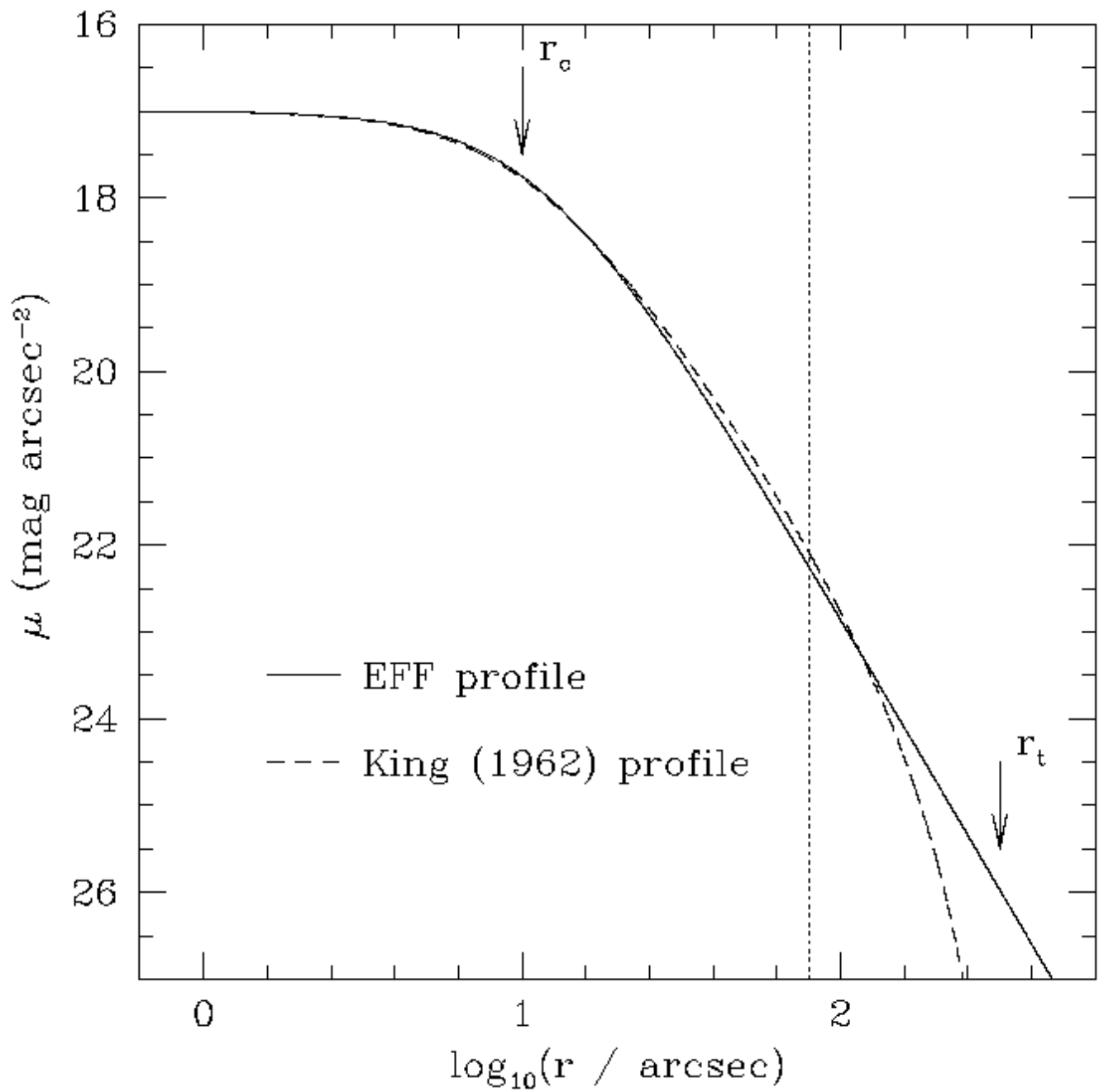
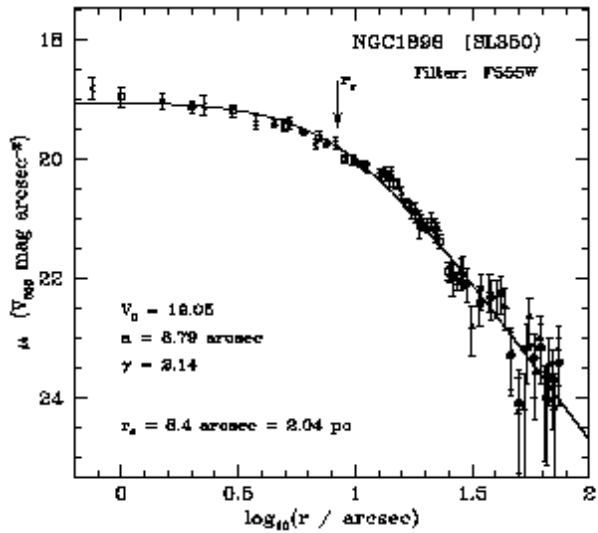
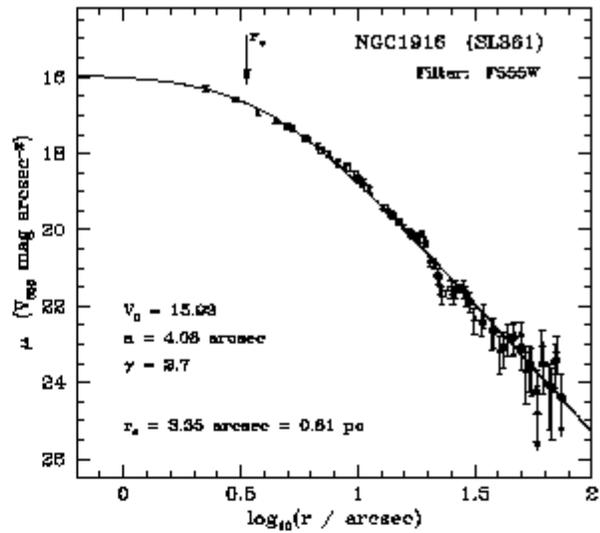
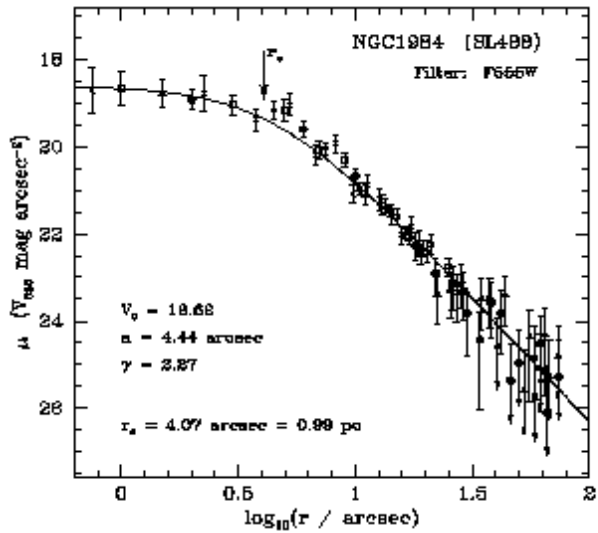
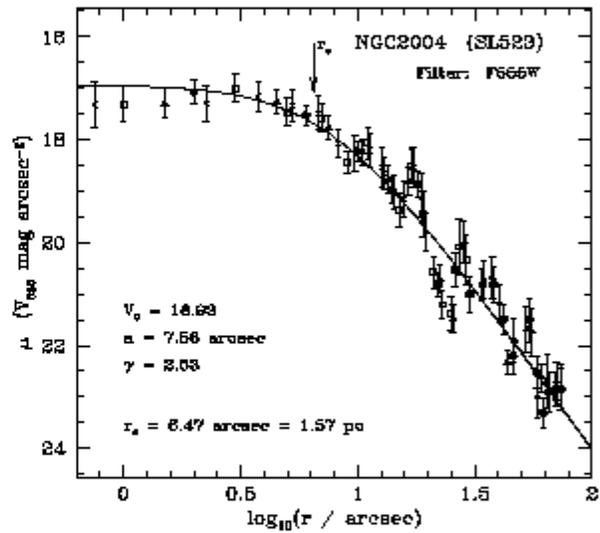
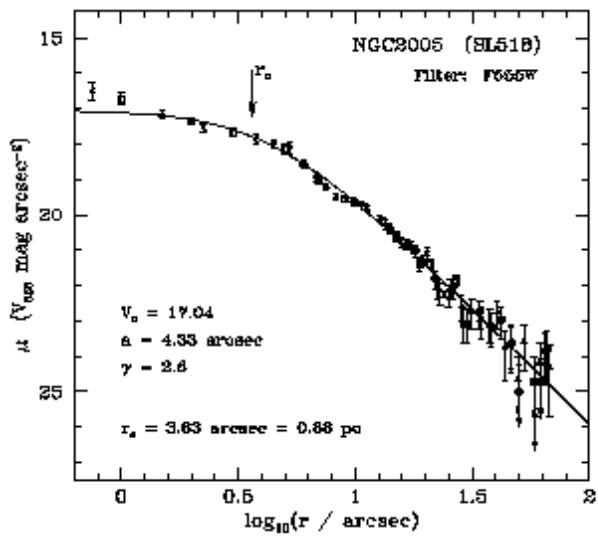
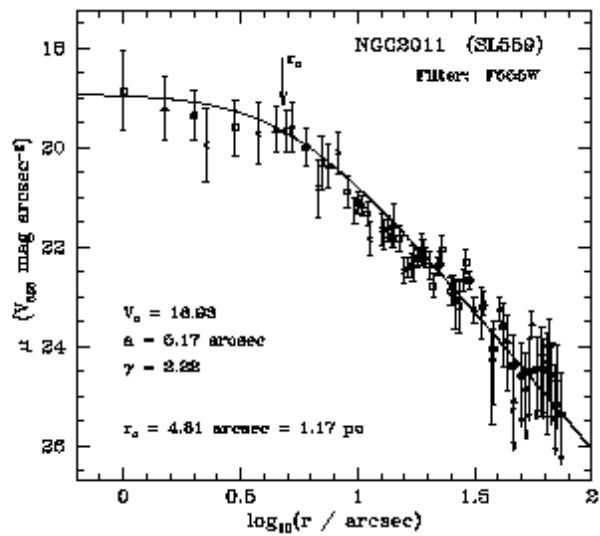
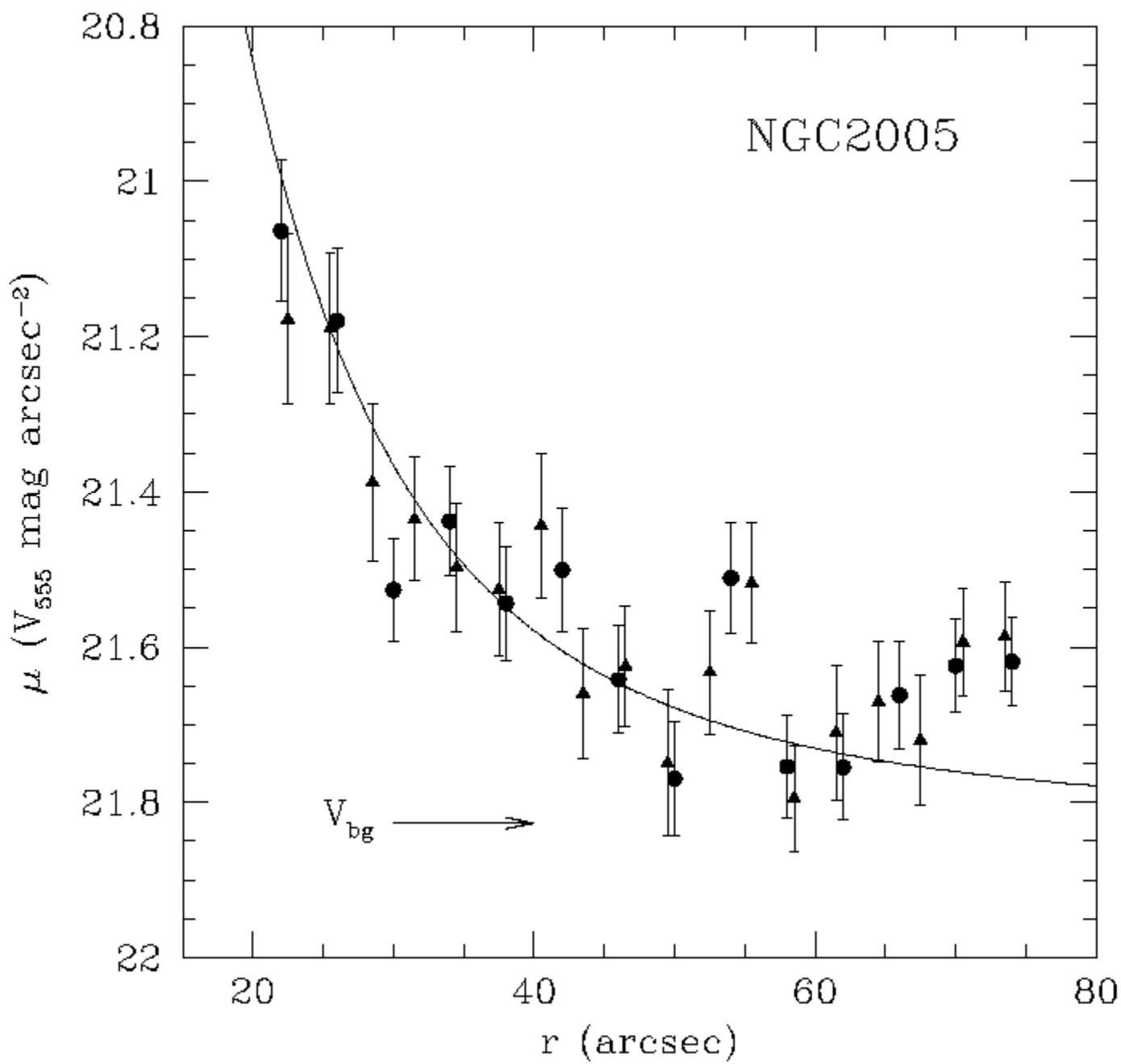
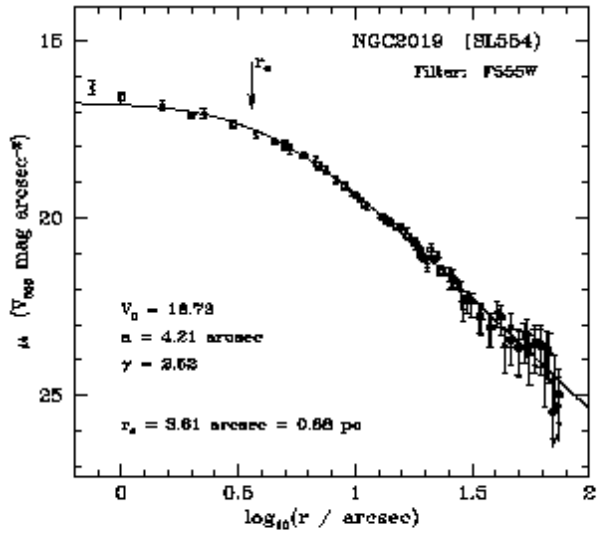
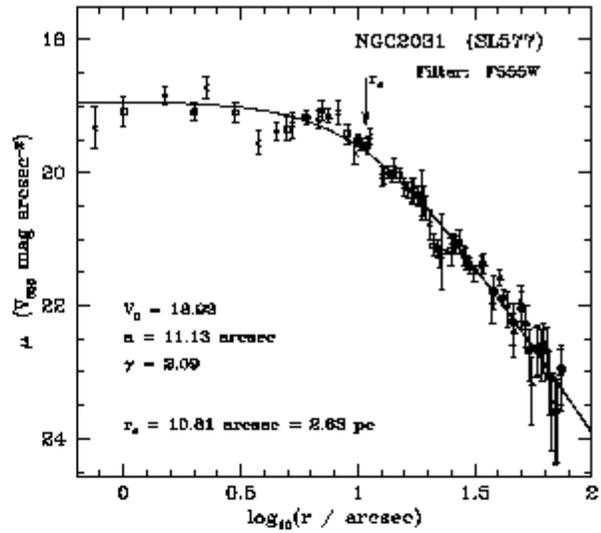
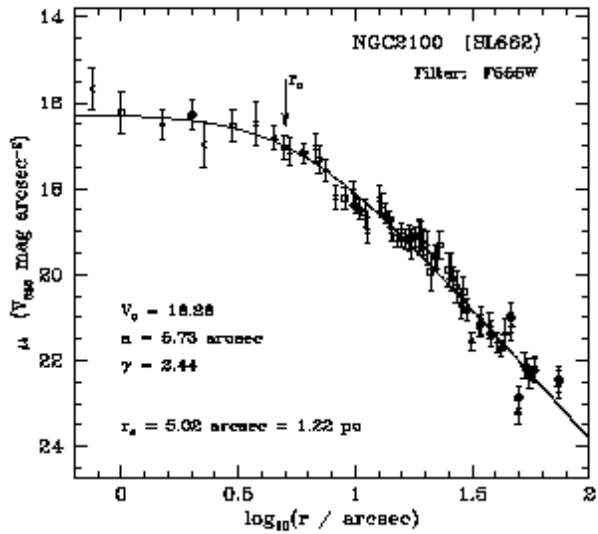
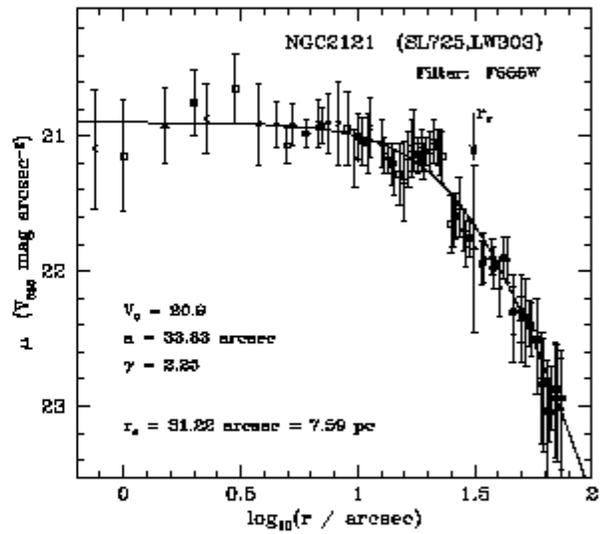
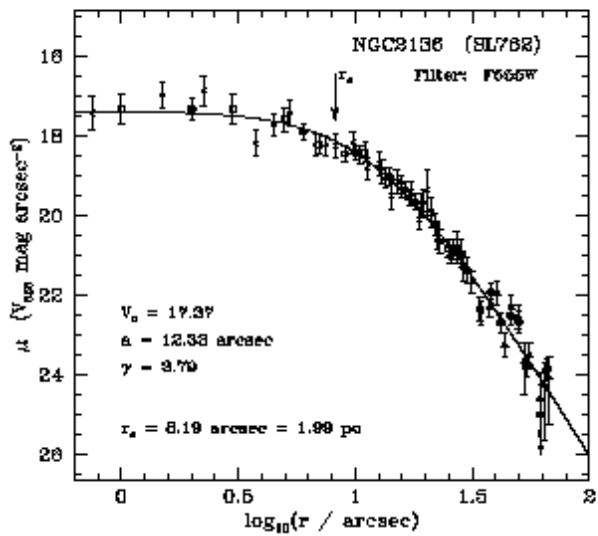
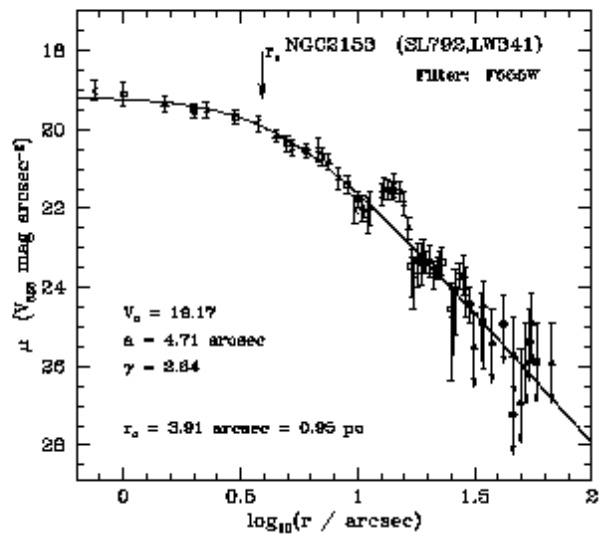
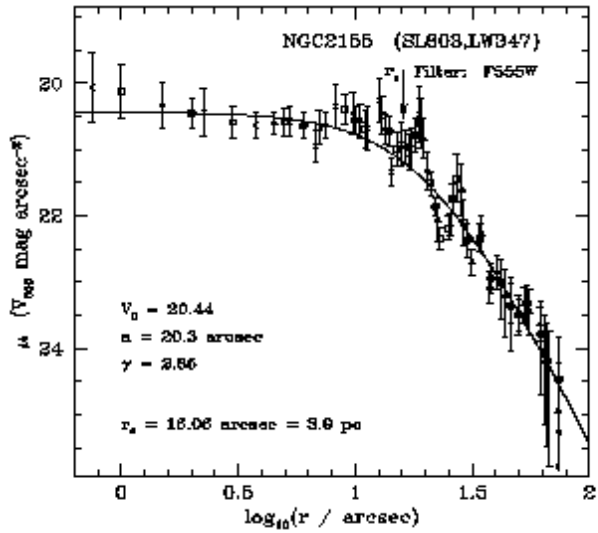
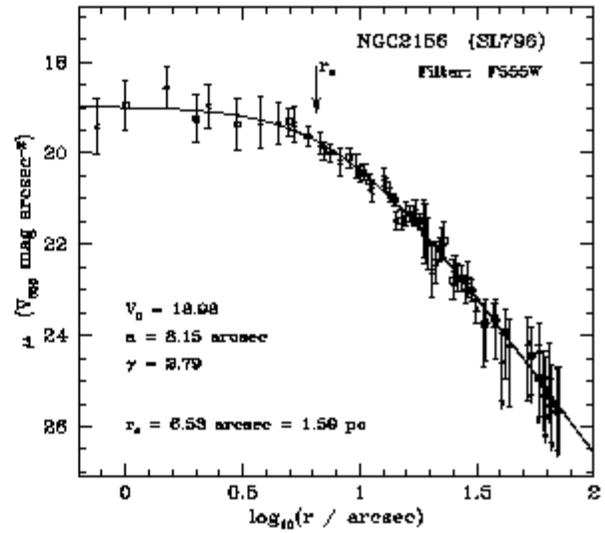
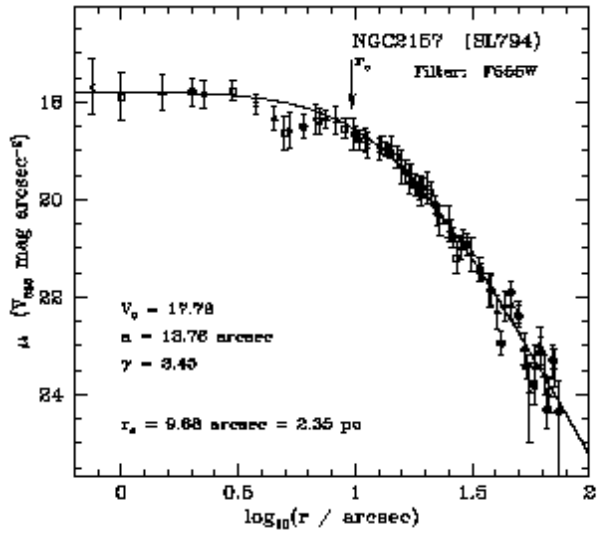
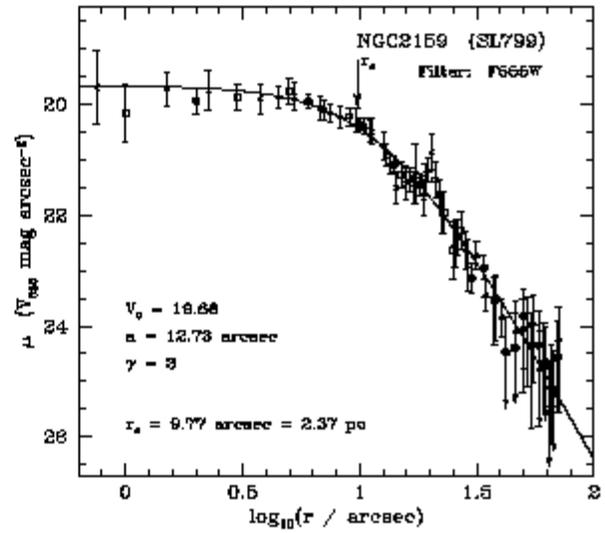
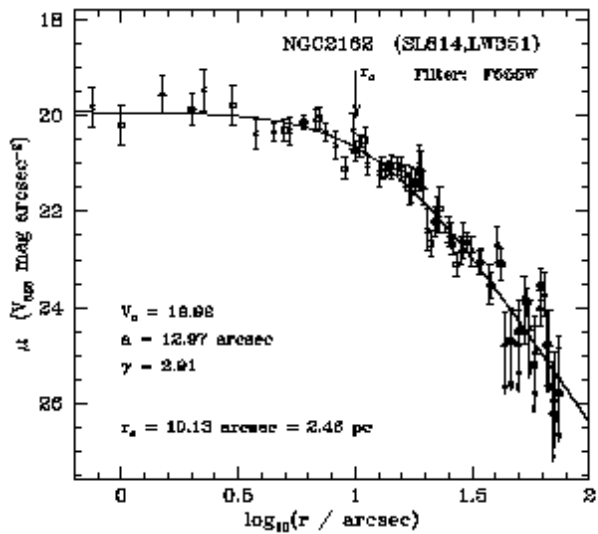
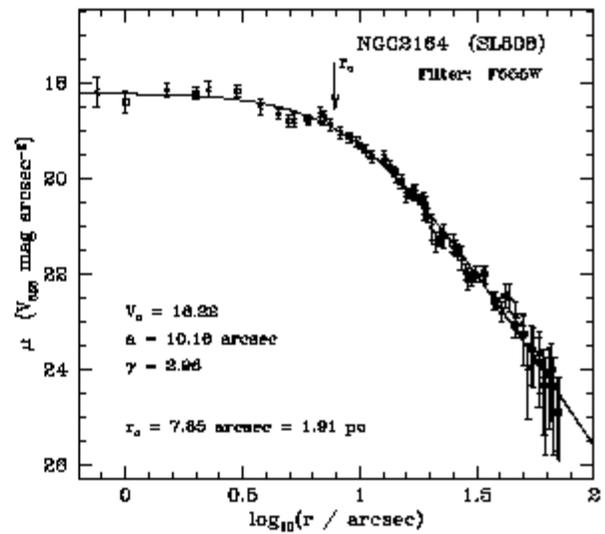
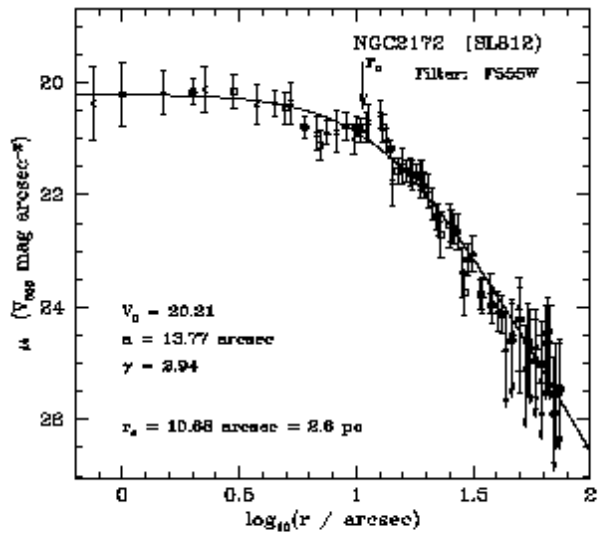
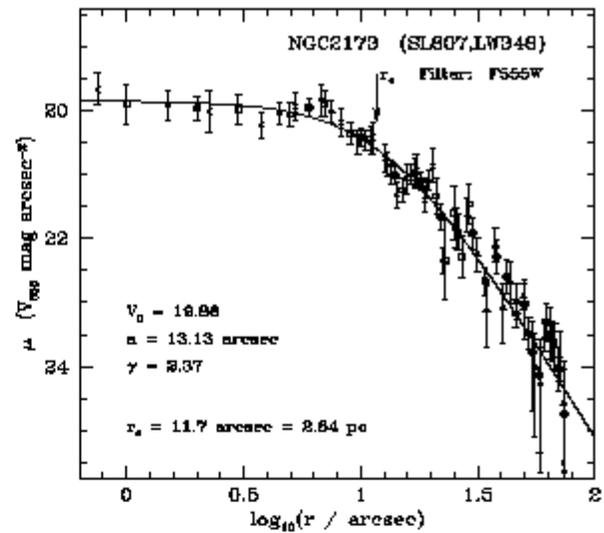
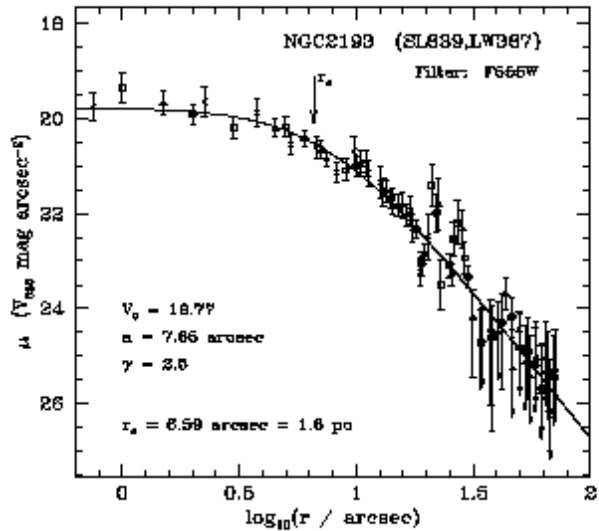
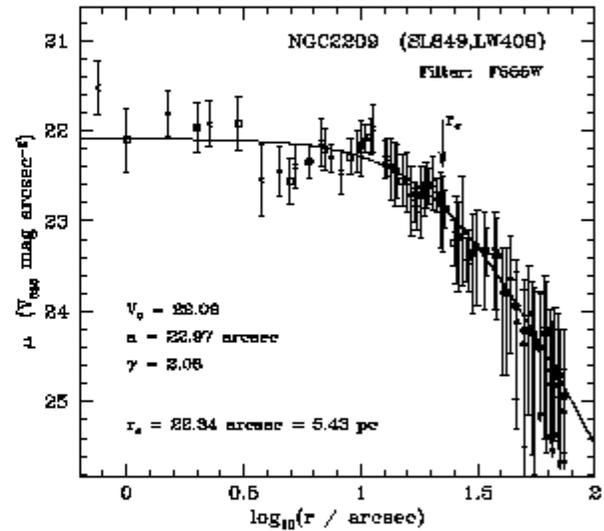
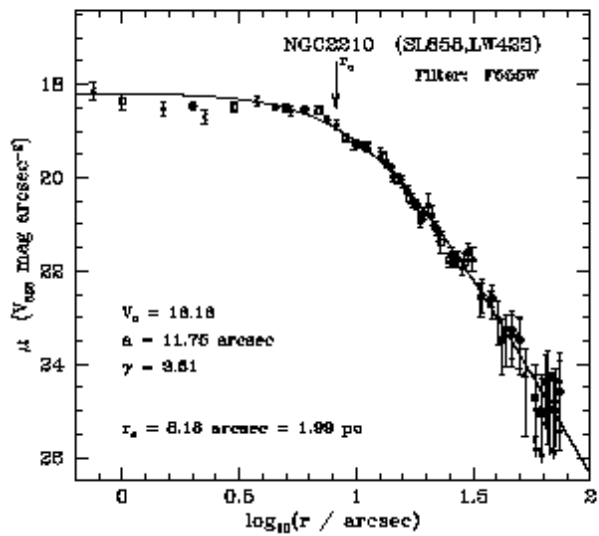
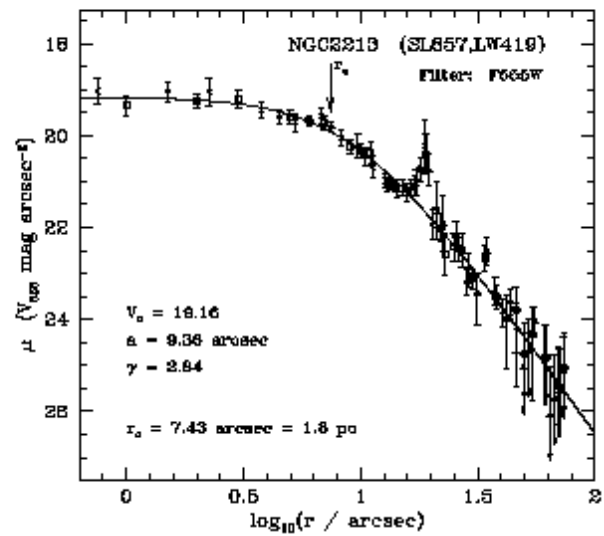
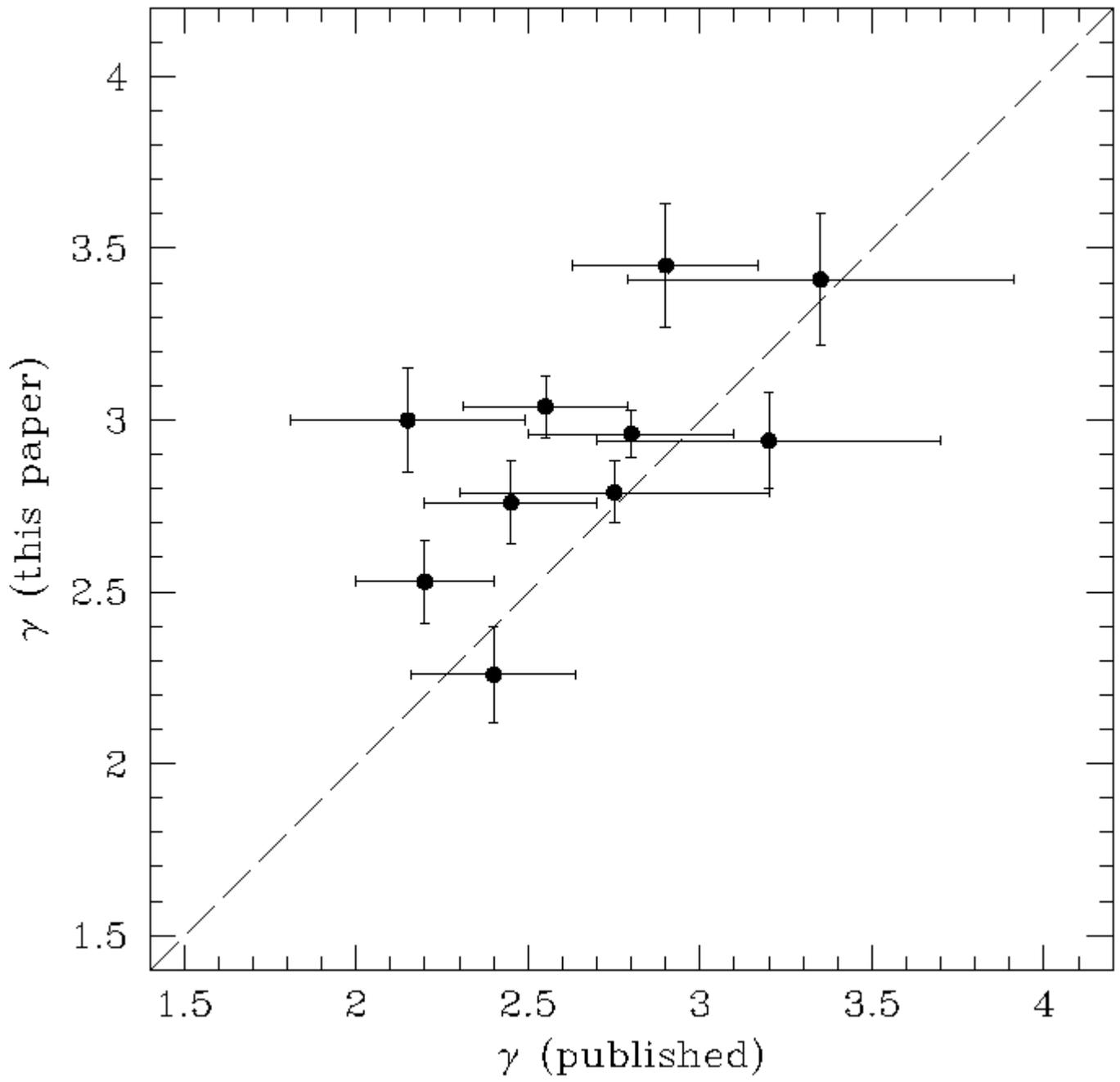
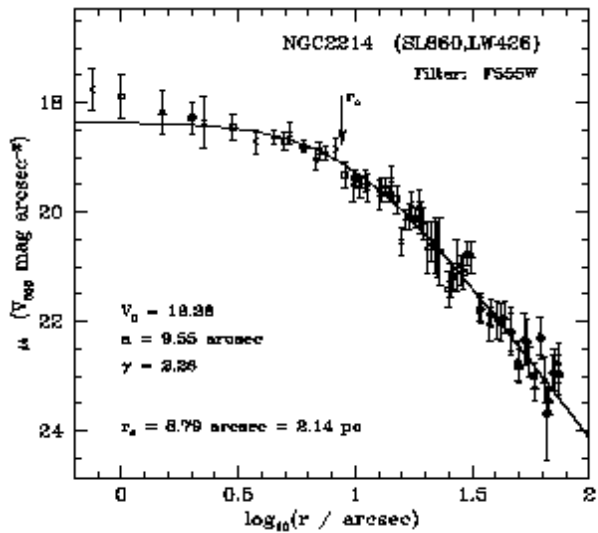
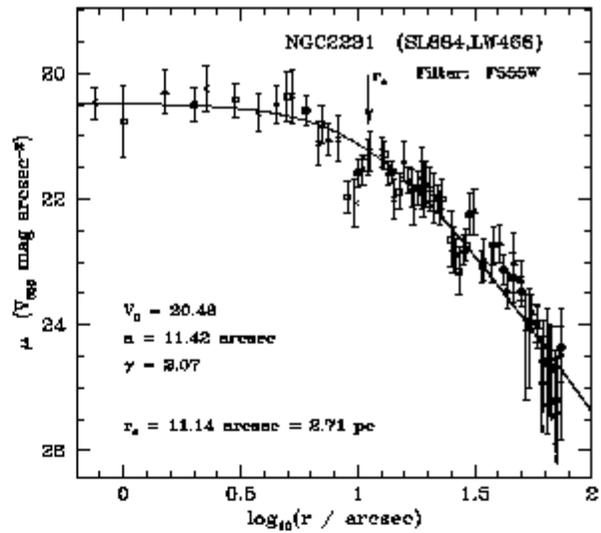
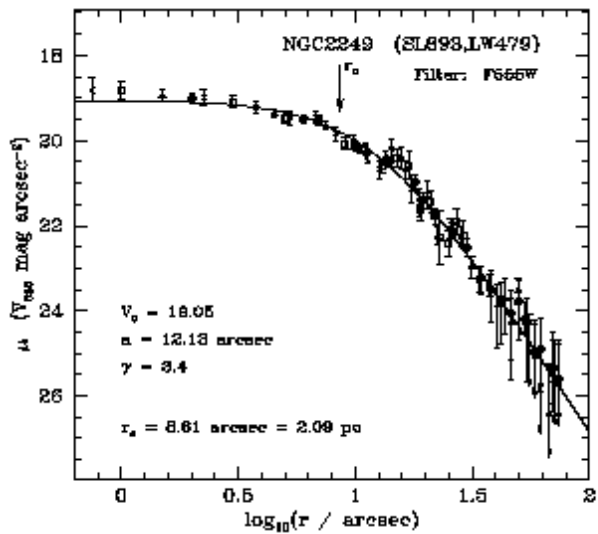
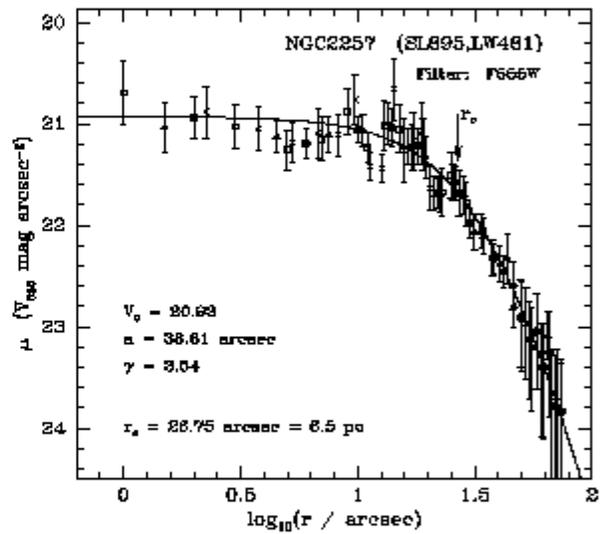
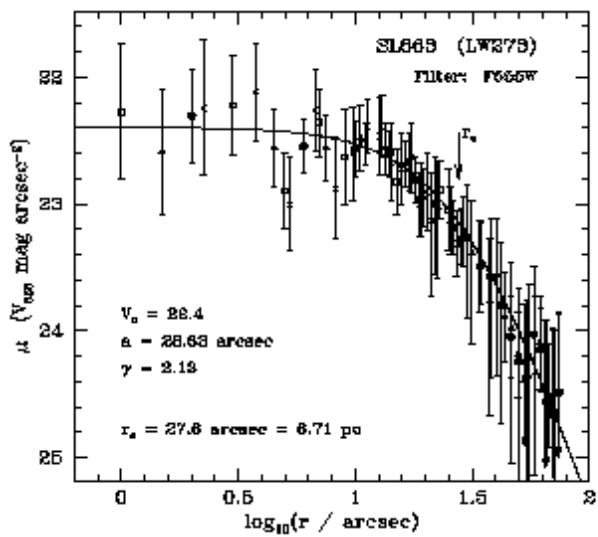
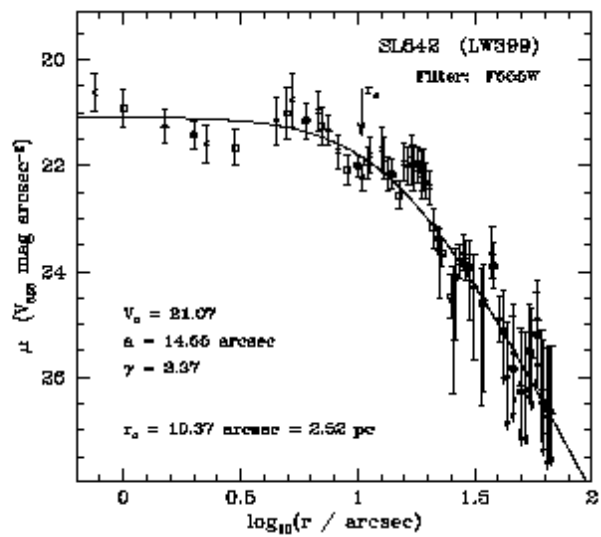
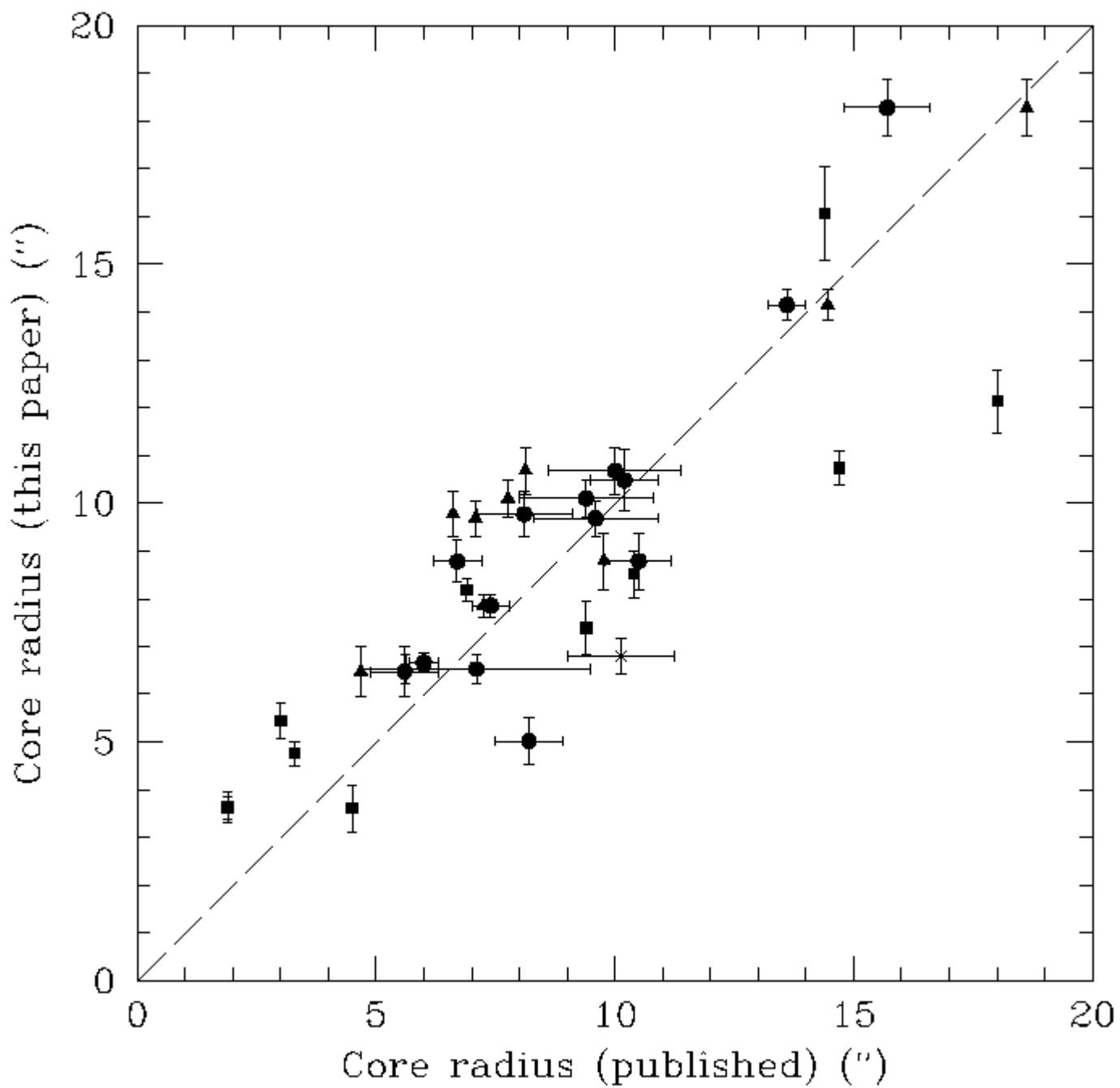
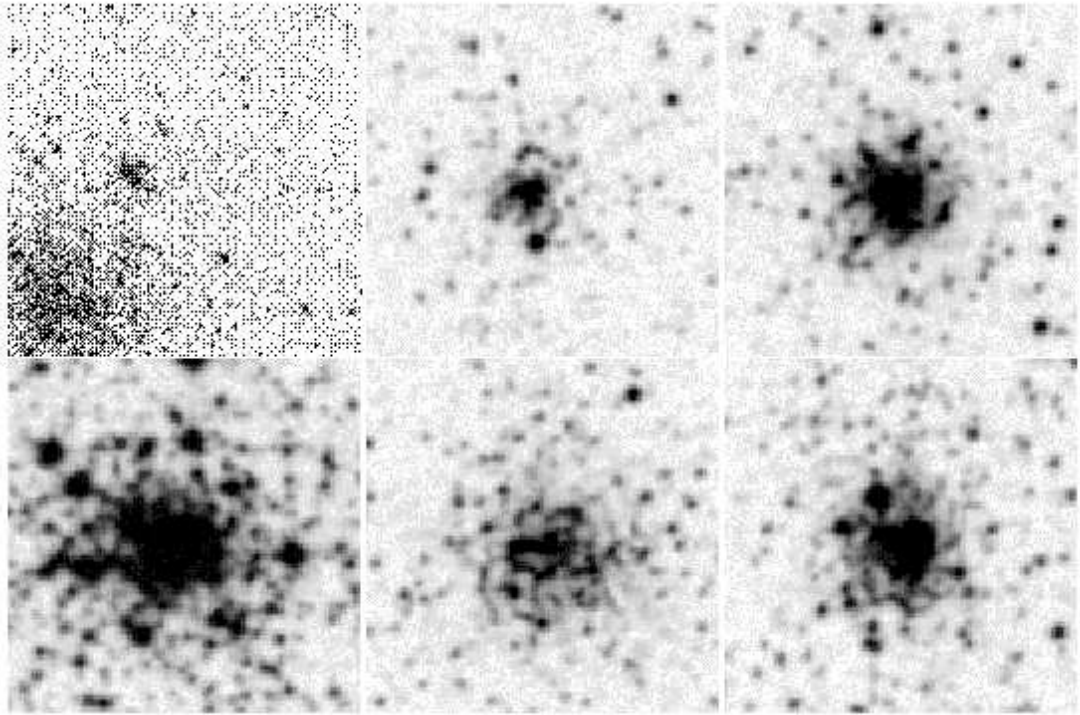
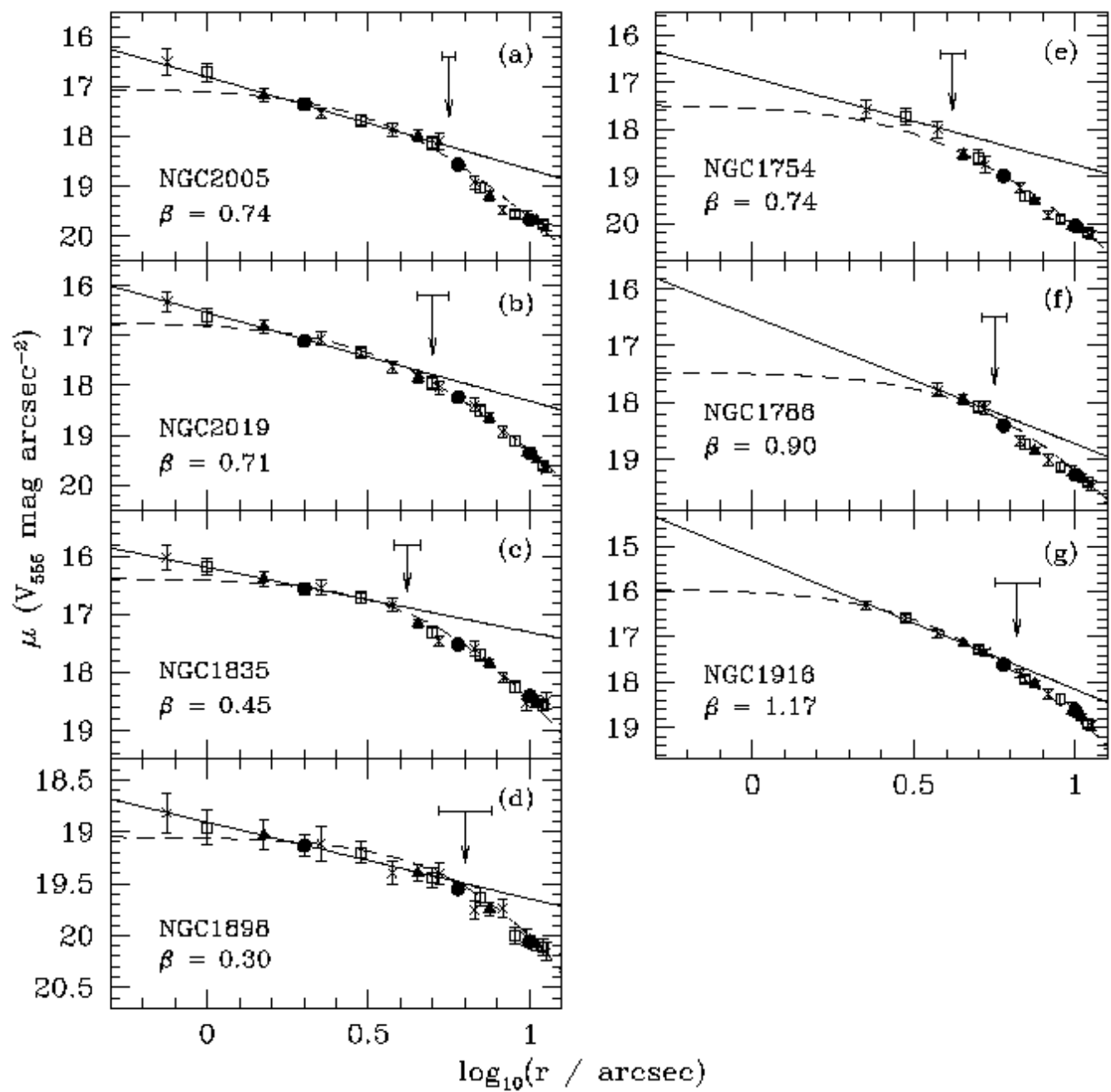
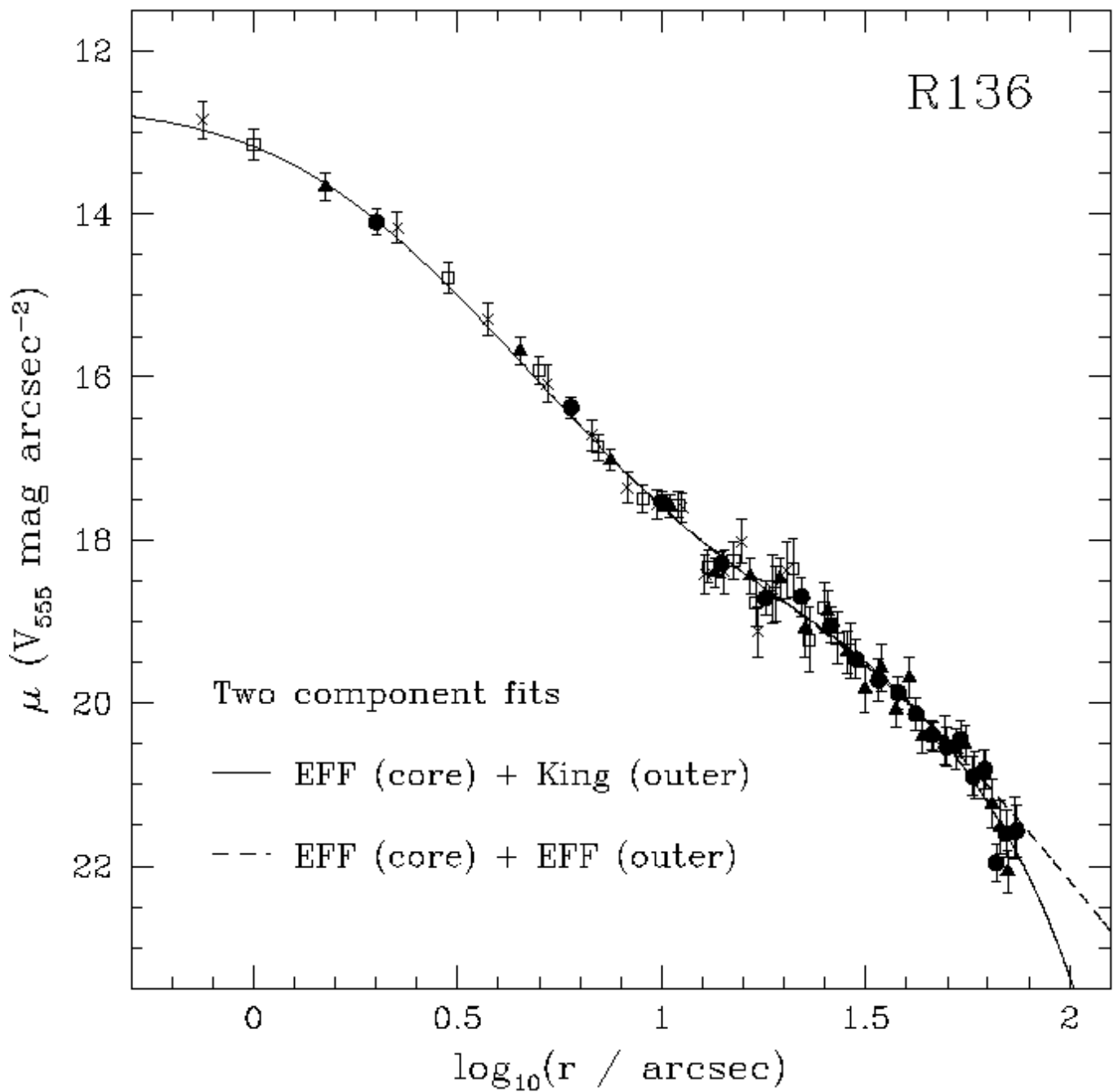
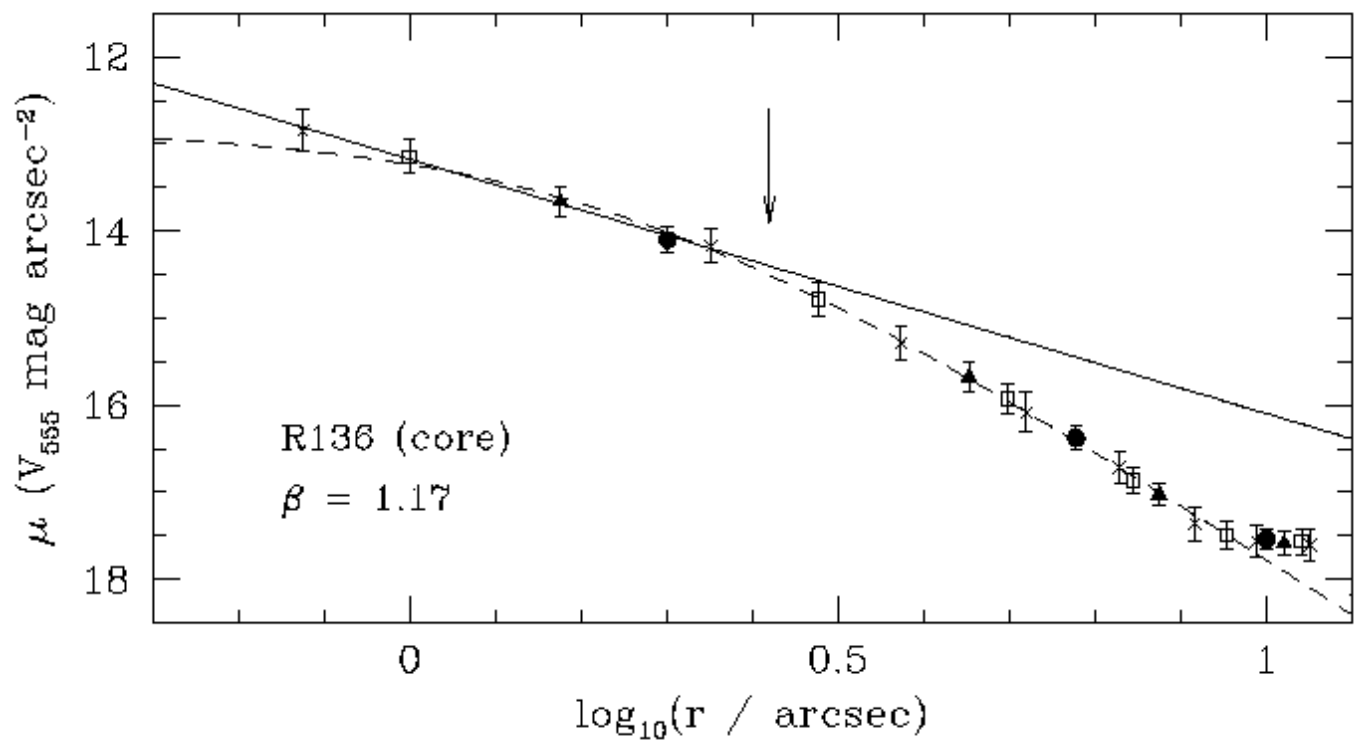
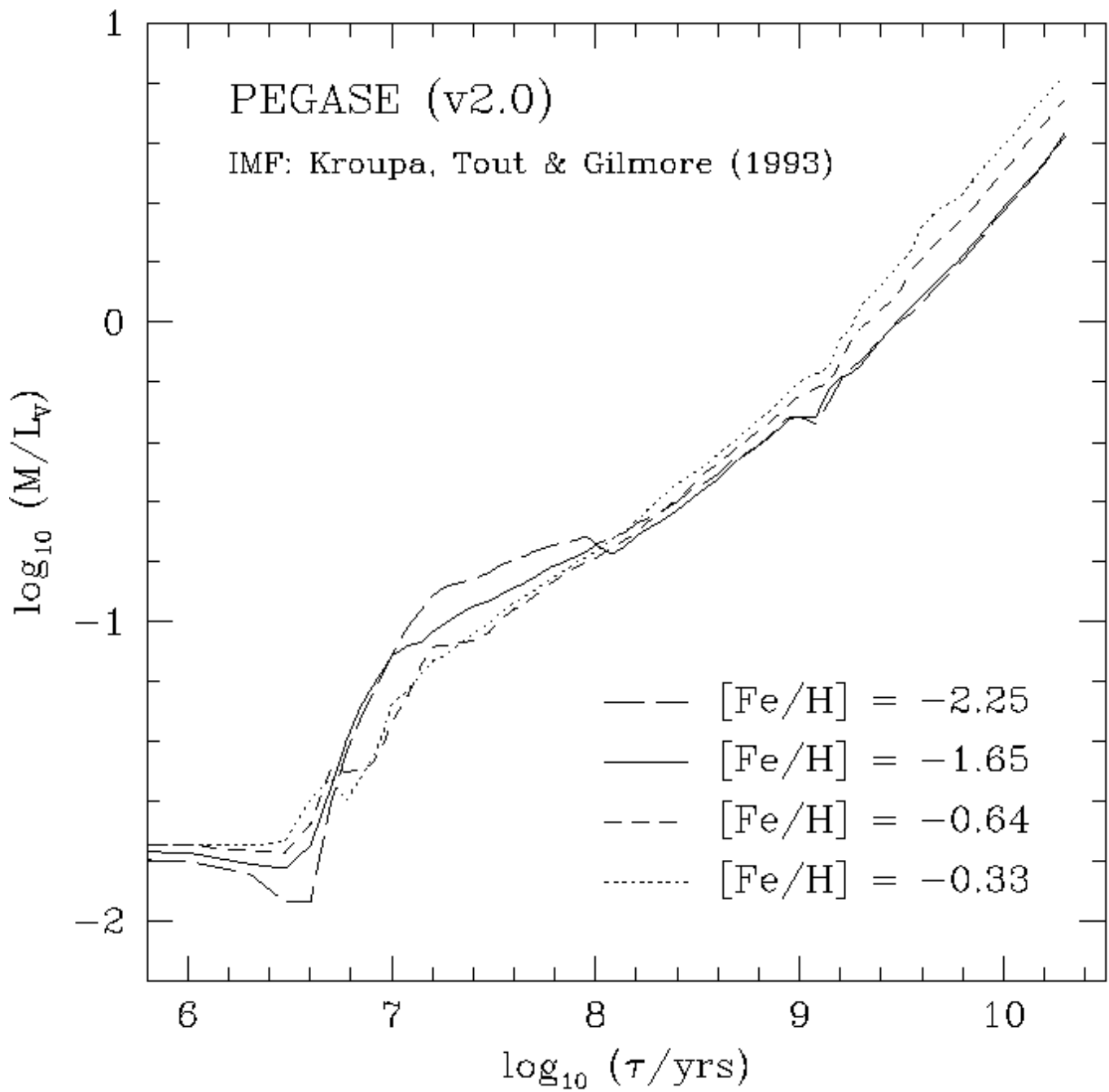
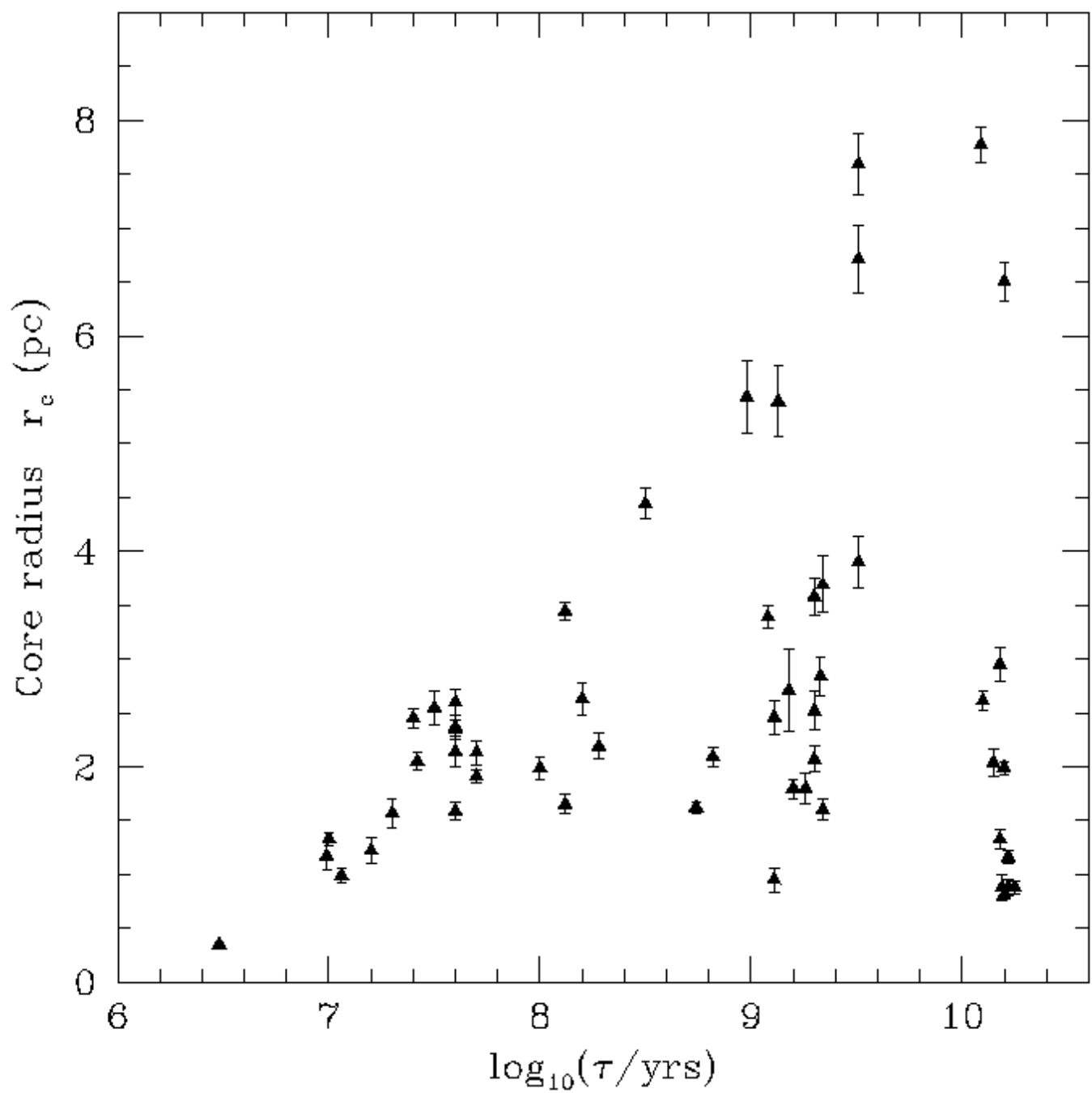
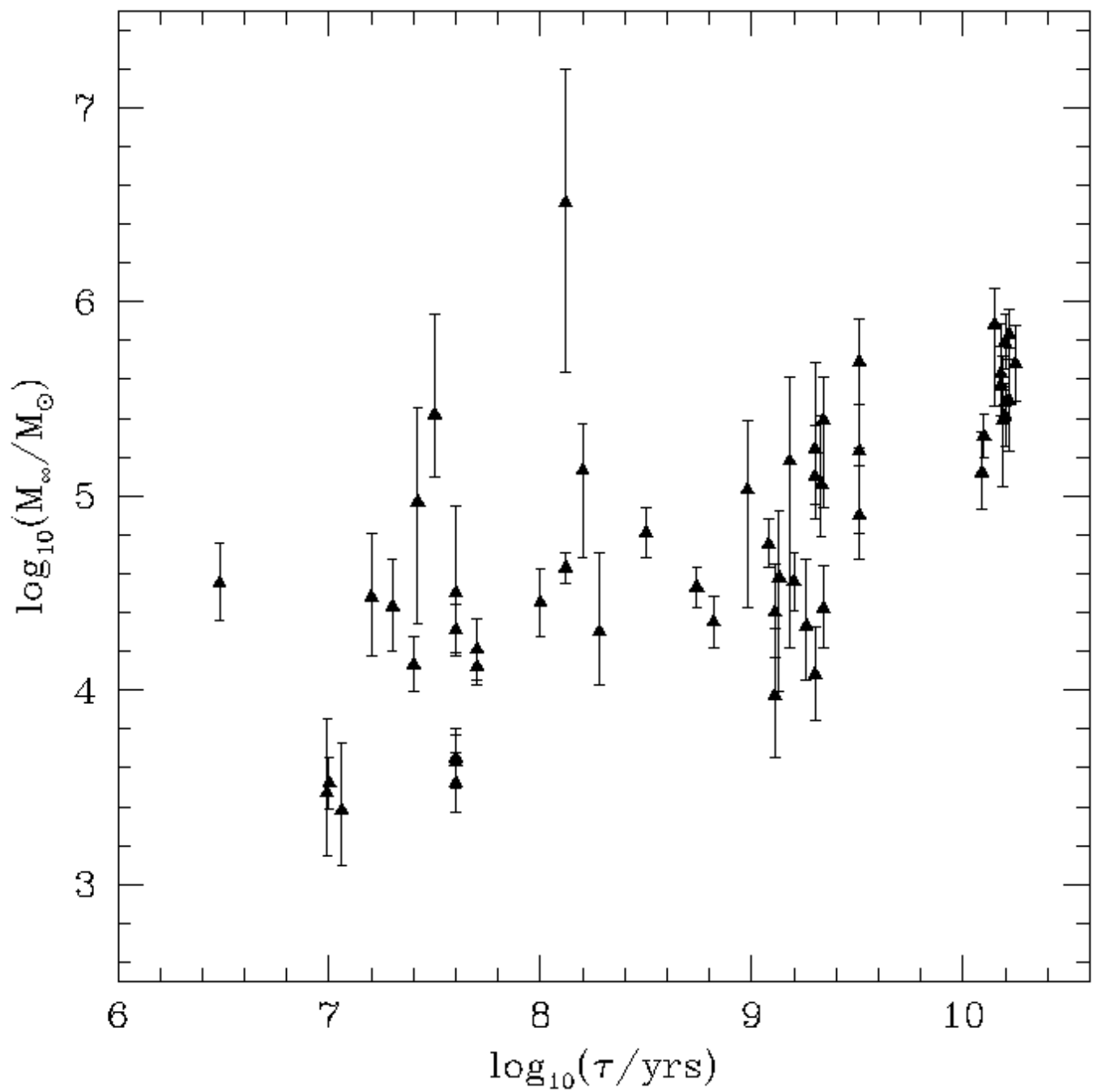
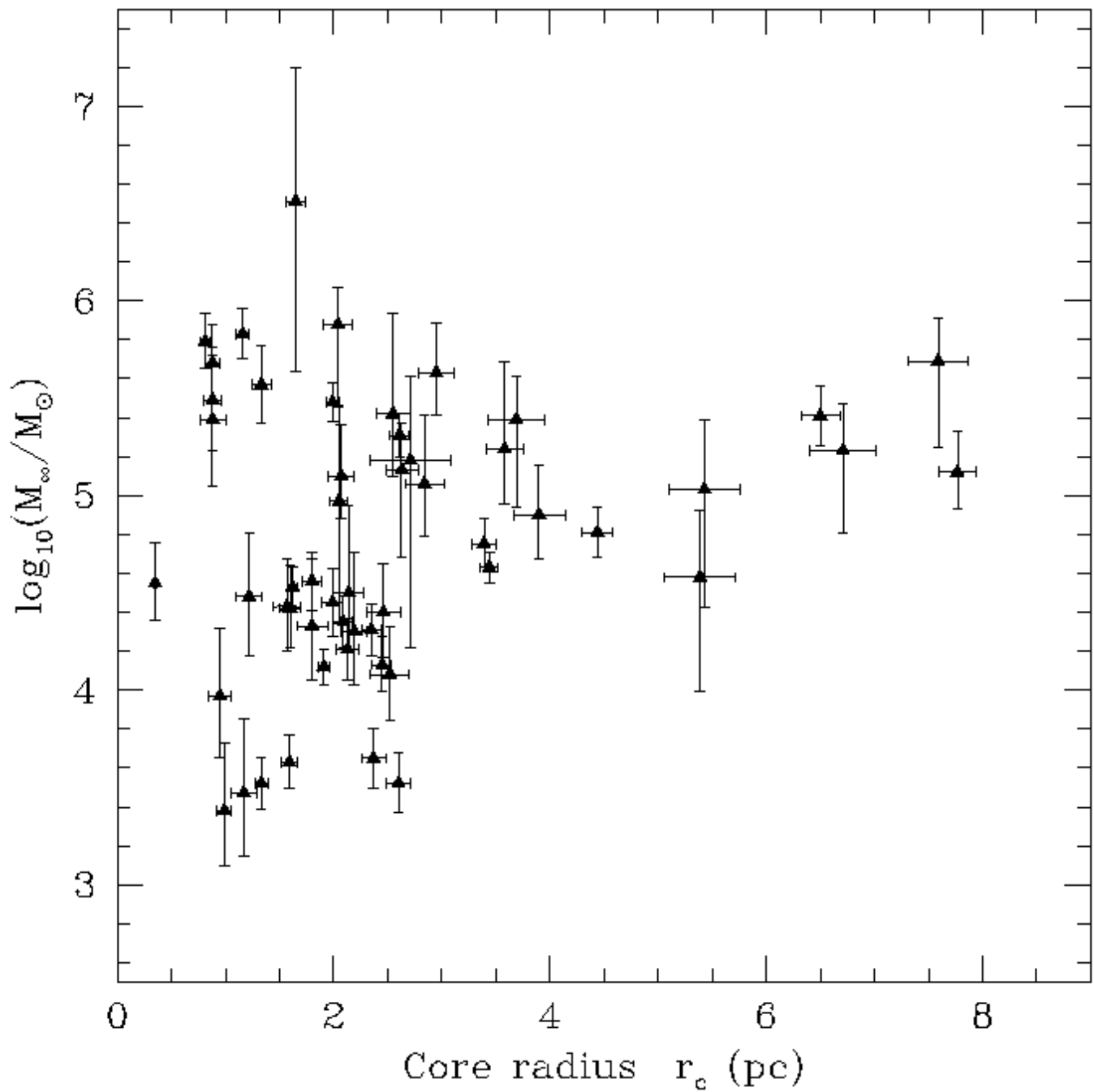
\end{document}